\documentclass[twocolumn]{aastex631}

\usepackage{amsmath} 
\usepackage{makecell} 

\newcommand{\ee}[1]{\mbox{${} \times 10^{#1}$}}    
\newcommand{\msun}{\mbox{M$_\odot$}}               

\newcommand{\coo}{\mbox{$^{13}$CO}}
\newcommand{\cotw}{\mbox{$^{12}$CO}}

\newcommand{\kms}{km s$^{-1}$}
\newcommand{\kkms}{K km s$^{-1}$}

\newcommand{\hh}{\mbox{{\rm H}$_2$}}
\newcommand{\hcop}{\mbox{{\rm HCO}$^+$}}

\newcommand{\hcn}{\mbox{{\rm HCN}}}

\newcommand{\hii}{\mbox{\ion{H}{2}}}
\newcommand{\jj}[2]{\mbox{$J = #1\rightarrow#2$}}
\newcommand{\gal}{\mbox{Galactocentric }}

\newcommand{\mdense}{\mbox{$M_{\rm dense}$}}

\newcommand{\sfr}{\mbox{\rm SFR}}

\newcommand{\av}{\mbox{$A_{\rm V}$}}

\newcommand{\ammonia}{\mbox{{\rm NH}$_3$}}

\newcommand{\lhcn}{\mbox{$L({\rm HCN})$}}          

\newcommand{\innclouda}{\mbox{G034.158+00.147}}
\newcommand{\inncloudb}{\mbox{G034.997+00.330}}
\newcommand{\inncloudc}{\mbox{G036.459-00.183}}
\newcommand{\inncloudd}{\mbox{G037.677+00.155}}
\newcommand{\inncloude}{\mbox{G045.825-00.291}}
\newcommand{\inncloudf}{\mbox{G046.495-00.241}}

\newcommand{\rg}{\mbox{$R_{\rm G}$}}
\newcommand{\rgal}{\mbox{$R_{\rm G}$}}
\newcommand{\rgalsun}{\mbox{$R_{\rm G, \sun}$}}
\newcommand{\gammasun}{\mbox{$\gamma_\sun$}}

\shorttitle {Outer Galaxy}    
\shortauthors{Patra et al.}   

\graphicspath{{./}{figures/}}

\begin{document}

\title{Variation of Dense Gas Mass-Luminosity conversion factor with metallicity in the Milky Way}

\correspondingauthor{Sudeshna Patra}
\email{inspire.sudeshna@gmail.com}
\author[0000-0002-3577-6488]{Sudeshna Patra}
\affiliation{Department of Physics, Indian Institute of Science Education and Research Tirupati, Yerpedu, Tirupati - 517619, Andhra Pradesh, India}

\correspondingauthor{Neal J. Evans II}
\email{nje@astro.as.utexas.edu}
\author[0000-0001-5175-1777]{Neal J. Evans II}
\affiliation{Department of Astronomy, The University of Texas at Austin, 2515 Speedway, Stop C1400 Austin, Texas 78712-1205, USA}

\author[0000-0003-2412-7092]{Kee-Tae Kim}
\affiliation{Korea Astronomy and Space Science Institute, 776 Daedeokdae-ro, Yuseong-gu, Daejeon 34055, Republic of Korea}
\affiliation{University of Science and Technology, Korea (UST), 217 Gajeong-ro, Yuseong-gu, Daejeon 34113, Republic of Korea}

\author[0000-0002-3871-010X]{Mark Heyer}
\affiliation{Department of Astronomy University of Massachusetts Amherst, Massachusetts 01003, USA}

\author[0000-0003-3869-6501]{Andrea Giannetti}
\affiliation{INAF - Istituto di Radioastronomia, Via P. Gobetti 101, I-40129 Bologna, Italy}

\author[0000-0002-9120-5890]{Davide Elia}
\affiliation{INAF-IAPS, Via del Fosso del Cavaliere 100, I-00133 Roma, Italy}

\author[0000-0003-4908-4404]{Jessy Jose}
\affiliation{Department of Physics, Indian Institute of Science Education and Research Tirupati, Yerpedu, Tirupati - 517619, Andhra Pradesh, India}

\author[0000-0002-5094-6393]{Jens Kauffmann}
\affiliation{Haystack Observatory MIT, 99 Milstone Rd. Westford, MA 01886 USA}

\author[0000-0002-9431-6297]{Manash R. Samal}
\affiliation{Astronomy \& Astrophysics Division, Physical Research Laboratory,
Ahmedabad, 380009, Gujarat, India}

\author[0000-0001-8913-925X]{Agata Karska}
\affiliation{Max Planck Institute for Radio Astronomy, Auf dem Hügel 69, 53121 Bonn, Germany}
\affiliation{Argelander-Institut fur Astronomie, Universität Bonn, Auf dem Hügel 71, 53121 Bonn, Germany}
\affiliation{Institute of Astronomy, Faculty of Physics, Astronomy and Informatics, Nicolaus Copernicus University, ul. Grudziadzka 5, 87-100 Toruń, Poland}

\author[0000-0001-7151-0882]{Swagat R. Das}
\affiliation{Departamento de Astronomia, Universidad de Chile, Las Condes, 7591245 Santiago, Chile}

\author{Gyuho Lee}
\affiliation{Korea Astronomy and Space Science Institute, 776 Daedeokdae-ro, Yuseong-gu, Daejeon 34055, Republic of Korea}
\affiliation{University of Science and Technology, Korea (UST), 217 Gajeong-ro, Yuseong-gu, Daejeon 34113, Republic of Korea}

\author[0000-0001-8467-3736]{Geumsook Park}
\affiliation{Telepix Co., Ltd., 17, Techno 4-ro, Yuseong-gu, Daejeon 34013, Republic of Korea}
\affiliation{Research Institute of Natural Sciences, Chungnam National University, 99 Daehak-ro, Yuseong-gu, Daejeon 34134, Republic of Korea}
\affiliation{Korea Astronomy and Space Science Institute, 776 Daedeokdae-ro, Yuseong-gu, Daejeon 34055, Republic of Korea}



\begin{abstract}

\hcn\ and \hcop\ are the most common dense gas tracers used both in the Milky Way and external galaxies. 
The luminosity of \hcn\ and \hcop\ \jj10 lines are converted to a dense gas mass by the conversion factor, $\alpha_{Q}$. 
Traditionally, this $\alpha_{Q}$ has been considered constant throughout the Galaxy and in other galaxies, regardless of the environment.
We analyzed 17 outer Galaxy clouds and 5 inner Galaxy clouds with metallicities ranging from 0.38 Z$_{\odot}$ to 1.29 Z$_{\odot}$. Our analysis indicates that $\alpha_{Q}$ is not constant; instead, it varies with metallicity. 
The metallicity-corrected $\alpha_{Q}$ derived from the \hcn\ luminosity of the entire cloud is almost three times higher in the outer Galaxy than in the inner galaxy. 
In contrast, \hcop\ seems less sensitive to metallicity. 
We recommend using the metallicity-corrected dense gas conversion factors 
$\alpha^{'}_{\rm tot, Gas}(\rm HCN) = 19.5^{+5.6}_{-4.4} Z^{(-1.53 \pm 0.59)}$ and  
$\alpha^{'}_{\rm tot, Gas}(\rm HCO^{+}) = 21.4^{+5.5}_{-4.4}  Z^{(-1.32\pm0.55)}$ for extragalactic studies.
Radiation from nearby stars has an effect on the conversion factor of similar magnitude as that of the metallicity.
If we extend the metallicity-corrected scaling relation for \hcn\ to the Central Molecular Zone, the value of $\alpha(\hcn)$ becomes $1/3$ to $1/2$ of the local values. 
This effect could partially account for the low star formation rate per dense gas mass observed in the CMZ.

\end{abstract}


\section{Introduction} \label{sec:intro}

Studies of molecular clouds within the Milky Way found that most of the dense cores and young stellar objects (YSOs) are associated with regions of high surface density, measured by visual extinction, $\av>8 \ \mathrm{mag}$ \citep{2010ApJ...723.1019H, 2010ApJ...724..687L, 2012ApJ...745..190L}. 
Also, \citet{2016ApJ...831...73V} found a nearly linear relationship between the star formation rate (SFR) and the mass of dense gas, as determined by millimeter continuum emission data from the Bolocam Galactic Plane Survey (BGPS, \citealt{2013ApJS..208...14G}), particularly for distant and massive clouds.
Some studies have indicated a decreased star formation efficiency for the most massive dense clumps \citep{2022MNRAS.516.4245W},
but
\cite{2024A&A...688A.163M} found that the star formation efficiency in dense gas is nearly constant. Their analysis of all southern molecular clouds within 3 kpc shows that the amount of dense gas is the primary indicator of the star formation rate.
On extragalactic scales, a pioneering study by \cite{2004ApJS..152...63G} identified a robust linear relationship between far-infrared luminosities, related  to the SFR, and \hcn\ line luminosities, which serve as measures of dense gas content. 
A consistent observation has emerged from several studies investigating entire galaxies 
\citep{2004ApJS..152...63G, 2014ApJ...784L..31Z, 2015ApJ...810L..14L} and spatially resolved nearby galaxies 
\citep{2015ApJ...810..140C, 2018ApJ...860..165T, 2019ApJ...880..127J,2022ApJ...930..170H, 2023MNRAS.521.3348N}, 
utilizing various dense gas tracers such as \hcn, \hcop, CS, and high-$J$ \cotw\ lines.
These studies suggest that the denser parts of the molecular clouds serve as the direct sources of fuel for the process of star formation. 
The correlation between star formation rate (\sfr) and the luminosity of \hcn\ \jj10\ emission (\lhcn) is tighter than the correlation of \sfr\ with \cotw\ \jj10\ emission, supporting the idea that dense parts of molecular clouds are more likely sites of star formation.

There are three distinct methods in use to measure the mass of gas associated with star formation: 
a surface density criterion ($\av > 8 \ \mathrm{mag}$) used in nearby clouds \citep{2010ApJ...723.1019H, 2010ApJ...724..687L, 2012ApJ...745..190L}, 
a millimeter continuum emission method used in more distant clouds in our Galaxy \citep{2016ApJ...831...73V, 2024MNRAS.528.4746U}, and a line luminosity method (usually HCN, but also \hcop ) \citep{Wu:2005, 2020ApJ...894..103E, 2022AJ....164..129P} used most often in other galaxies. 
While the first method is based on column density, the last two methods are more sensitive to volume density because of the limited spatial scales available to ground-based millimeter continuum observations or the excitation criterion for molecules with large dipole moments \citep{2015PASP..127..299S}.
Millimeter continuum dust emission is an optically thin tracer of gas column density in dense molecular clouds, effectively identifying molecular clumps where star formation may occur. However, due to the methods used to remove atmospheric fluctuations, the BGPS emission traces volume density, with the characteristic density probed decreasing with distance, averaging between 
$5 \times 10^3$ to $1 \times 10^{4}$  $\rm cm^{-3}$ for distances of 2-6 kpc \citep{2011ApJ...741..110D}.
We will compare these three methods in this paper.

Clouds in our Galaxy provide opportunities to study in greater detail the relation of these dense gas tracers to the mass of dense gas. 
\citet{Wu:2005} showed that the relation between star formation rate and the luminosity of HCN seen in other galaxies persists into dense cores in Galactic clouds. However, subsequent studies that mapped \hcn\ emission over larger regions found that a very substantial fraction of \lhcn\ arises in lower density parts of the cloud 
\citep{2017A&A...604A..74S, 2017A&A...605L...5K, 2017A&A...599A..98P, 2020ApJ...894..103E, 2020MNRAS.497.1972B, 2022AJ....164..129P}.
In part, the ability of relatively diffuse parts of the cloud to produce significant contributions to \lhcn\  arises from the good sensitivity of newer receivers and the fact that the area and mass of molecular clouds is dominated by low-density gas.
Trapping of photons in optically thick lines and other excitation effects allows gas at densities far below the so-called critical density to produce detectable emission \citep{2020ApJ...894..103E, 2022AJ....164..129P, 2015PASP..127..299S}. In addition, photo-ionization by ultraviolet photons  can cause electron collisions to contribute to the excitation of \hcn\
\citep{2017ApJ...841...25G}.
Taken together, these results indicate that the relationship between \lhcn\ and the mass of dense gas is more complicated than usually assumed.

A complication that has not been explored so far is the likelihood that the \hcn\ abundance and hence emission will be affected by metallicity.
It has become increasingly recognized that the mass-luminosity conversion factor for \cotw\ is not constant, but varies with metallicity \citep{2020ApJ...903..142G, 2022ApJ...931...28H}.
In extragalactic studies, the dependency of the \cotw\ conversion factor on metallicity has already been explored \citep{2013ARA&A..51..207B, 2020ApJ...901L...8S, 2021MNRAS.504.2360J, 2021ApJ...907...29C, 2024ApJ...968...97L}, and a super-linear dependence on metallicity is commonly assumed.  
This raises an important question: What about the effect of metallicity on the mass-luminosity conversion factor for dense gas as traced by \hcn\ or  \hcop?
This question is the central focus of this paper.

The outer parts of the Milky Way provide a convenient laboratory for answering this question, as there is a well established gradient of decreasing metallicity with increasing Galactocentric radius 
\citep{2022MNRAS.510.4436M, 2024ApJ...973...89P}.
It also provides a unique opportunity to study the impact of metallicity on star formation at spatial resolution sufficient to resolve individual clumps. 
The metallicity declines in the outer Galaxy \citep{2017MNRAS.471..987E, 2018MNRAS.478.2315E} extending down to values relative to solar (denoted $Z$) as low as 0.3, which overlaps with that of the Large Magellanic Cloud (LMC)($Z\sim0.3-0.5$ Z$_\odot$; \citealt{1992ApJ...384..508R}). The abundance of \hcn\ appears to decline in a similar way in the outer Galaxy and the LMC \citep{2024arXiv241104451S}.

By combining observations of 17 outer Galaxy star-forming regions with previous studies of regions in the Solar neighborhood and the inner Galaxy, we will examine the effects of metallicity on emission from \hcn\ and \hcop.

This paper is organized as follows: Sections \ref{sec:sample} and \ref{sec:obs} describe the target and the observation dataset used, respectively. In Section \ref{sec:analysis}, we estimate the dense gas mass conversion factors for \hcn\ and \hcop\ using various methods. Section \ref{solar} summarizes the mass-luminosity conversion factors in the Solar Neighborhood based on information from the literature. In Section \ref{sec:discussion}, we discuss the results and their implications, followed by the conclusion in Section \ref{sec:conclude}. 


\section{Sample} \label{sec:sample}
We have selected 17 clouds in the outer part of the Milky Way, i.e., with the \gal\ distance \rg\  greater than the Solar radius ($\rgalsun=8.178 \pm 0.013_{stat.} \pm 0.022_{sys}$, \citealt{2019A&A...625L..10G}). 
Among them, 10 clouds are newly observed and 7 clouds are taken from our previous study \citep{2022AJ....164..129P}.    
These 17 clouds exhibit diversity in their physical characteristics such as size, mass, gas density, metallicity and the associated massive stars, as well as distance, as detailed in Table \ref{tab:cloud details}. 
All these regions are associated with young clusters, most of which are analyzed in \citealt{2024ApJ...970...88P}, and the clusters' ages range from $0.9$ to $2.1$ Myr.
 
Our selected star-forming regions are situated in the outer Galaxy with Galactic longitude in the range $97.45^{\circ} \leq l \leq 196.83^{\circ}$ and \rg\ ranging from 9.4 to 17 kpc.
The spatial distribution plot of the star-forming regions of our sample  within the Milky Way  (shown in \citealt{2023pcsf.conf...63P}) indicates that approximately half are located in or near spiral arms and half are not.
The heliocentric distance ($D$) distribution among our sample indicates a varied landscape: four targets lie within 2 kpc of Sun, while five targets cover the $2< D <4$ kpc range. Moreover, four targets fall within the $4<D<6$ kpc range from the Sun, and an additional four targets are located beyond 6 kpc from the Sun.
The distance information based on Gaia EDR3 (Early Data Release 3) parallaxes data are taken from \cite{2022MNRAS.510.4436M} except Sh2-142, Sh2-148, Sh2-242 and Sh2-269, for which we performed the following procedure for uniformity.
For these targets, we first identified the ionizing sources (O, early B or Wolf-Rayet-type stars) and we placed their coordinates in the Gaia EDR3 catalogs to finally obtain their parallax-based distances \citep{2018AJ....156...58B, 2021AJ....161..147B} information.
We have taken the metallicity information i.e., the oxygen abundance with respect to hydrogen ($\mathrm{12+log[O/H]}$) without correction for temperature variations from \cite{2022MNRAS.510.4436M} and \cite{2018PASP..130k4301W}. 
Table \ref{tab:cloud details} also includes the $\mathrm{12+log[O/H]}$ values for each target, alongside their corresponding references.
Table \ref{tab:cloud details} also includes an estimate for the total luminosity of the associated stars, as discussed in \S \ref{dispersion}. 

We have also added five inner Galaxy clouds from \citet{2020ApJ...894..103E} to include regions with $Z > 1$ in our study. 
We have excluded \innclouda\ from all the analysis, because this cloud is highly affected by the self-absorption of \hcn\ and \hcop\ (see the spectra of \hcn\ and \hcop\ in Figure A.1 of \citealt{2020ApJ...894..103E}).
Distances of these five clouds are also taken from \citet{2020ApJ...894..103E}. 
To determine the metallicity of these inner Galaxy clouds, we utilized the radial gradient equation obtained for the \rg\ range of $4.88-17$ kpc from \cite{2022MNRAS.510.4436M}. 
This equation delineates the oxygen abundance gradient with a reported slope of $\mathrm{-0.044 \pm 0.009 \ dex \ kpc^{-1}}$ if temperature variation in the \hii~ region is negligible.
%
%
This slope is similar to that observed in radio recombination lines \citep{2011ApJ...738...27B}  and is also consistent with the value observed in double-mode pulsating Cepheids, both at $-0.045\ \rm dex \ kpc^{-1}$ \citep{2018A&A...618A.160L}.
As discussed by  \citet{2022ApJ...929L..18E}, this dependence is both the most conservative (in the sense of the weakest dependence on \rgal) and the most well determined over a wide range of \rgal.

All the clouds were mapped in \hcn\ and \hcop\ (see more details in Section \ref{sec:obs}), and the mapping areas are listed in Table \ref{tab:cloud details}.

\begin{deluxetable*}{l r r c c c c c l}
    \tablenum{1}
    \tabletypesize{\footnotesize}
    \tablecaption{Target Details \label{tab:cloud details}}
    \tablewidth{0pt}
    \tablehead{ 
    \colhead{Target} &  \colhead{$l$}   & \colhead{$b$}   & \colhead{\rg}   & \colhead{$D$}           & \colhead{\mbox{12+log(O/H)} }  &  \colhead{Log $L_{*}$}    &  \colhead{Map Size}   \\
    \colhead{}       & \colhead{(deg)}  & \colhead{(deg)} & \colhead{(kpc)} & \colhead{(kpc)}         & \colhead{[Ref]}       &   \colhead{(L$_{\odot}$)}   &  \colhead{($\ \ '\ \times \ \ '$)}  
    }
    \startdata
        & \multicolumn{5}{c}{New Outer Galaxy Targets in this study}\\
        \hline
        Sh2-128 \tablenotemark{$\ddagger$}     & $97.4485$     & $3.1798$   & $11.78^{+0.78}_{-0.64}$    & $7.27^{+1.10}_{-0.82}$      &  $8.21^{+0.05}_{-0.05}$ [1]   & 5.14  & $27 \times 17$    \\
        Sh2-132 \tablenotemark{$\dagger$}      & $102.9270$    & $-0.6563$  & $10.21^{+0.28}_{-0.26}$    & $4.53^{+0.29}_{-0.27}$      &  $8.37^{+0.14}_{-0.15}$ [1]   & 5.81  & $52 \times 30$    \\
        Sh2-142 \tablenotemark{$\dagger$}      & $107.2209$    & $-0.9265$  & $10.60^{+0.09}_{-0.08}$    & $2.96^{+0.18}_{-0.15}$      &  $8.44^{+0.08}_{-0.08}$ [2]   & 5.54  & $15 \times 15$    \\
        Sh2-148 \tablenotemark{$\dagger$}      & $108.3675$    & $-1.0552$  & $\ 9.79^{+0.12}_{-0.10}$    & $3.36^{+0.20}_{-0.17}$     &  $8.30^{+0.08}_{-0.08}$ [2]   & 5.03  & $90 \times 48$    \\
        Sh2-156\tablenotemark{*}      & $110.1027$    & $0.0153$   & $\ 9.40^{+0.21}_{-0.22}$    & $2.56^{+0.22}_{-0.19}$     &  $8.34^{+0.10}_{-0.10}$ [1]   & 5.14  & $32 \times 44$    \\
        Sh2-209 \tablenotemark{$\ddagger$}     & $151.5999$    & $-0.3602$  & $17.00^{+0.70}_{-0.70}$    & $9.33^{+0.70}_{-0.70}$      &  $8.08^{+0.26}_{-0.12}$ [1]   & 5.43  & $12 \times 20$    \\
        Sh2-242\tablenotemark{*}      & $182.1487$    & $0.3028$   & $10.58^{+0.04}_{-0.03}$    & $1.52^{+0.04}_{-0.03}$      &  $8.59^{+0.06}_{-0.06}$ [2]   & 4.47  & $46 \times 20$    \\
        Sh2-266\tablenotemark{*}      & $195.7051$    & $-0.1344$  & $12.69^{+0.62}_{-0.64}$    & $4.60^{+0.53}_{-0.51}$      &  $8.19^{+0.16}_{-0.16}$ [1]   & 4.19  & $21 \times 18$    \\
        Sh2-269 \tablenotemark{$\dagger$}      & $196.4510$    & $-1.6606$  & $12.57^{+0.25}_{-0.21}$    & $4.06^{+0.37}_{-0.31}$      &  $8.42^{+0.08}_{-0.08}$ [2]   & 4.89  & $14 \times 14$    \\
        Sh2-270 \tablenotemark{$\ddagger$}     & $196.8288$    & $-3.1458$  & $16.10^{+1.40}_{-1.40}$    & $8.08^{+1.29}_{-1.29}$      &  $8.09^{+0.26}_{-0.26}$ [1]   & 4.47  & $11 \times 15$    \\
        %
        \hline
        & \multicolumn{5}{c}{Outer Galaxy Targets from \cite{2022AJ....164..129P}}\\
        \hline
        Sh2-206 \tablenotemark{$\ddagger$}     &  $150.5886$    &  $-0.8570$ & $10.88^{+0.25}_{-0.26}$    & $2.96^{+0.17}_{-0.15}$     & $8.37^{+0.08}_{-0.08}$  [2]     &  5.67  & $15 \times 15$    \\
        Sh2-208 \tablenotemark{$\ddagger$}     &  $151.3098$    &  $1.9045$  & $11.87^{+0.37}_{-0.34}$    & $4.02^{+0.27}_{-0.25}$     & $8.43^{+0.07}_{-0.07}$  [2]     &  4.57  & $10 \times 20$    \\
        Sh2-212 \tablenotemark{$\dagger$}      &  $155.3375$    &  $2.6345$  & $14.76^{+1.30}_{-1.30}$    & $6.65^{+1.36}_{-1.26}$     & $8.34^{+0.26}_{-0.13}$ [1]      &  5.14  & $13 \times 14$    \\
        Sh2-228\tablenotemark{*}      &  $169.1432$    &  $-1.0475$ & $10.72^{+0.19}_{-0.19}$    & $2.56^{+0.09}_{-0.09}$     & $8.39^{+0.07}_{-0.07}$  [2]     &  4.96  & $11 \times 33$    \\
        Sh2-235 \tablenotemark{$\dagger$}      &  $173.6682$    &  $2.7799$  & $9.85^{+0.16}_{-0.17}$     & $1.66^{+0.07}_{-0.07}$     & $8.42^{+0.07}_{-0.07}$  [1]     &  4.68  & $30 \times 30$    \\
        Sh2-252\tablenotemark{*}      &  $189.8418$    &  $0.3156$  & $10.09^{+0.12}_{-0.11}$    & $1.92^{+0.12}_{-0.11}$     & $8.51^{+0.08}_{-0.08}$   [2]    &  5.23  & $30 \times 30$    \\
        Sh2-254\tablenotemark{*}      &  $192.7935$    &  $0.0290$  & $10.13^{+0.21}_{-0.21}$    & $1.96^{+0.12}_{-0.09}$     & $8.41^{+0.10}_{-0.10}$   [1]    &  4.89  & $50 \times 34$    \\
        \hline
        & \multicolumn{5}{c}{Inner Galaxy Targets from \cite{2020ApJ...894..103E}}\\
        \hline
        G034.997+00.330\tablenotemark{*} &  $35.0697$    & $0.3682$   & $5.99^{+0.24}_{-0.24}$     & $10.43^{+0.38}_{-0.41}$     & $8.60^{+0.14}_{-0.14}$ [3]   &  6.46   & $10 \times 20$   \\
        G036.459-00.183\tablenotemark{*} &  $36.5175$    & $-0.1529$  & $5.30^{+0.24}_{-0.21}$     & $8.68^{+0.56}_{-0.60}$      & $8.63^{+0.14}_{-0.14}$ [3]   &  5.83   & $14 \times 22$   \\
        G037.677+00.155\tablenotemark{*} &  $37.6765$    & $0.0201$   & $5.01^{+0.11}_{-0.12}$     & $6.60^{+0.13}_{-0.14}$      & $8.64^{+0.14}_{-0.14}$ [3]   &  5.73   & $10 \times 26$   \\
        G045.825-00.291\tablenotemark{*} &  $45.8250$    & $-0.2910$  & $6.43^{+0.20}_{-0.17}$     & $8.31^{+0.46}_{-0.62}$      & $8.58^{+0.15}_{-0.15}$ [3]   &  6.26   & $30 \times 30$   \\
        G046.495-00.241\tablenotemark{*} &  $46.4000$    & $-0.2410$  & $6.25^{+0.21}_{-0.18}$     & $3.71^{+0.68}_{-0.60}$      & $8.58^{+0.15}_{-0.15}$ [3]   &  5.15   & $20 \times 10$   \\
        \hline
    \enddata
    \tablenotemark{$l$= Galactic Longitude, $b$= Galactic Latitude, \rg=Galactocentric Distance, }
    \tablenotemark{$D$=Heliocentric Distance, $12+log(O/H)$=Metallicity, $L_{*}$=Stellar Luminosity}
    \tablecomments{[1] \cite{2022MNRAS.510.4436M}; [2] \cite{2018PASP..130k4301W}; [3] Calculated by using radial gradient equation from \cite{2022MNRAS.510.4436M};    \\
    \textbf{Data availability:} \tablenotetext{\ddagger}{\cotw, \coo, \hcn, \hcop} \tablenotetext{\dagger}{\cotw, \coo, \hcn, \hcop, BGPS} \tablenotetext{*}{\cotw, \coo, \hcn, \hcop, BGPS, \textit{Herschel} } }
\end{deluxetable*}

\section{Observations and Data sets} \label{sec:obs}

\subsection{Molecular line observations with the TRAO}

We conducted observations of the ten outer Galaxy clouds using the \hcn (\jj10, 88.631847 GHz) and \hcop (\jj10, 89.1885296 GHz) lines simultaneously. Additionally, we acquired data for eight of these targets in the \coo (\jj10) and \cotw (\jj10) lines. All these observations were obtained at the Taeduk Radio Astronomy Observatory (TRAO) between January and March 2021. The TRAO facility is equipped with the SEcond QUabbin Optical Image Array (SEQUOIA-TRAO), a multi-beam receiver featuring a 16-pixel array arranged in a $4 \times 4$ configuration, operating within the 85-115 GHz range. Its back-end utilizes a 2G FFT (Fast Fourier Transform) Spectrometer. 
Comprehensive details regarding the instrumental specifications are available in \citet{2019JKAS...52..227J}. 
The main-beam sizes of TRAO 14-m telescope at frequencies of 86.243 GHz, 110.201 GHz, and 115.271 GHz are $57 \pm 2$\arcsec, $49\pm1$\arcsec, and $47\pm1$\arcsec\ respectively, as detailed in \citet{2019JKAS...52..227J}. 
Furthermore, the main-beam efficiency ($\rm \eta_{mb}$) of TRAO is characterized by values of $45\pm2 \%$, $46\pm2 \%$, and $41\pm2 \%$ at the aforementioned frequencies. We used the main-beam corrected temperature ($\rm T_{X}= T_{A}^{*}/\eta_{mb}$) of the species $X$ (\hcn, \hcop) for luminosity calculations.
The individual SEQUOIA beams maintain consistency in beam size and efficiency, exhibiting variations of only a few arcseconds and a few percent respectively. 
Moreover, these beams demonstrate a high degree of circularity, with beam efficiencies fluctuating by less than $3\%$ with changes in elevation angle.

Using the OTF (On-The-Fly) mode, we mapped the clouds, determining the areas for each target based on their $^{13}$CO (\jj10) line emission maps.
We estimated the mapping area for \hcn\ and \hcop\ using a threshold value of 5 times the RMS noise in the \coo\ map. All targets were mapped in Galactic coordinates, and the mapped areas for each target are listed in Table \ref{tab:cloud details}.
During \hcn\ and \hcop\ observations, the system temperature was around $ 180\  \rm K$, and we used $\rm SiO$ maser sources at 86 GHz every 3 hours for telescope pointing and focusing. 
We applied a boxcar function to smooth the data, achieving a velocity resolution of approximately 0.2 \kms, which led to an RMS sensitivity of 0.13-0.18 K on the main-beam corrected temperature $( \rm T_{X})$ scale for all targets.
System temperatures during \cotw\ and \coo\ observations were approximately $500\  \rm K$ and $ 250\ \rm K$ respectively. 
Total observing time was approximately $ 500\ \rm hrs$ of observations.

\begin{deluxetable*}{l| c c c c}
    \tablenum{2}
    \tabletypesize{\footnotesize}
    \tablecaption{TRAO Telescope Parameters \label{tab:telescope}}
    \tablewidth{0pt}
    \tablehead{
    \colhead{\makecell[l]{Molecular Lines \\ }} & \colhead{\cotw} & \colhead{\coo} & \colhead{\hcn} & \colhead{\hcop} }
    \startdata
    Frequency [GHz]                           &   115.271204  & 110.201353   & 88.631847 & 89.188526  \\
    Main-beam Size ($\rm \theta_{M}$) [\arcsec]   &   $47\pm1$    & $49\pm1$     & $57\pm2$  & $57\pm2$   \\
    Main-beam Efficiency ($\rm \eta_{B}$) [\%]    &   $41\pm2$    & $46\pm2$     & $45\pm2$  & $45\pm2$   \\
    System Temperature ($\rm T_{sys}$) [K]        &   550         & 250          & 180       & 180        \\
    RMS Sensitivity in $\rm T_{X}$ [K] &   0.73        & 0.43         & 0.13-0.18  & 0.13-0.18 \\
    Velocity Resolution [\kms]                &   0.16        & 0.31         & 0.20       & 0.20       \\
    \tableline
    \enddata
\end{deluxetable*}

To process the data, we executed the following steps (same as \citealt{2022AJ....164..129P}).
First, we utilized the OTFTOOL to read the raw 
OTF data for each map of each tile and subsequently converted them into the CLASS format map after subtracting a first-order baseline.
We also applied noise weighting and opted for an output cell size of $20\arcsec$ using OTFTOOL.
We did not apply any convolution function during the mapping process, ensuring that the effective beam size matched the main-beam size.
We proceeded with further data reduction using the CLASS format of GILDAS, available at \url{https://www.iram.fr/IRAMFR/GILDAS/}. 
Comprehensive details of the reduction process are outlined in \S Appendix \ref{appenA}, Table \ref{tab:reduction}. 
Finally, we subtracted a second-order polynomial baseline and converted the resulting files to FITS format using GREG.
For all analysis, we corrected the antenna temperature to the main-beam temperature by using the main beam efficiencies, given above.

 \subsection{Millimeter Dust Continuum Emission} 
 We utilized 1.1 mm continuum emission data from the Bolocam Galactic Plane Survey (BGPS) conducted with Bolocam on the Caltech Submillimeter Observatory. 
 The survey covers four specific regions unevenly: IC1396, a region toward the Perseus arm, W3/4/5, and Gem OB1, with an effective resolution of $33\arcsec$. 
 Due to non-uniform coverage, not all targets in the outer Galaxy regions have BGPS data, as indicated in Table \ref{tab:cloud details}.
 Sh2-128, Sh2-206, Sh2-208, Sh2-209 and Sh2-270  are not covered by the BGPS survey.
 The mask files of the sources are  in  V2.1 table\footnote{\url{https://irsa.ipac.caltech.edu/data/BOLOCAM_GPS/}}, where pixel values above the $2\sigma$ ($\sigma$ is the rms noise of the BGPS image) threshold are cataloged, otherwise set to zero \citep{2013ApJS..208...14G}. 
 Bolocat V2.1 provides photometry for 8594 sources at aperture radii of $20\arcsec$, $40\arcsec$, and $60\arcsec$, alongside integrated flux densities.

\subsection{Far Infrared data from Herschel}
We utilized the dust temperature maps obtained from \textit{Herschel} data \citep{2017MNRAS.471.2730M}, which are accessible on the Herschel infrared Galactic Plane (Hi-GAL) survey site\footnote{\url{http://www.astro.cardiff.ac.uk/research/ViaLactea/}} \citep{2010PASP..122..314M}. 
These maps are created using the PPMAP procedure \citep{2015MNRAS.454.4282M} applied to the continuum data ranging from 70 to 500 $\rm \mu$m from the Hi-GAL survey. 
Note that the Hi-GAL survey does not cover all the targets in our list. The data availability for each source is indicated in the notes for Table \ref{tab:cloud details}.

\section{Estimation of Mass-Luminosity Conversion Factor}\label{sec:analysis}

Our aim is to obtain the mass-luminosity conversion factor which is used to estimate the mass of dense gas ($ M_{\rm dg}$) from the line luminosity of a molecule, generically labeled Q. 
The conversion factor is defined as the ratio between the gas mass and the line luminosity: 
\begin{equation}
    \alpha_{Q} = M_{\rm dg}/L_{Q},     
\end{equation}
where $\alpha_{Q}$ is typically measured in units of $\mathrm{\msun (K \ km \ s^{-1}\ pc^{2})^{-1}}$. 
Here, we computed $M_{dg}$ using two approaches, one based on gas analysis and the other one using dust. 
For $L_{Q}$, we used the line luminosities from \hcn\ and \hcop, widely utilized tracers of dense gas among the extragalactic community \citep{2016ApJ...822L..26B,2019ApJ...880..127J}.

Two categories of mass-luminosity conversion factors exist. One uses the tracer luminosity, $L_{\rm X,tot}$, obtained from a map of the whole molecular cloud, where $X$ indicates the tracer. This is generally the only measure available for extragalactic observations, in which most clouds are not resolved. It provided the basis for the original 
\citet{2004ApJS..152...63G} relation. 
In contrast, Galactic clouds have large angular extent, and most early work (e.g., \citealt{Wu:2005, 2010ApJS..188..313W})
mapped only the regions of relatively strong emission. 
Because substantial luminosity can arise from weak, extended emission (e.g., \citealt{2020ApJ...894..103E, 2022AJ....164..129P, 2023A&A...679A...4S}), estimates of luminosity based on partial maps can be substantially smaller, leading to larger mass-luminosity conversion factors. To distinguish these,
\citet{2020ApJ...894..103E}
introduced the notation $L_{\rm X,in}$ to represent the luminosity inside a contour of gas above some level of volume or column density, either $\av = 8$ mag or the region of mm-wave emission in the BGPS survey. This measure is more relevant to Galactic clouds, for which most maps of HCN and \hcop\ are incomplete. 
The ratio $f_l = L_{\rm X,in}/L_{\rm X,tot}$ ranged from 0 to 0.54, with a mean of $0.172\pm 0.185$ for HCN and the $\av = 8$ mag criterion.
\citet{2022AJ....164..129P}
found somewhat higher ratios for six outer Galaxy clouds ($f_l \equiv L_{\rm X,in}/L_{\rm X,tot} = 0.343\pm 0.225$), but the general trend is the same: on average, most of the line luminosity arises from outside the boundaries used to measure dense gas mass. In fact, much of the luminosity can arise from regions where individual spectra do not detect lines
(Fig. 11 in \citealt{2020ApJ...894..103E}).

We compare the luminosities of \hcn\ and \hcop\ \jj10 transitions to other conventional tracers of dense gas, in particular extinction thresholds and millimeter-wave emission from dust. However, extinction maps based on background stars are not available for distant clouds, so we translate the extinction threshold to one of gas column density.
Visual extinction (\av) is related to the column density of molecular hydrogen (\hh) in molecular clouds through the equation $N_{\rm H_{2}}= 1 \times 10^{21} \av \ \rm cm^{-2}$ leading to a dense gas criterion of 8\ee{21} cm$^{-2}$ to correspond to the $\av = 8$ mag criterion locally \citep{2010ApJ...723.1019H, 2010ApJ...724..687L, 2012ApJ...745..190L}.
To determine gas column density, we used maps derived from \coo\ and \cotw. We refer to this method as the `Gas based analysis'.

For the dust emission, we utilize the mask file of the BGPS data, which indicates the presence of gas with relatively high volume density \citep{2011ApJ...741..110D}. 
We refer to this method as the `Dust based analysis'.
We conducted the analysis both with and without accounting for metallicity correction. 
We also calculated $\alpha_{Q}$ both for line luminosities inside the dense gas regions for comparison to other resolved studies and for the total line luminosity over the whole cloud, more relevant to extragalactic studies. 
Using line luminosities from \hcn\ and \hcop\ separately, we obtain 8 distinct estimates of $\alpha_{Q}$. 

\subsection{Gas based Analysis}\label{subsec:gas}

Our first approach is to use the molecular hydrogen column density ($N_{\rm{H_{2}}}$) map to  measure the line luminosities of \hcn\ and \hcop\ \jj10 emission based on the threshold criterion and estimate the mass of the dense gas. Finally, we derive the mass-luminosity conversion factor.

\subsubsection{Column Density Maps}\label{subsubsec:columndensity}
In this section, we outline the approach for determining the luminosity of line tracers originating from both the dense region and the entire cloud, relying on the $N_{\rm{H_{2}}}$ map derived from \coo\ column density. 
First, we generate the maps of molecular hydrogen column density ($N_{\rm{H_{2}}}$) for each target. 
This analysis is conducted for both cases, with and without metallicity correction.

To derive the molecular hydrogen column density map, we have used the \cotw\ and \coo\ data of each target. 
We assume that both molecular lines originate from the same regions within the clouds.
We selected the velocity range by examining the  spectrum  of \coo\ emission averaged over the whole mapped region of each target. 
The \cotw\ line is optically thick, whereas its isotopologue \coo\ is optically thin. We assume that both lines share the same excitation temperature $(T_{ex})$. Therefore, we determine the $T_{ex}$ using the peak brightness temperature $(T_{12;pk})$ of the optically thick \cotw\ line.
The procedure for generating column density maps of \coo\ is thoroughly explained in Appendix C of \citet{2022AJ....164..129P}. Our approach involves applying equations C1-C5 from the same appendix in \citet{2022AJ....164..129P} to create the total column density map of \coo\ ($N(\coo)$), utilizing data from both \coo\ and \cotw.
Following that, we perform convolution and regridding on the \coo\ column density map to align it with the 
\hcn\ and \hcop\ \jj10 emission
map from TRAO. Subsequently, we convert it into a molecular hydrogen column density ($N_{\rm{H_{2}}}$) map using the following equation
\begin{equation}\label{eq:nh2}
    {N_{\rm H_{2}}}={N(\rm \coo)} \left[\frac{\rm{H_{2}}}{\rm {\cotw}}\right]_{\odot} \left[\frac{\rm{\cotw}}{\coo} \right].
\end{equation}
Here $\left[\frac{\rm{H_{2}}}{\rm {\cotw}}\right]_{\odot}$ corresponds to the local ISM value of $\left[\frac{\rm{H_{2}}}{\rm{\cotw}}\right]$.
We employ the isotopic ratio $[^{12}$C]/[$^{13}$C] as determined by \citet{2020A&A...640A.125J}, which exhibits a linear dependency on the \gal\ distance of the observed target

\begin{equation}
    \frac{[^{12} \mbox{C}]}{[^{13}\mbox{C}]} = 5.87\times \frac{\rg}{\mathrm{kpc}} +13.25.
\end{equation}
 We assume a fractional abundance of \cotw\ from  \citet{2017ApJ...838...66L}:

\begin{equation}
\left[\frac{\rm{H_{2}}}{\rm {\cotw}}\right]_{\odot}  = 6000 .
\end{equation}

To incorporate the metallicity correction factor in the hydrogen column density estimation, we have introduced a multiplication factor of $\mathrm{Z^{-1}}$ in equation \ref{eq:nh2}, assuming $\mathrm{CO}$ abundance is proportional to metallicity \citep{2020ApJ...903..142G}. 
The dependence of the \cotw\ abundance on metallicity can be complicated, depending on radiation environment, how dust grain properties depend on metallicity, etc. \citep{2011MNRAS.412..337G, 2023MNRAS.522.4612H}, but the dependence is generally stronger than linear. Even for the optically thick \cotw\ \jj10\ emission intensity, the dependence is nearly linear \citep{2020ApJ...903..142G}, and here we need the actual abundance since we are using the optically thin \coo. We take a conservative approach in assuming a linear dependence on $Z$.

The column density map after metallicity correction ($N_{\rm H_{2}}^{'}$) is as follows
%
\begin{equation}
    \begin{split}
        {N_{\rm H_{2}}^{'}} & = \rm{N(\coo)} \left[\frac{\rm{H_{2}}}{\rm{\cotw}}\right] \left[\frac{\rm{\cotw}}{\coo} \right] \\
                            & = \rm{N(\coo)} \left[\frac{\rm{H_{2}}}{\rm{\cotw}}\right]_{\odot} \left[\frac{1}{\textit{Z}} \right] \left[\frac{\rm{\cotw}}{\coo} \right].
    \end{split}
\end{equation}
The metallicity $Z$ is relative to the metallicity in the local interstellar medium (ISM). We have used the following equation 
\begin{equation}
    Z = \rm{\frac{Z_{*}}{Z_{\odot}}} = \rm{10^{[M/H]}},
\end{equation}
where $\mathrm{[M/H]}$ is the metallicity difference between the target and solar neighborhood ISM measured by oxygen ($\mathrm{(12+log[O/H])_{*} - (12+log[O/H])_{local\ ISM}}$).
We have used the value for $\mathrm{(12+log[O/H])_{local\ ISM}=8.50}$ based on \citet{2022ApJ...931...92E}.
The metallicity of individual targets is represented by $\mathrm{(12+log[O/H])_{*}}$
and the values are taken from \citet{2022MNRAS.510.4436M}, \citet{2018PASP..130k4301W} and from the Galactic metallicity gradient, as mentioned in Table \ref{tab:cloud details}. We use oxygen as our proxy for metallicity as it is the best determined over the widest range of galactocentric distances.

The column density maps are employed to delineate areas of column density corresponding to the threshold of $N_{\rm{H_{2}}} (\mathrm{and} \ N_{\rm{H_{2}}}^{'}) \geq 8 \times 10^{21}\ \rm{cm^{-2}}$ utilized in the surface density criterion. 
Figure \ref{fig:figure0} shows the integrated intensity maps of \hcn\ and \hcop\ for the new targets studied in this work.
The contours in Figure \ref{fig:figure0} indicate the regions satisfying the column density threshold criterion  with (red contours) and without (white contours) correction for metallicity. Red contours are always bigger than white contours for the sub-solar metallicity clouds and opposite for super-solar metallicity clouds. The cyan contours in Figure \ref{fig:figure0} represent the BGPS ``in" regions for the new targets having dust continuum emission data. These cyan contours generally align well with the $N_{\rm H_{2}}$-based contours. However, the dense regions indicated by the BGPS contours typically correspond to lower extinction regions (see more discussion in Section \ref{sec_4.2.1}).
The integrated intensity maps of \hcn\ and \hcop\ show  that those emissions extend beyond the `dense gas' contours as seen previously \citep{2022AJ....164..129P, 2020ApJ...894..103E}.
To maintain a consistent analysis for the entire cloud, we set $N_{\rm{H_{2}}} (\mathrm{and} \ N_{\rm{H_{2}}}^{'}) \geq 1.5 \times 10^{21}\ \rm{cm^{-2}}$ as the outer boundary for the clouds. 
This threshold corresponds to the largest value of 5 times the RMS noise in all the \coo\ maps \citep{2022AJ....164..129P}. 


\begin{figure*}
    \centering
    \includegraphics[width=1.0\linewidth]{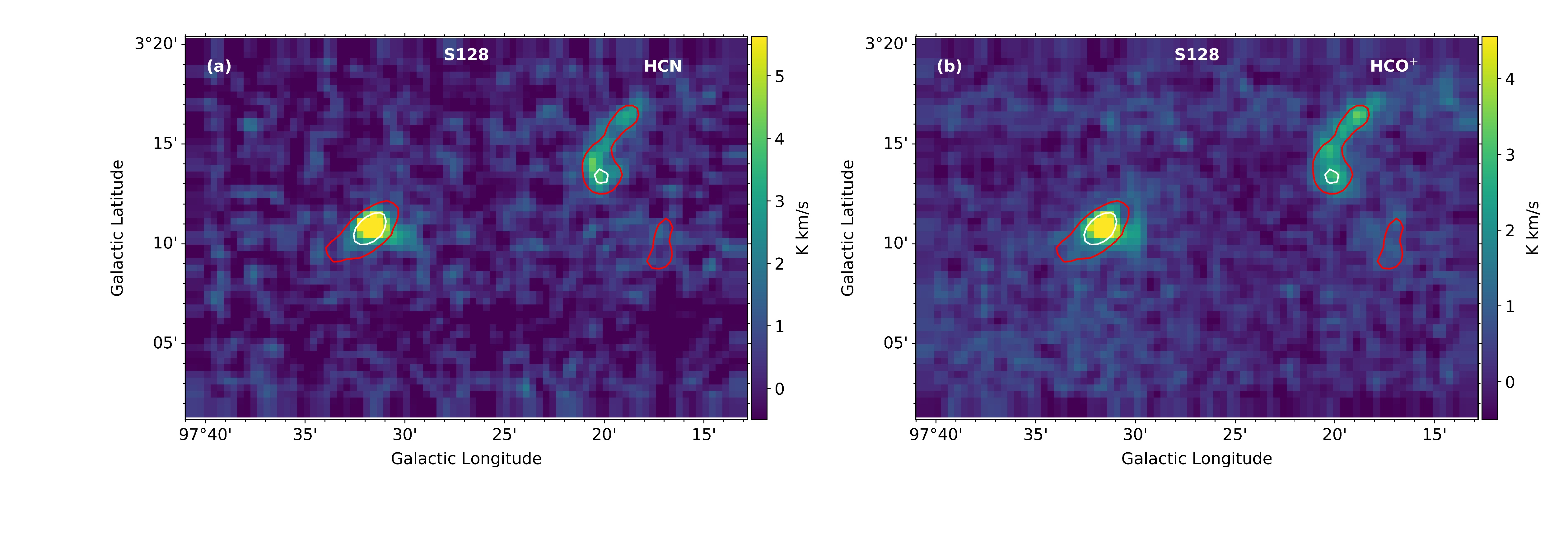}
    \includegraphics[width=1.0\linewidth]{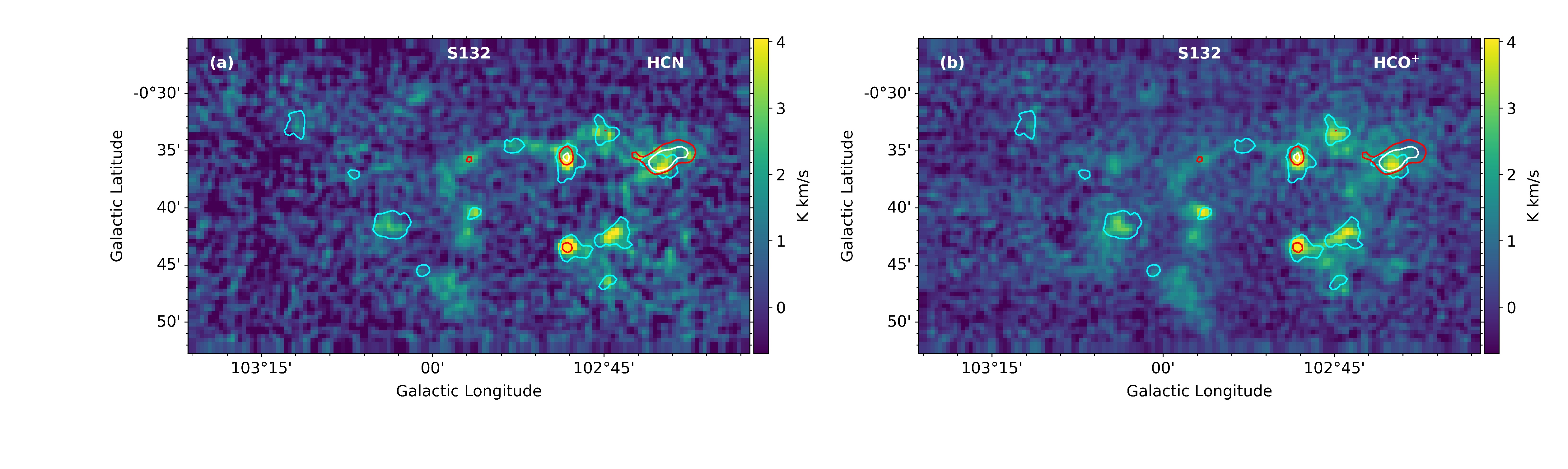}
    \includegraphics[width=1.0\linewidth]{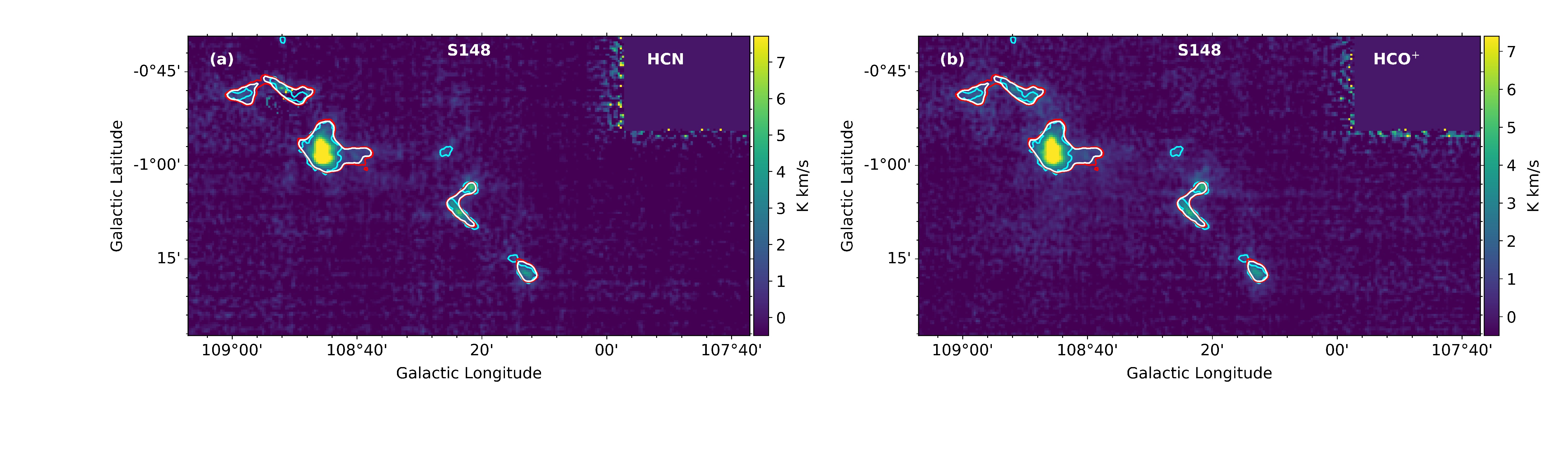}
    \includegraphics[width=1.0\linewidth]{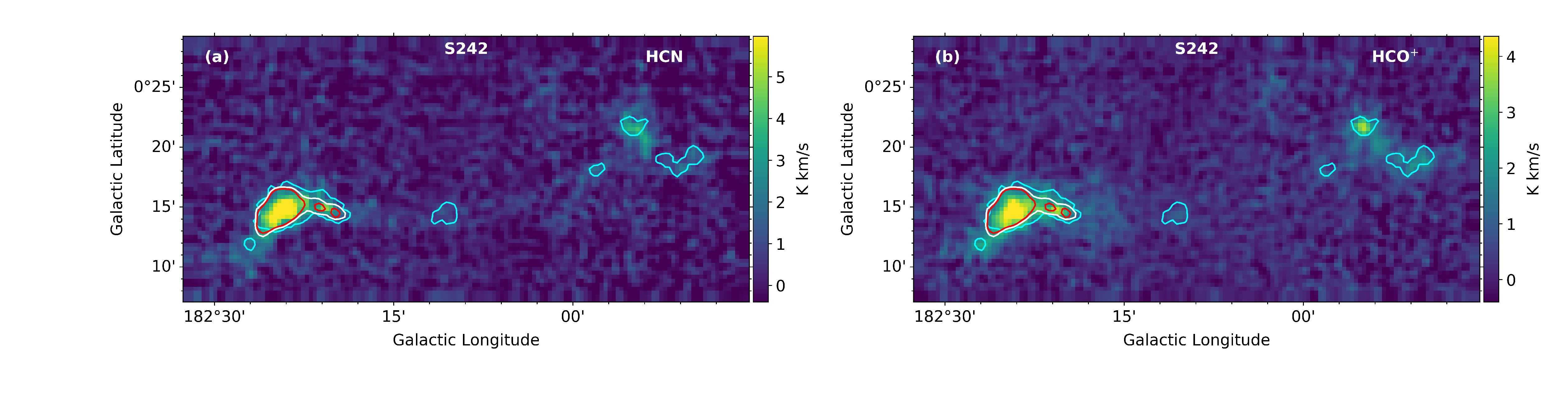}
    \caption{The integrated intensity map of (a) \hcn\ and (b) \hcop, respectively, for the new outer Galaxy targets studied in this paper, as identified in each panel.  The white contours indicate the $N_{\rm{H_{2}}}  \geq 8 \times 10^{21}\ \rm{cm^{-2}}$ regions, red contours show the $Z-$corrected column density, $N_{\rm{H_{2}}}^{'}  \geq 8 \times 10^{21}\ \rm{cm^{-2}}$ regions, derived from \coo (see Section \ref{subsubsec:columndensity}), and cyan contours indicate the BGPS mask regions. There is no cyan contour in Sh2-128 as it lacks BGPS data.}
    \label{fig:figure0}
\end{figure*}

\begin{figure*}
    \figurenum{1}
    \centering
    \includegraphics[width=0.9\linewidth]{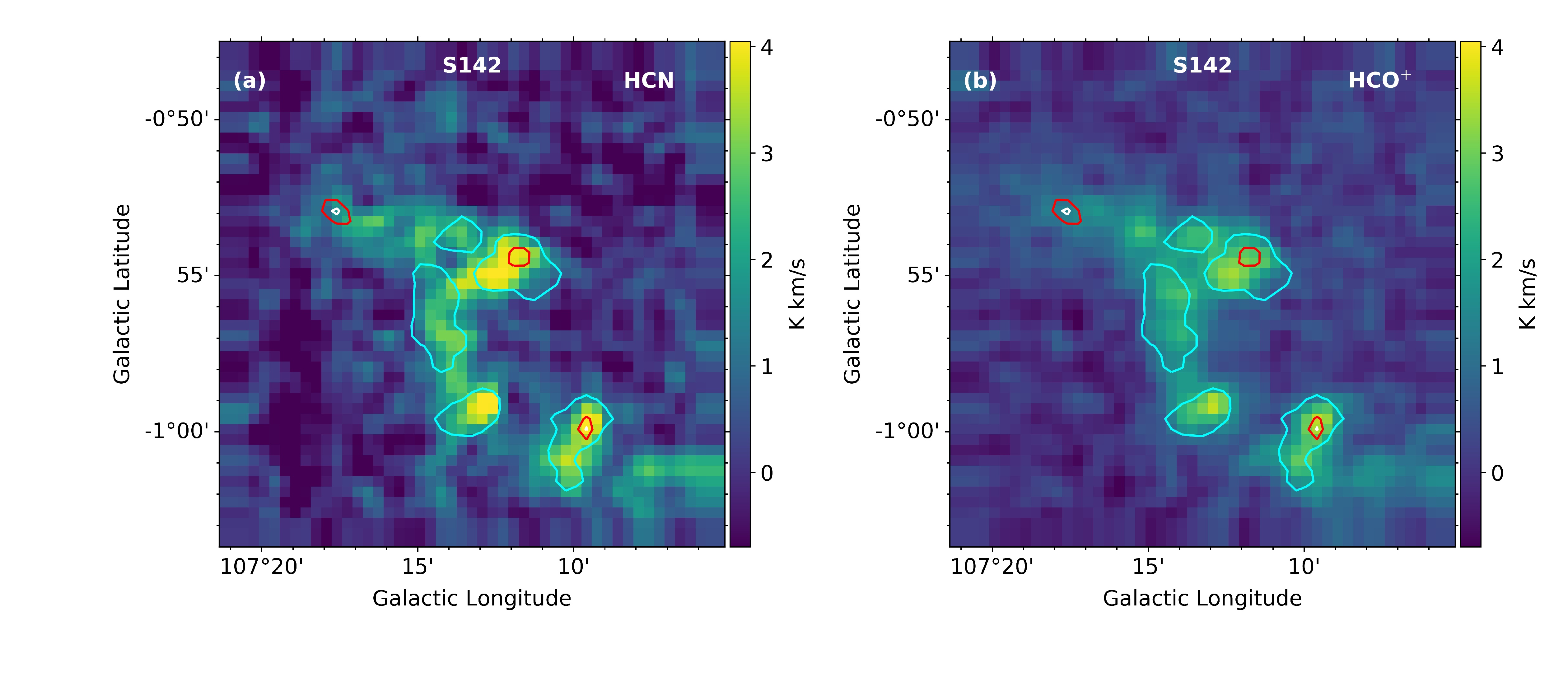}
    \includegraphics[width=0.8\linewidth]{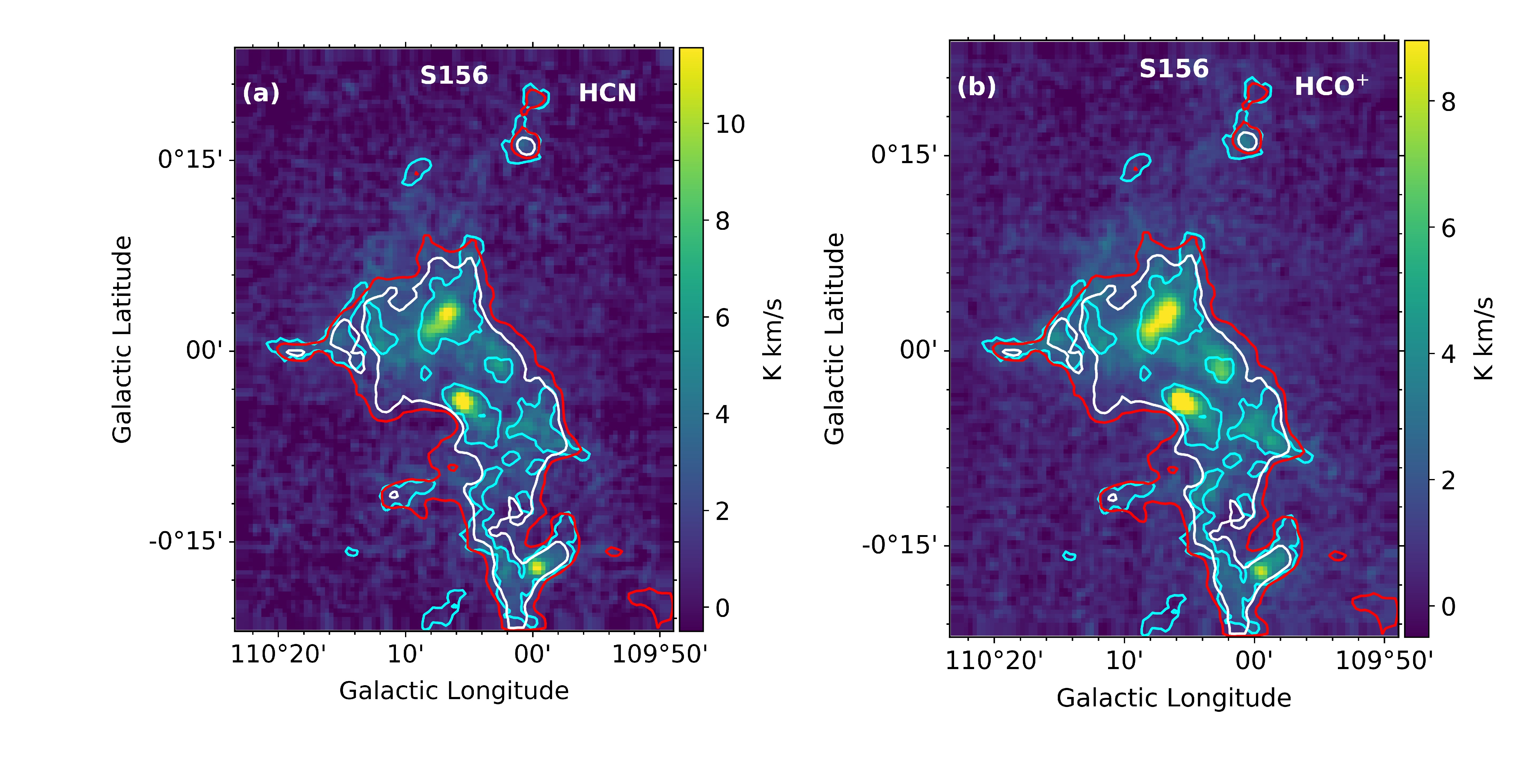}
    \includegraphics[width=0.9\linewidth]{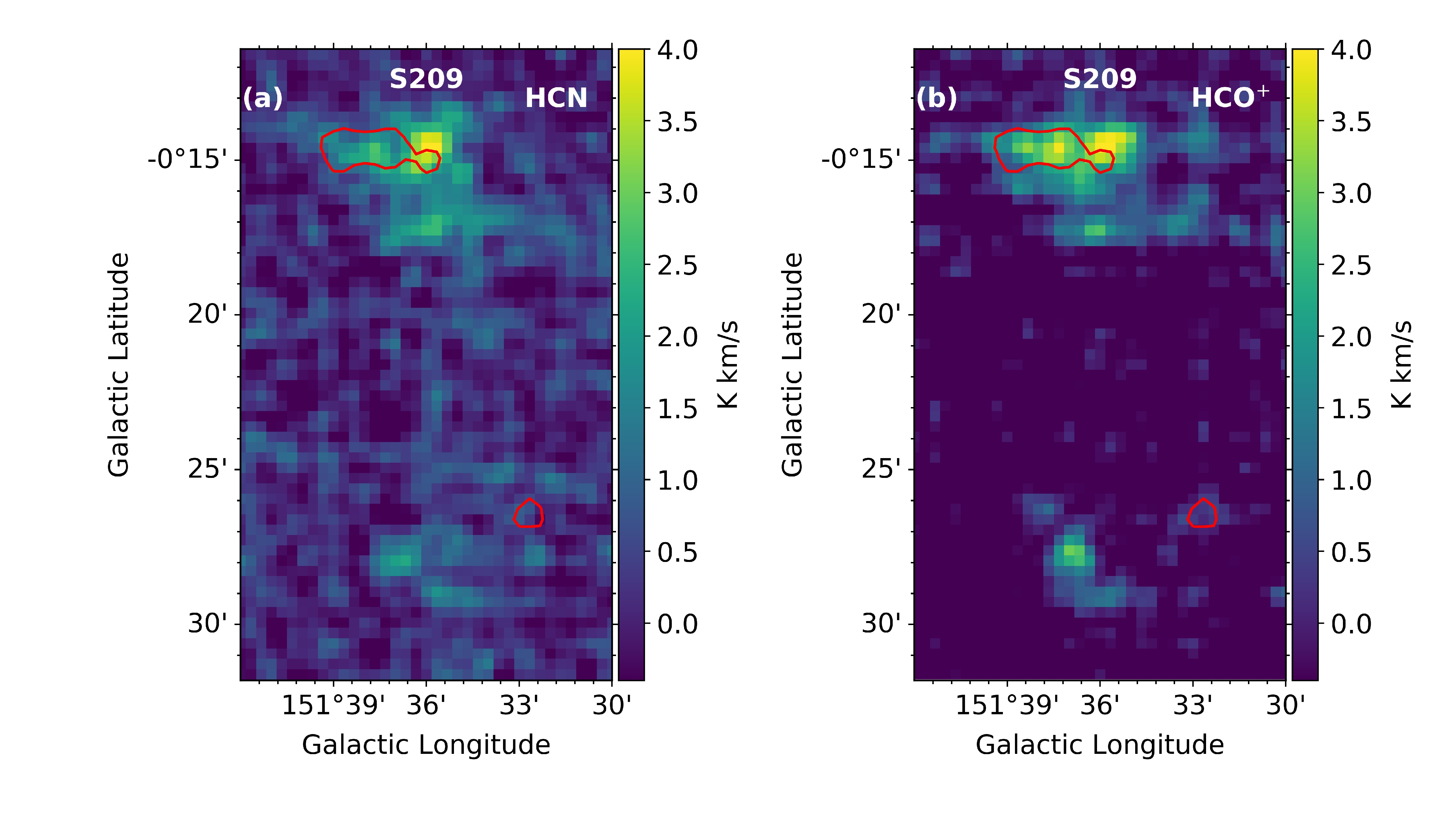}
    \caption{(continued) same as described before. There is no cyan contour in Sh2-209 as it lacks BGPS data and also it has no white contours as there is no region satisfying $N_{\rm{H_{2}}}  \geq 8 \times 10^{21}\ \rm{cm^{-2}}$. }
\end{figure*}

\begin{figure*}[htbp]
    \figurenum{1}
    \centering
    \includegraphics[width=0.9\linewidth]{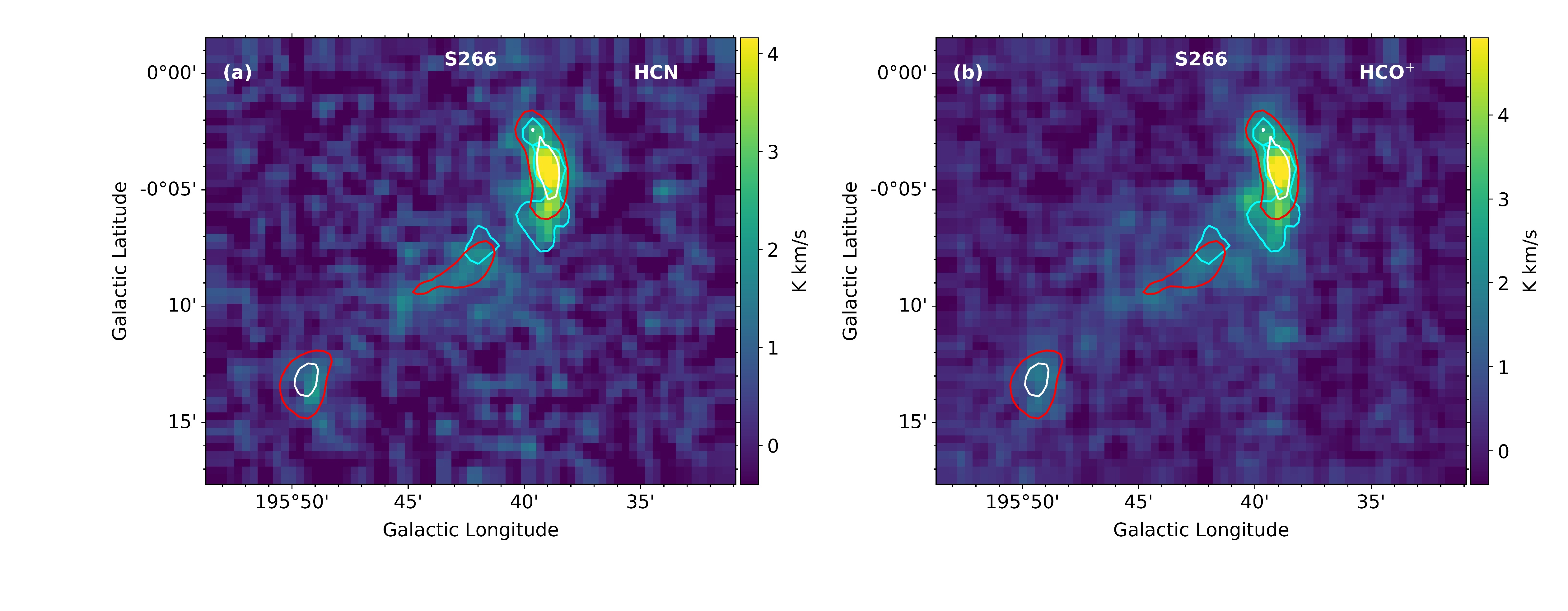}
    \includegraphics[width=0.9\linewidth]{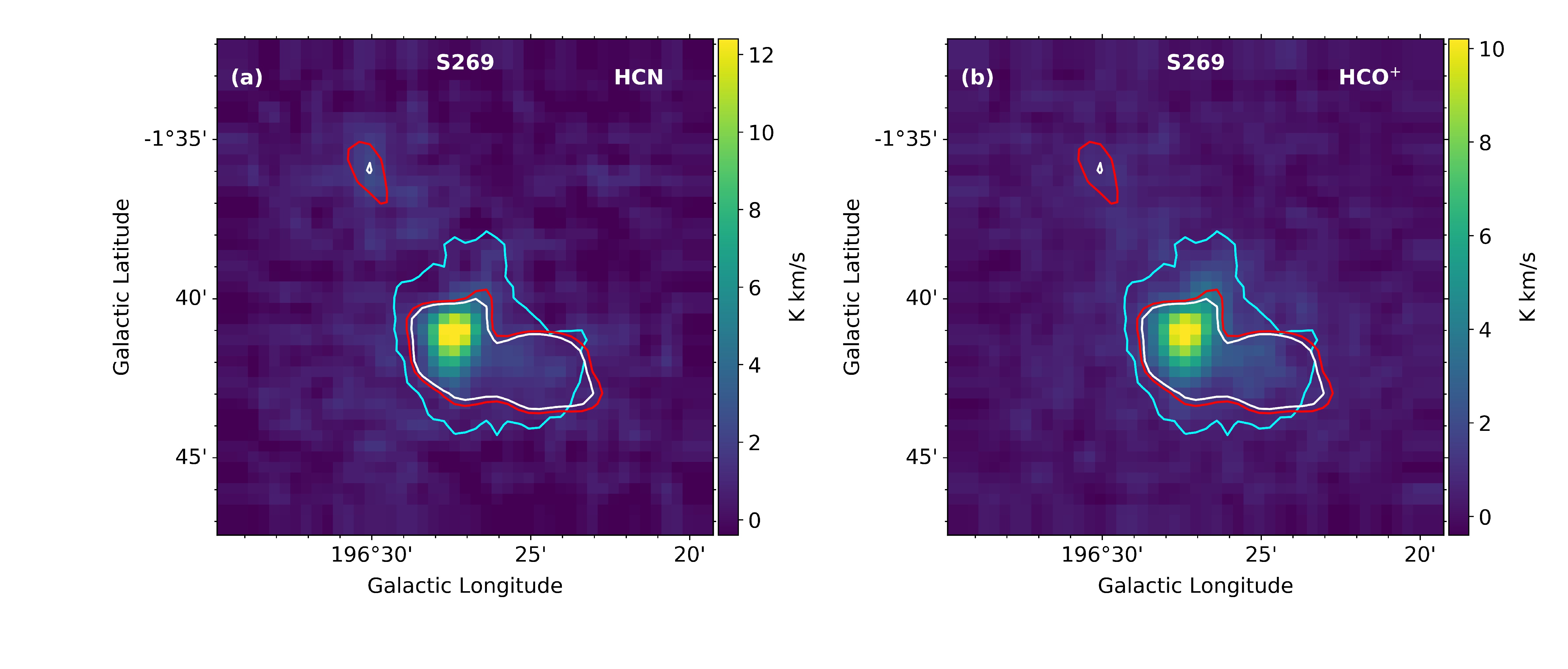}
    \includegraphics[width=0.9\linewidth]{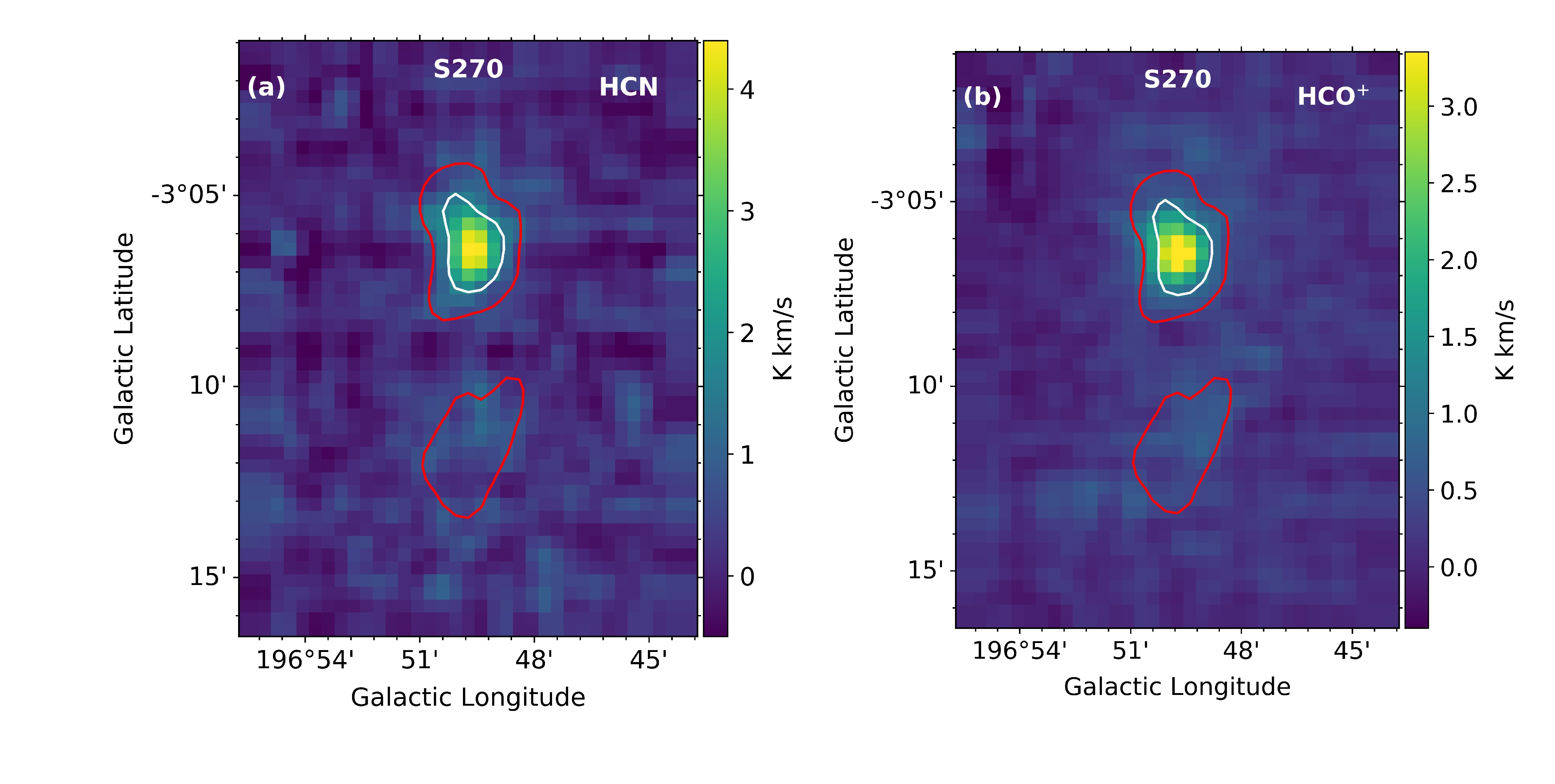}
    \caption{(continued) same as described before. There is no cyan contour in Sh2-270 as it lacks BGPS data. }
\end{figure*}


\begin{figure*}
    \centering
    \includegraphics[width=0.49\linewidth]{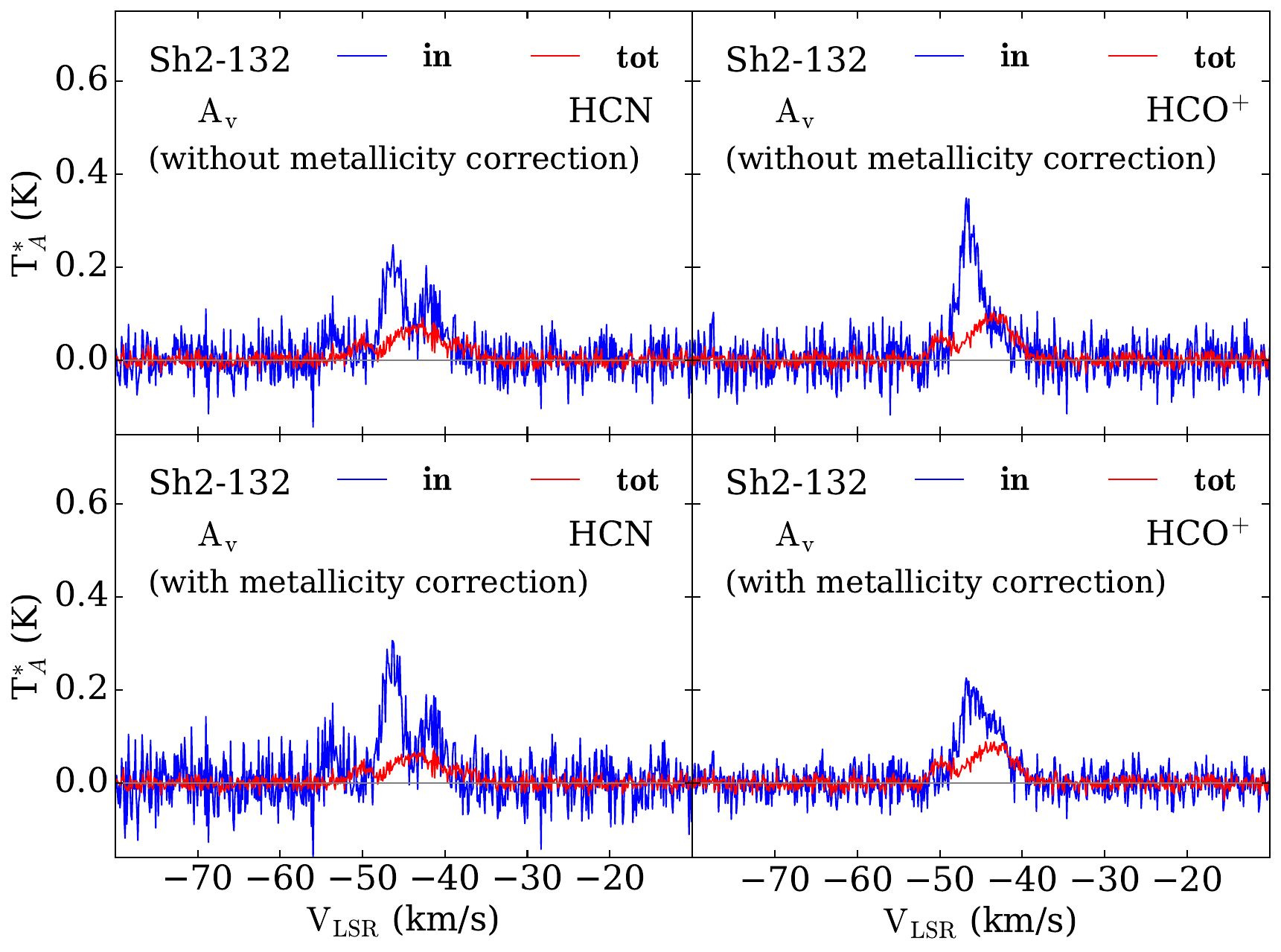}
    \includegraphics[width=0.49\linewidth]{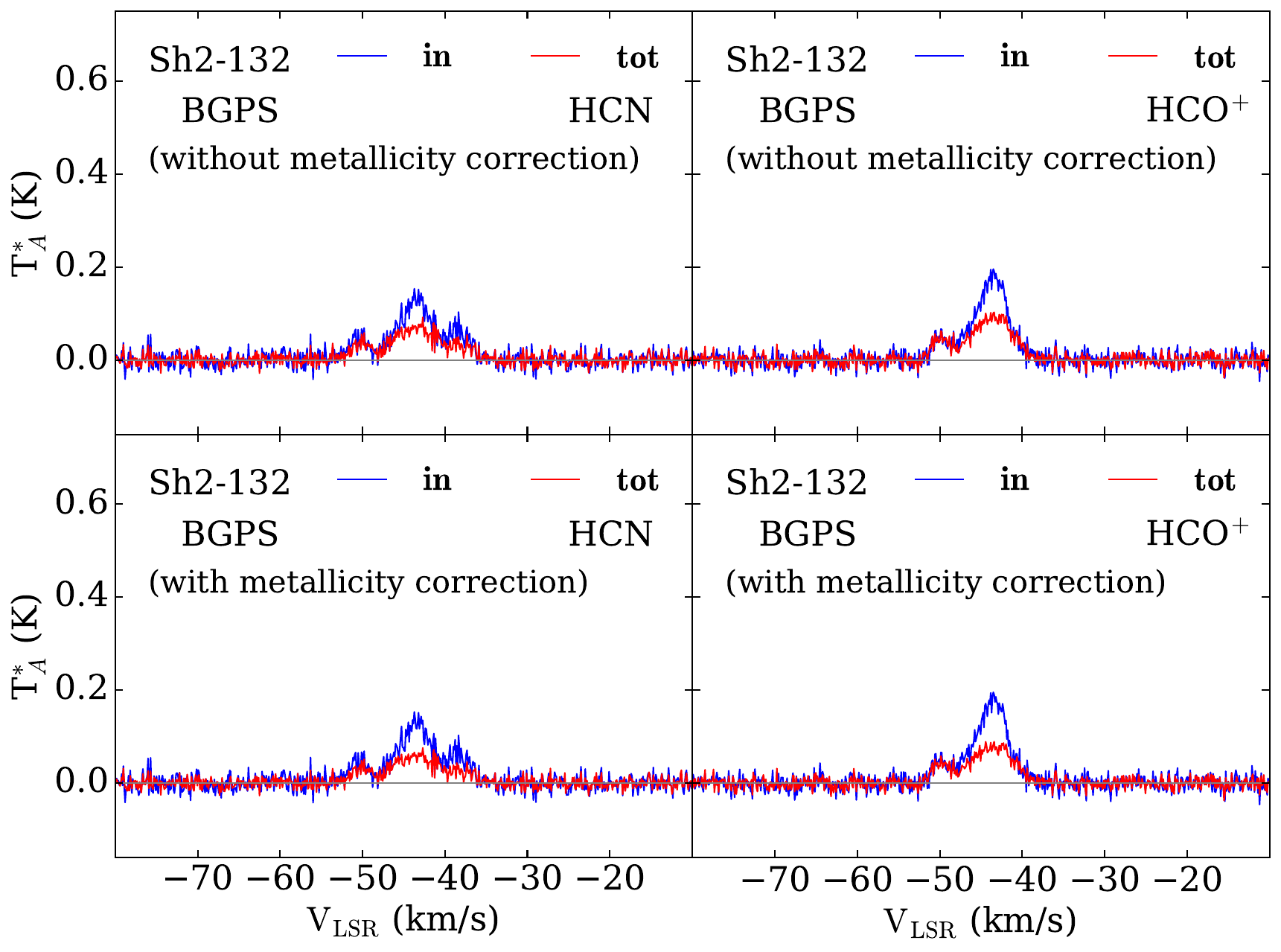}
    \includegraphics[width=0.49\linewidth]{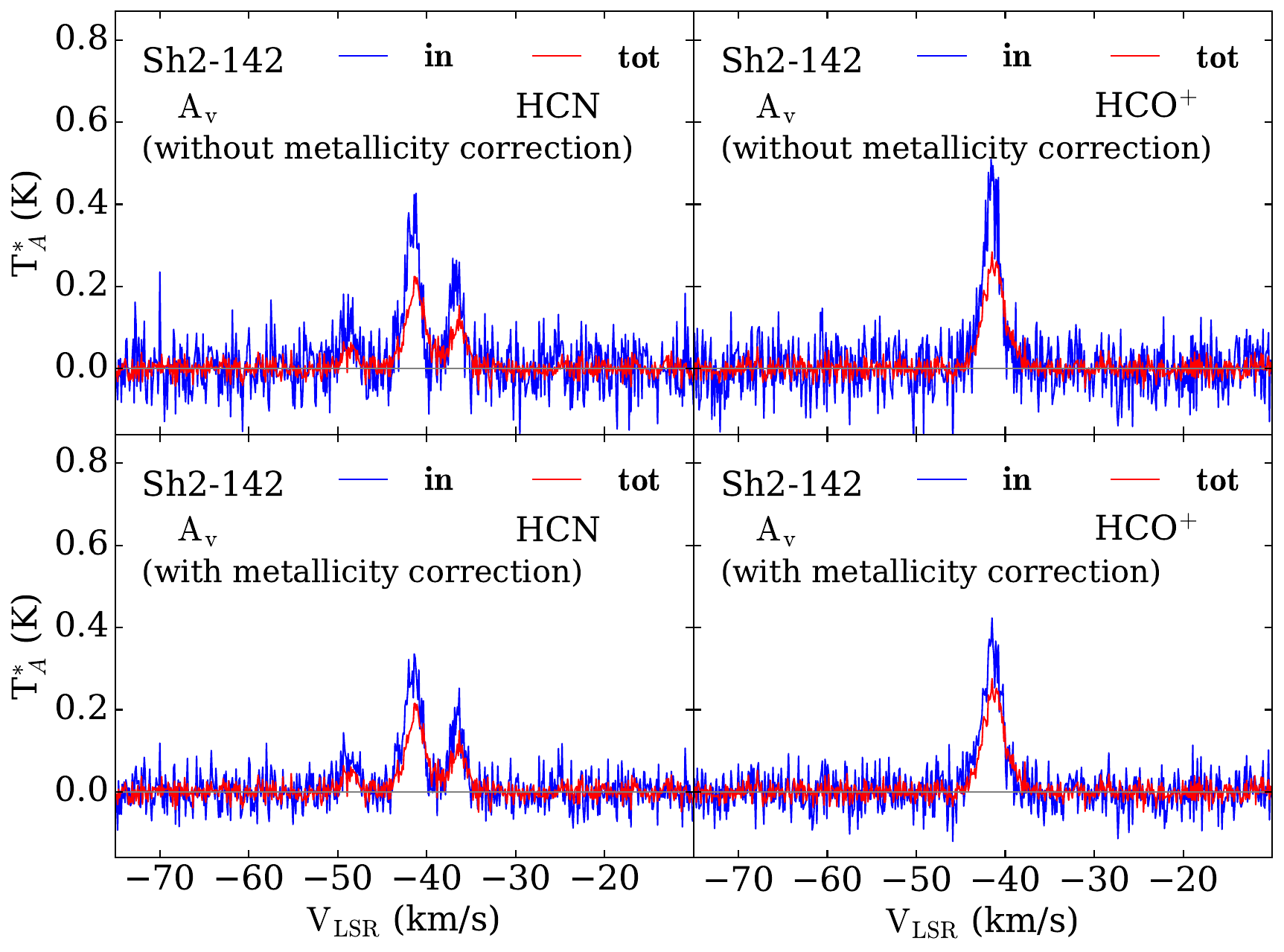}
    \includegraphics[width=0.49\linewidth]{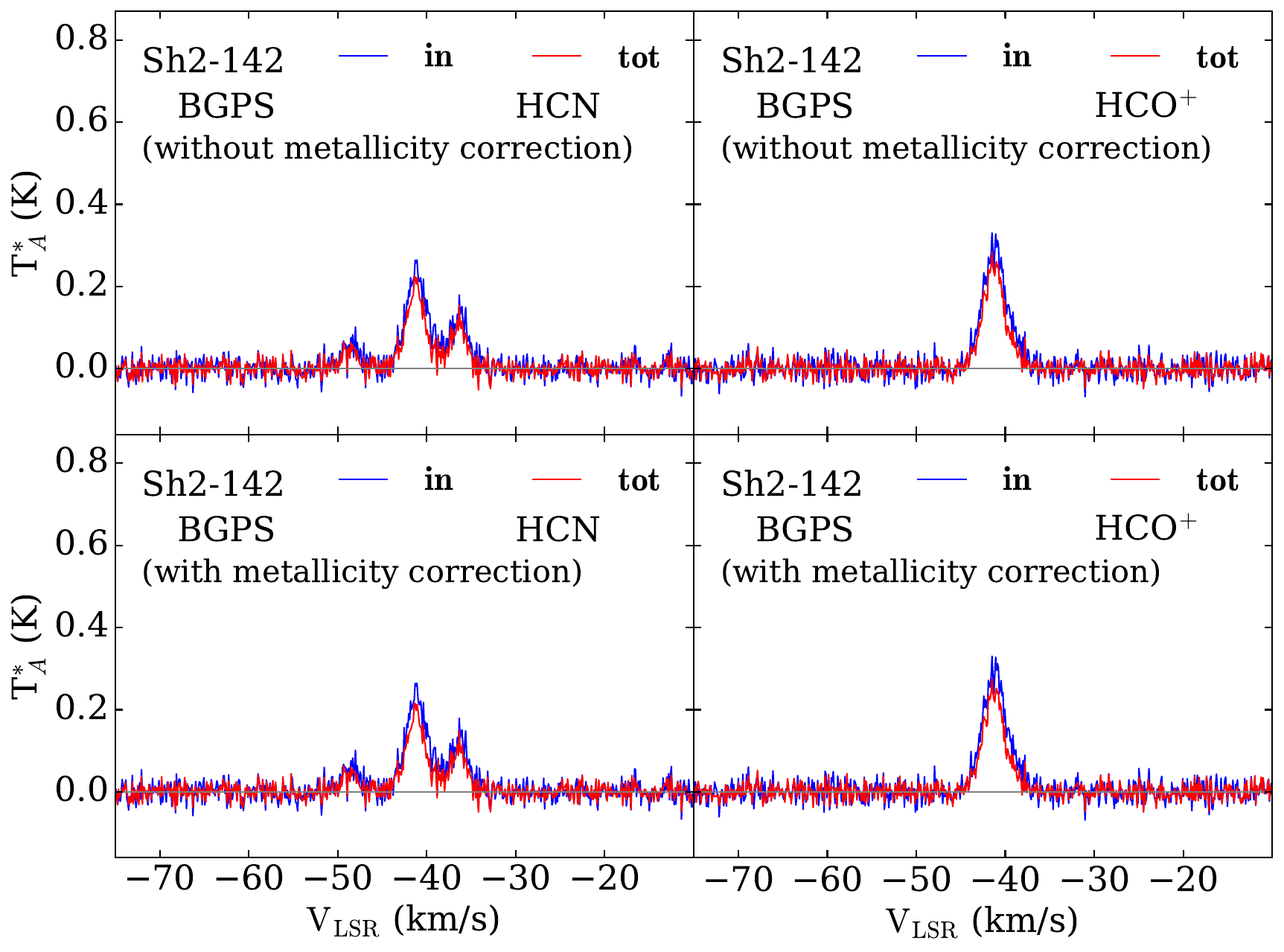}
    \includegraphics[width=0.49\linewidth]{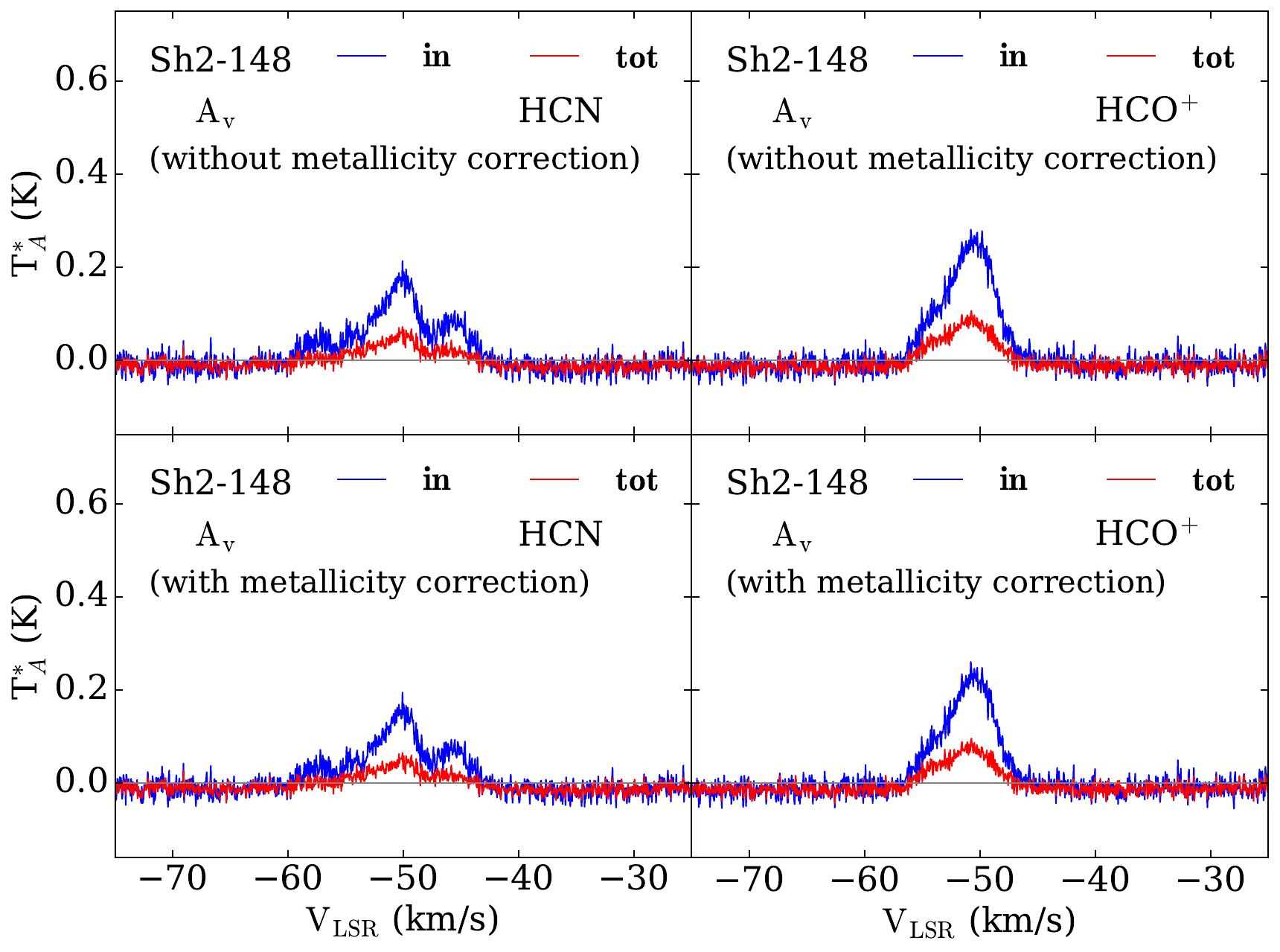}
    \includegraphics[width=0.49\linewidth]{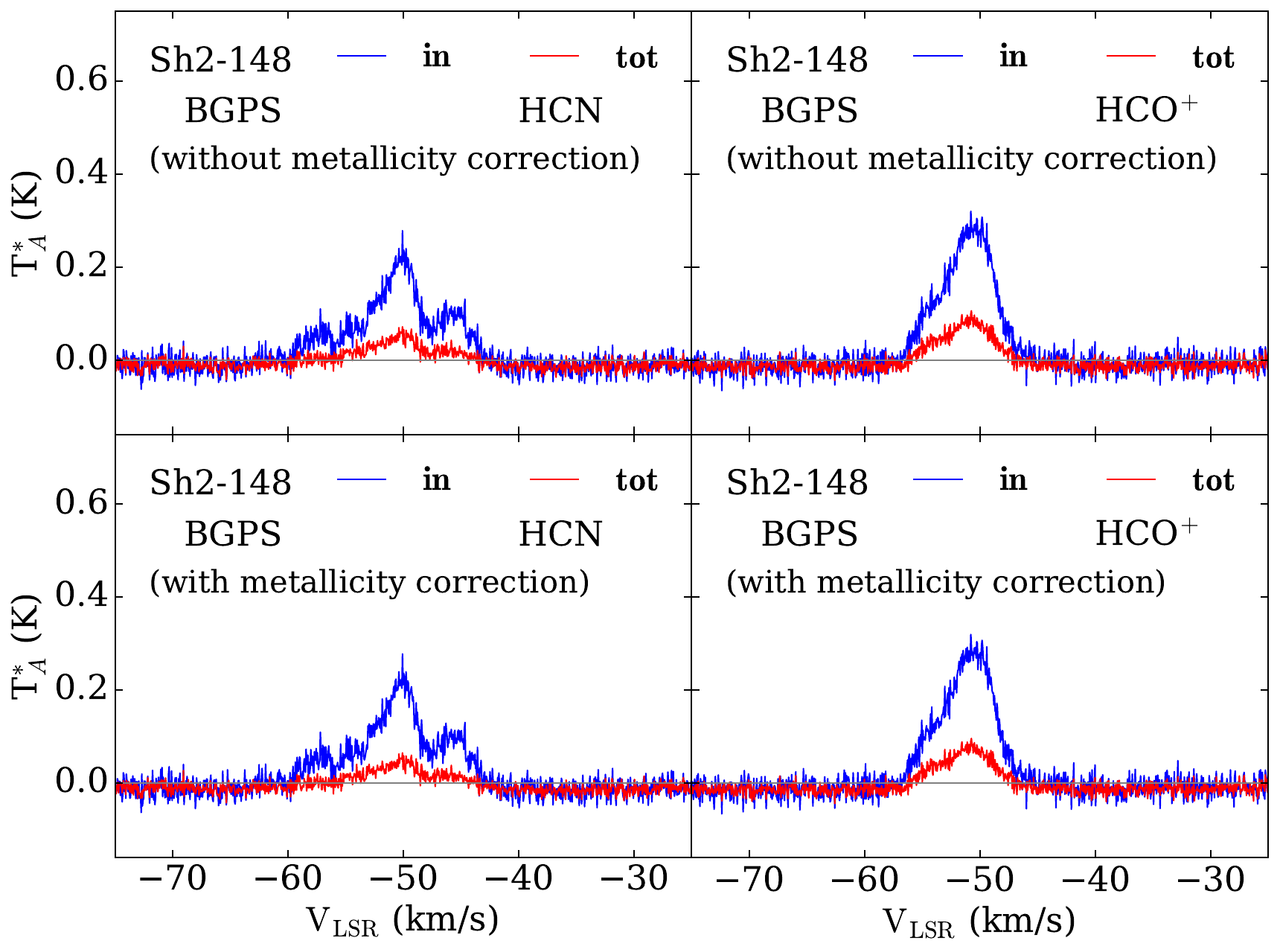}
    \caption{The average spectrum plot for the new outer Galaxy targets studied in this paper, as identified in each panel, based on `Gas based analysis' and 'Dust based analysis', respectively, with and without metallicity correction. The left panel in all images are for \hcn\ and right panels are for \hcop.
    The blue line indicates the line emission coming from the \textit{``in"} regions based on gas and dust analysis in the respective images. The red line indicates the emission from the \textit{``total"} regions. The upper panels of each image show the average spectrum for without metallicity correction case and the bottom panels depict the case after applying the metallicity correction, respectively.     }
    \label{fig:spectra_figure}
\end{figure*}

\begin{figure*}[htbp]
    \figurenum{2}
    \centering
    \includegraphics[width=0.45\linewidth]{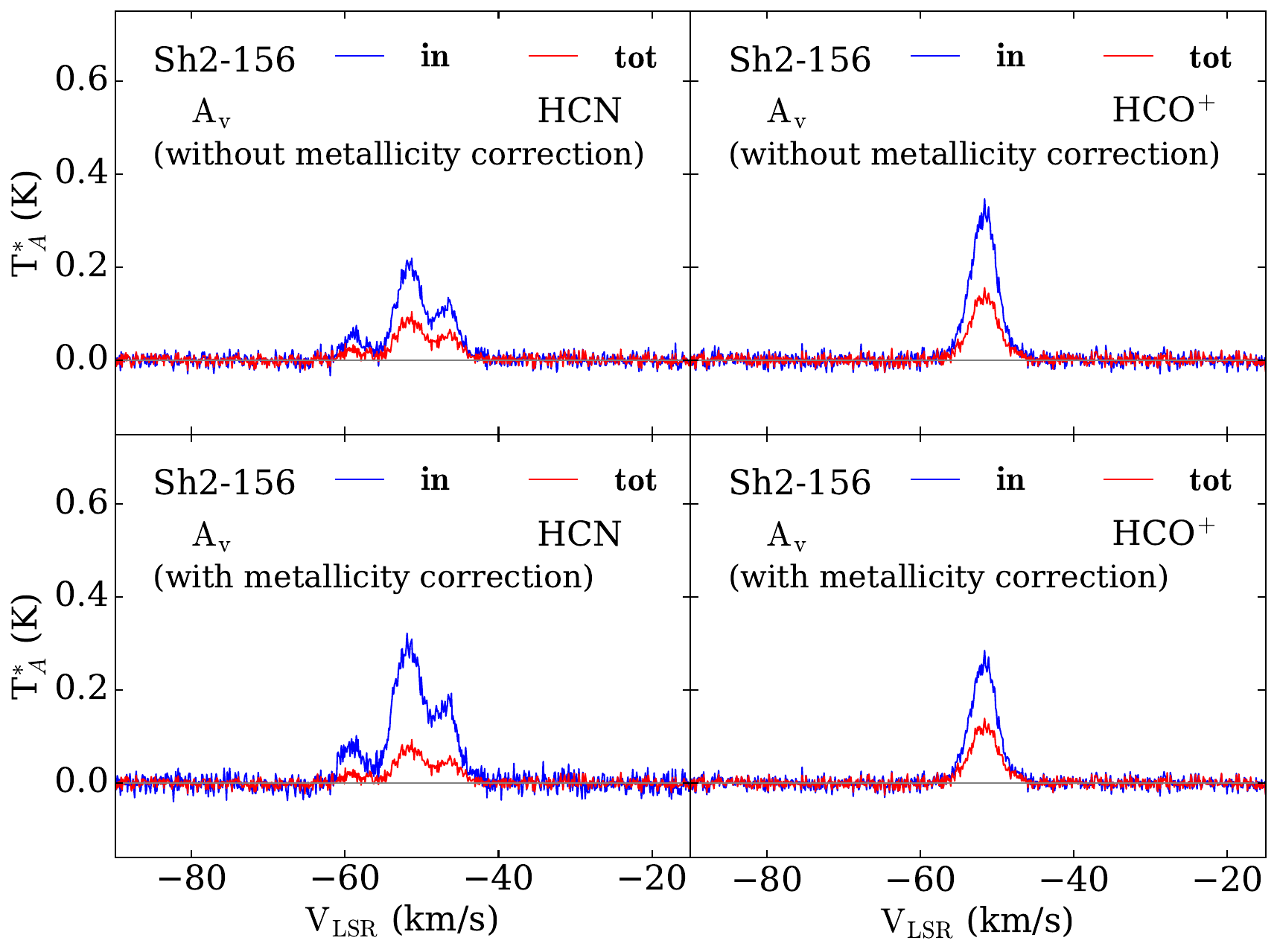}
    \includegraphics[width=0.45\linewidth]{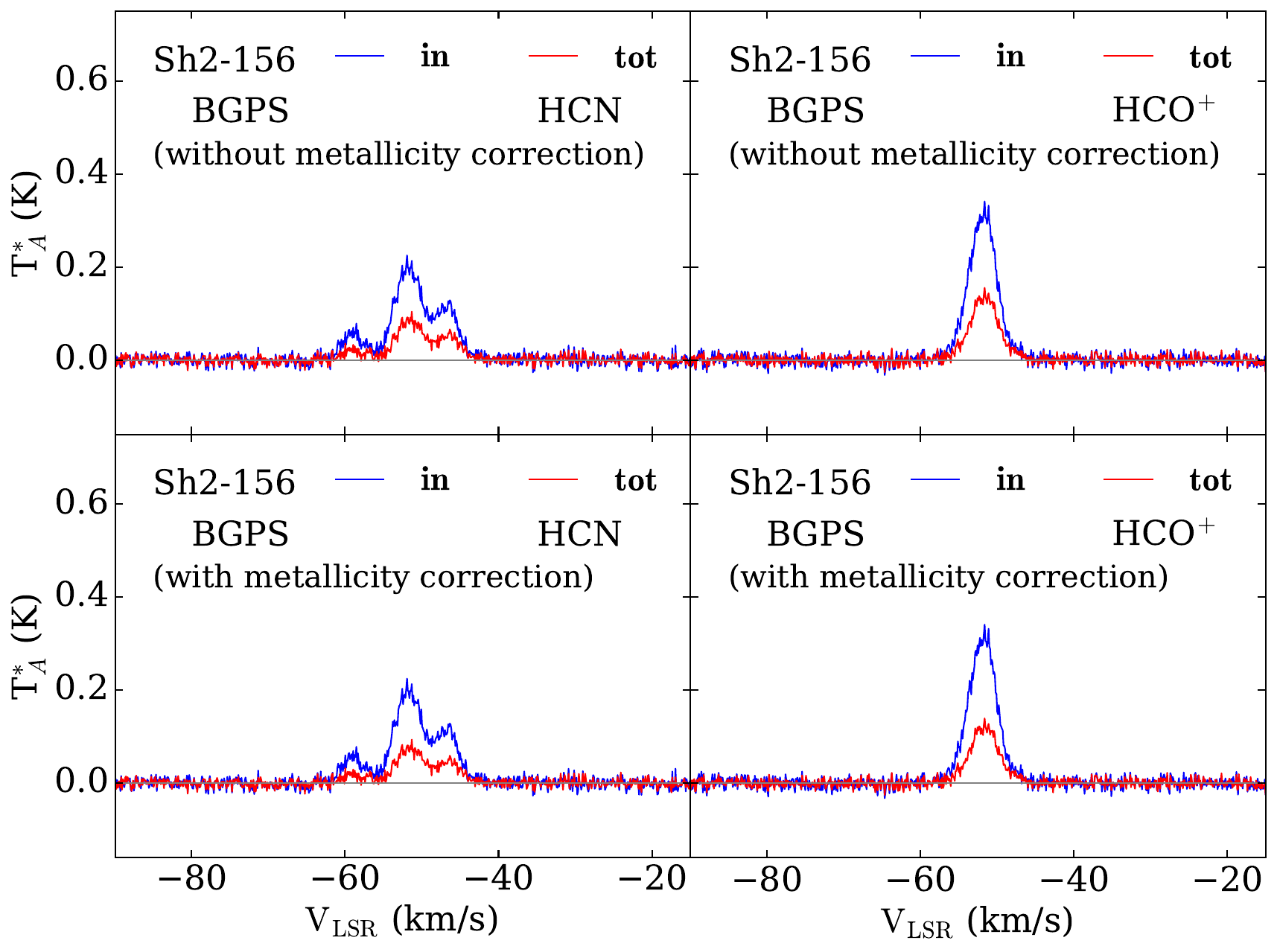}
    \includegraphics[width=0.45\linewidth]{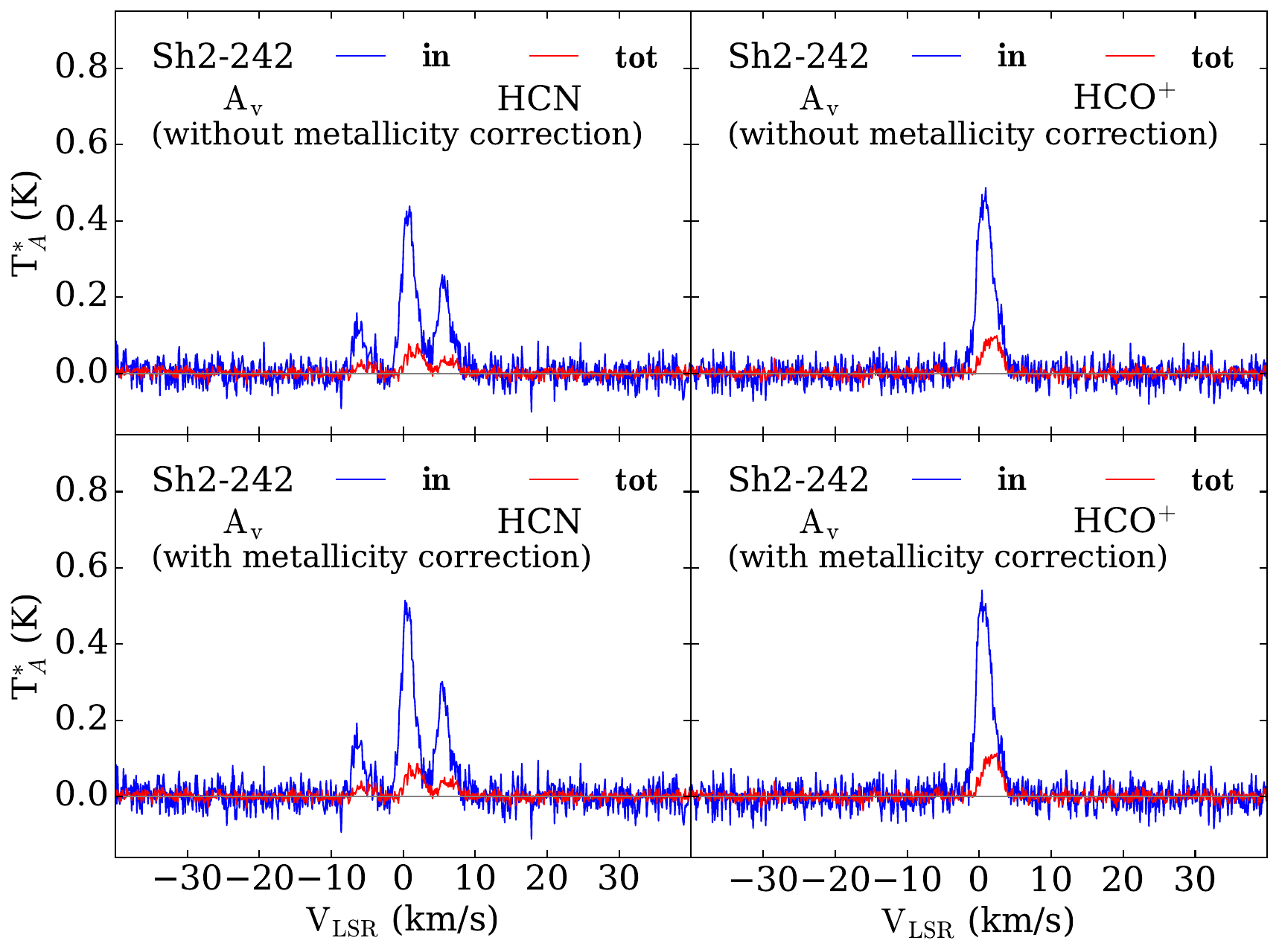}
    \includegraphics[width=0.45\linewidth]{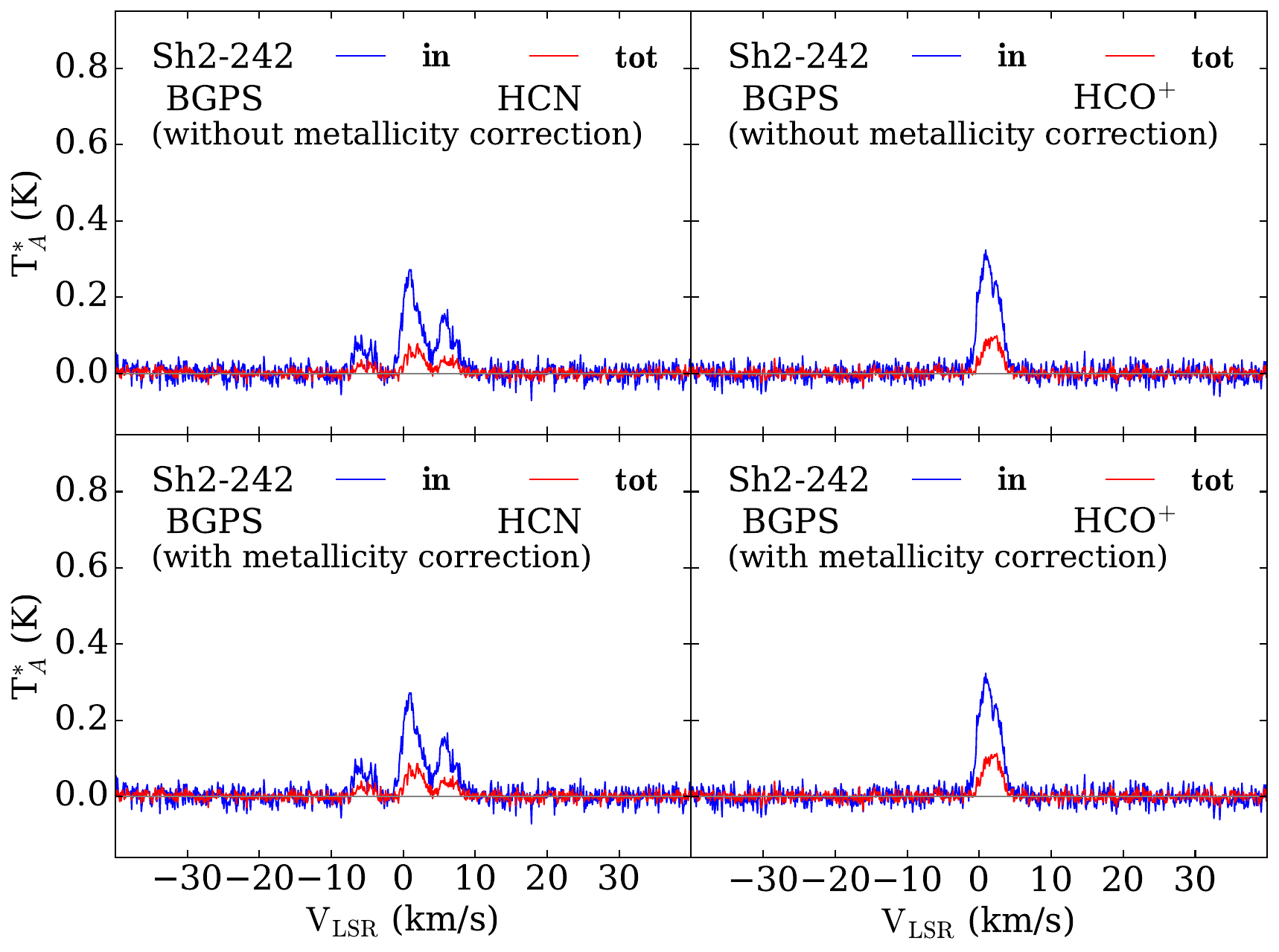}
    \includegraphics[width=0.45\linewidth]{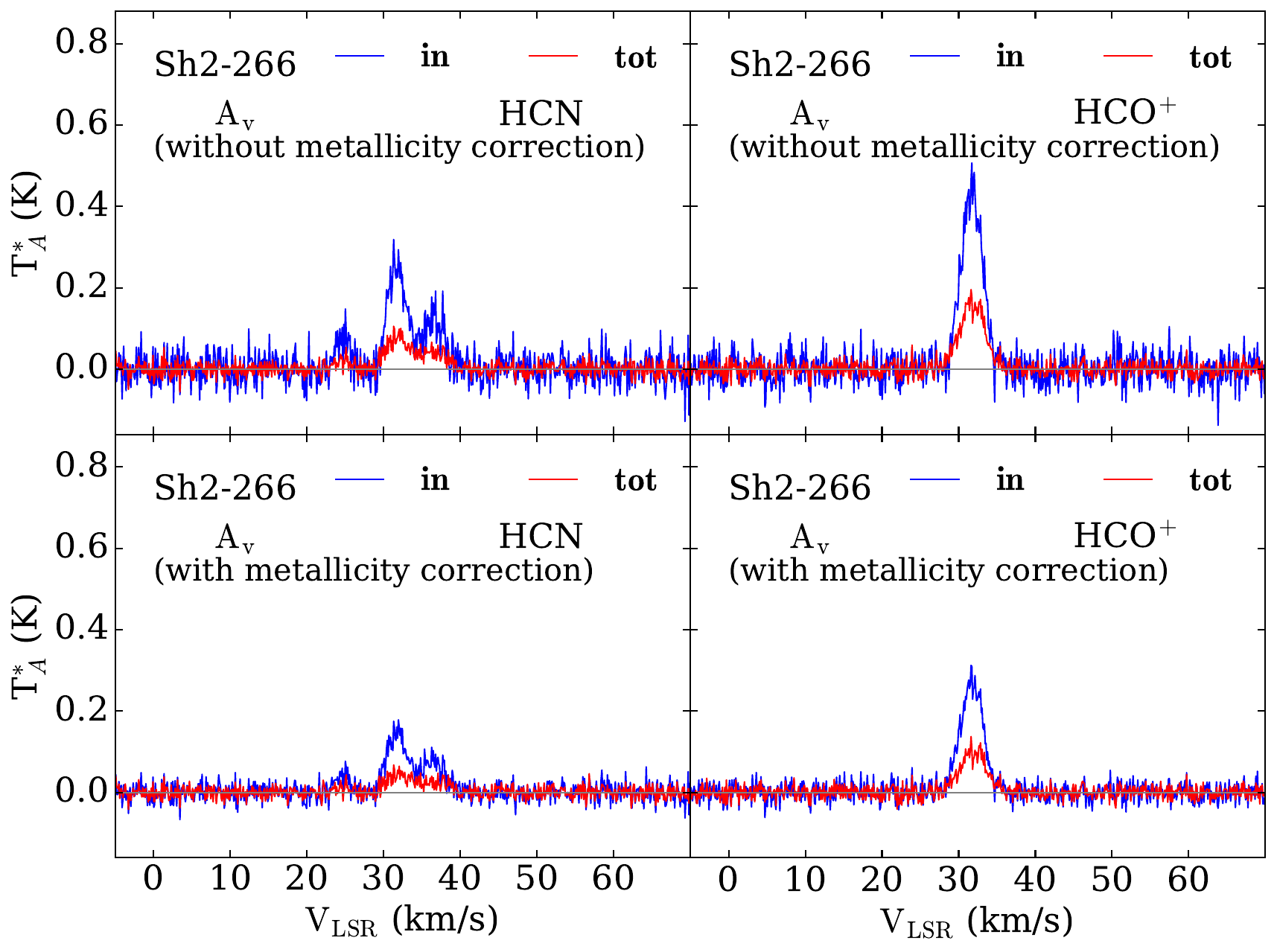}
    \includegraphics[width=0.45\linewidth]{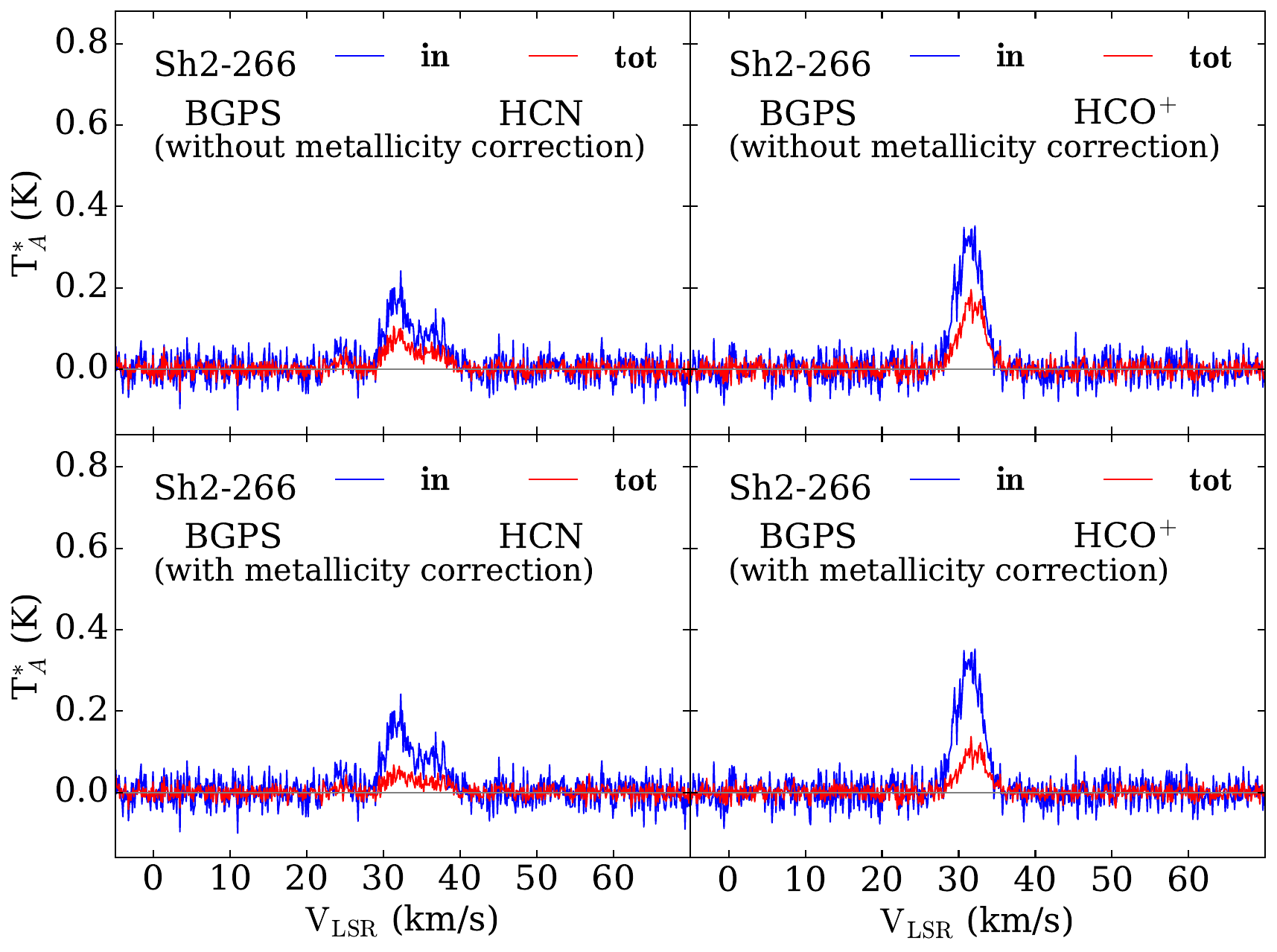}
    \includegraphics[width=0.45\linewidth]{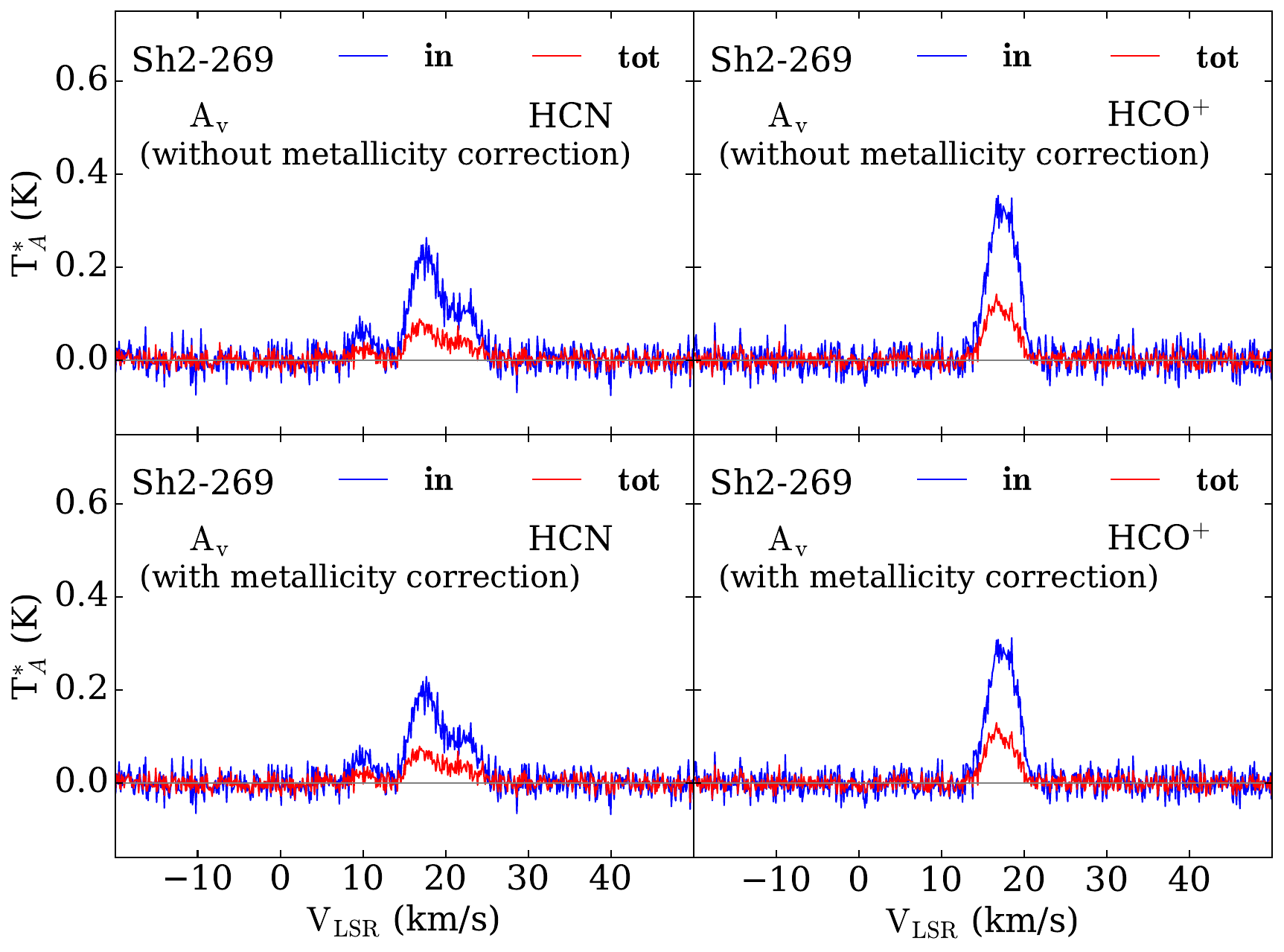}
    \includegraphics[width=0.45\linewidth]{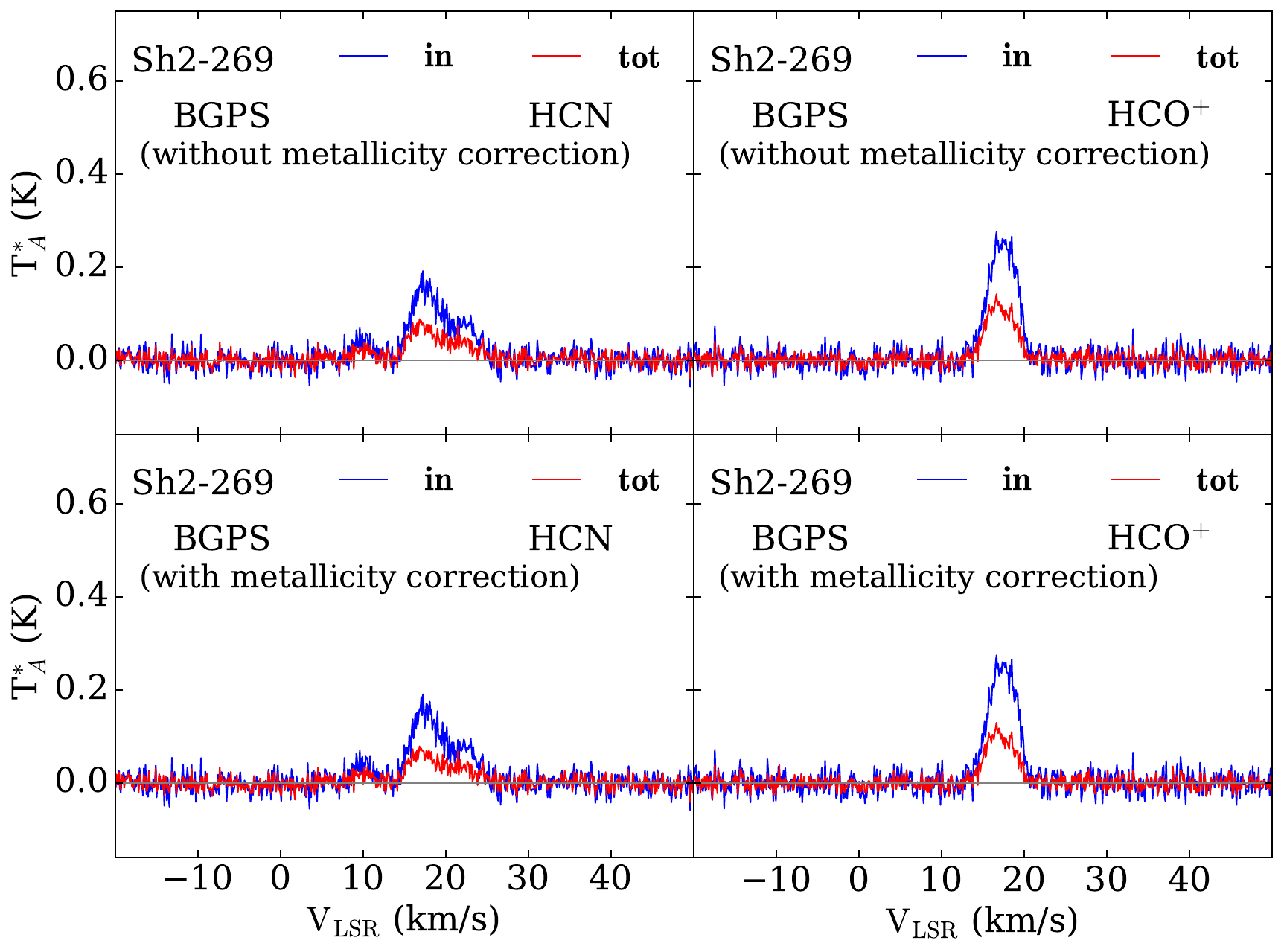}
    \caption{(continued) Same as described earlier for the clouds Sh2-156, Sh2-242, Sh2-266, and Sh2-269.}
\end{figure*}

\begin{figure*}[htbp]
    \figurenum{2}
    \centering
    \includegraphics[width=0.50\linewidth]{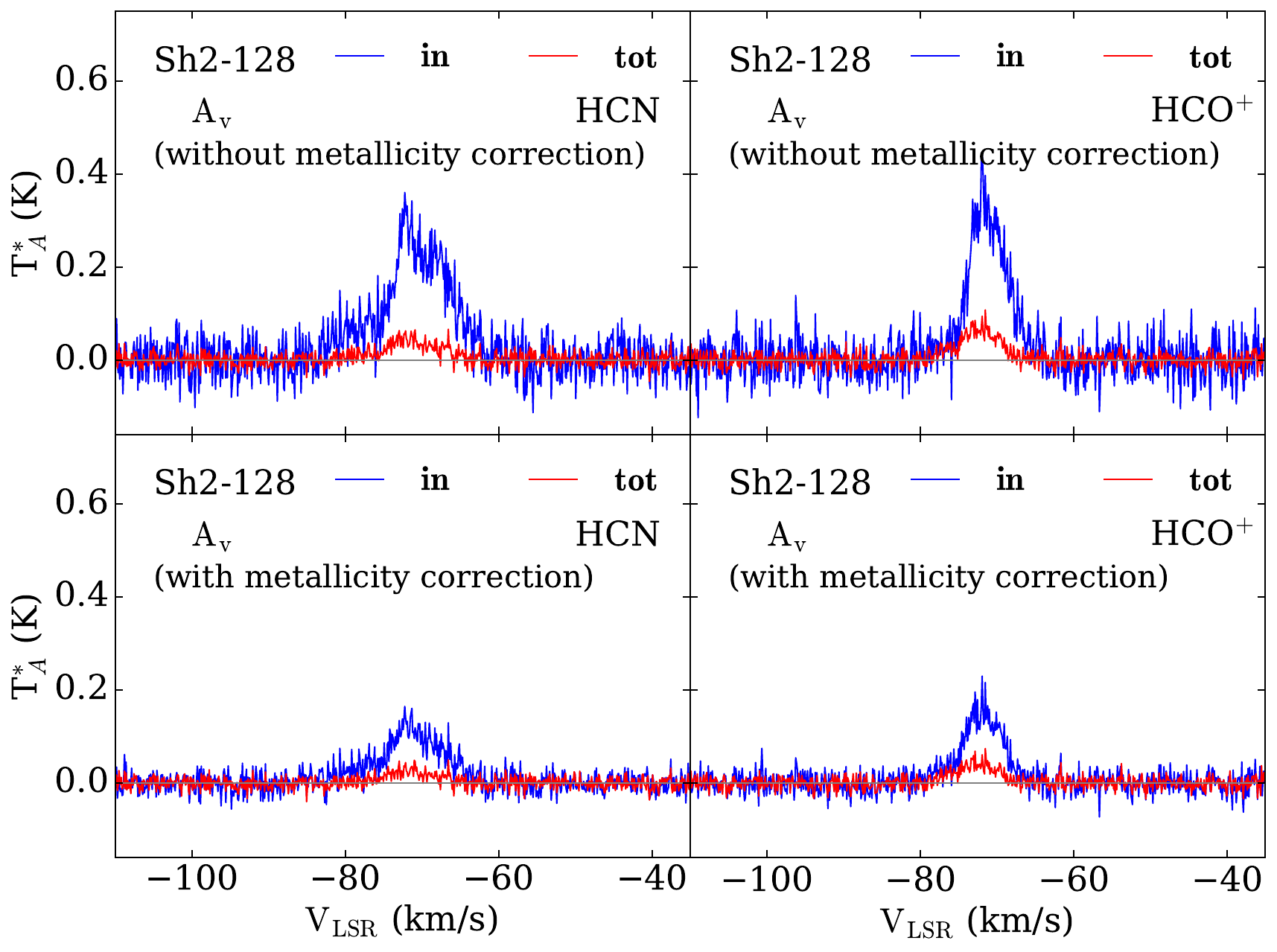}
    \includegraphics[width=0.50\linewidth]{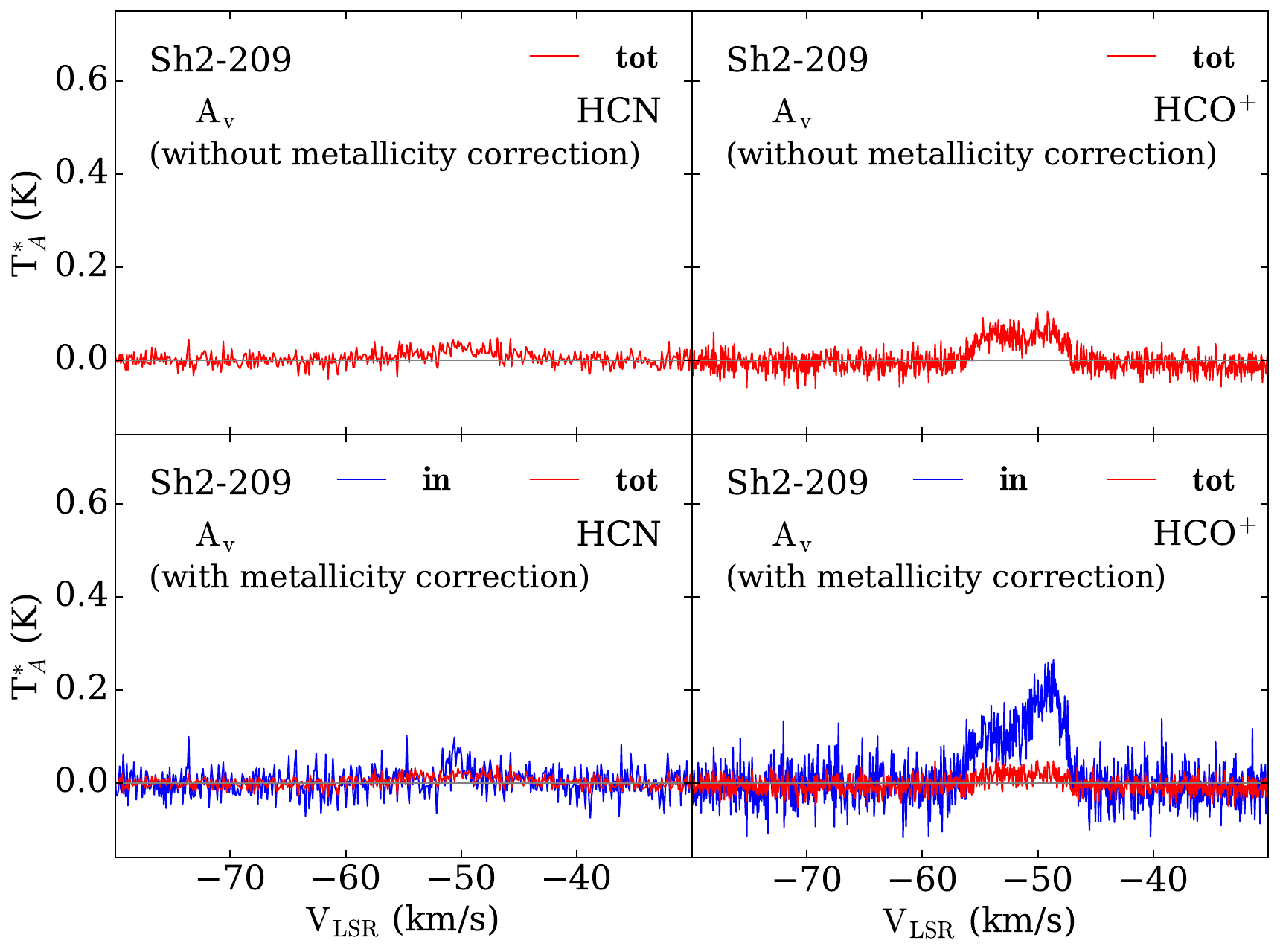}
    \includegraphics[width=0.50\linewidth]{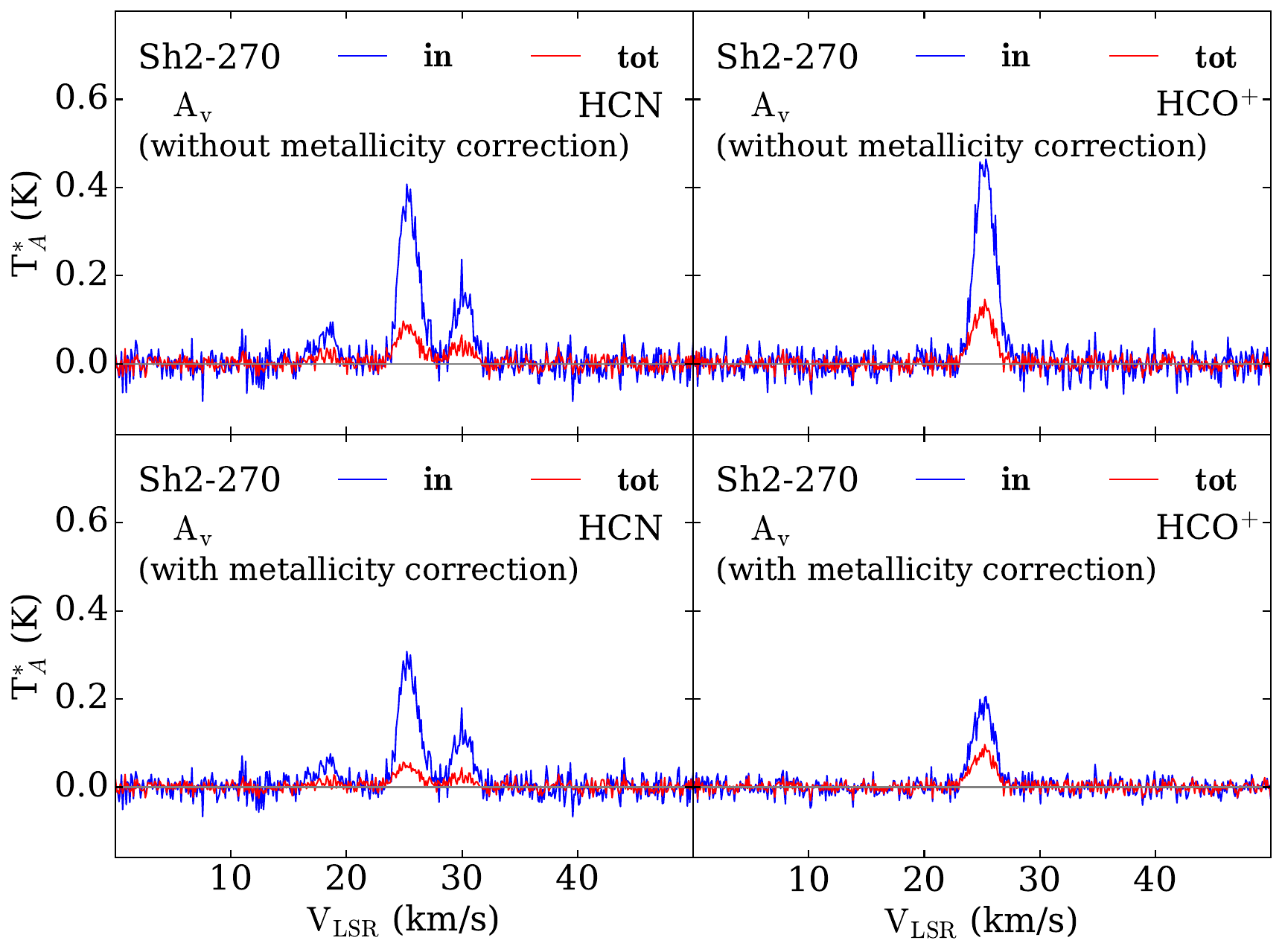}
    \caption{(continued) Same as described earlier for the clouds Sh2-128, Sh2-209, and Sh2-270, based on `Gas-based analysis'. The upper panel of Sh2-209 has no blue line because there is no region above $N_{\rm{H_{2}}}  \geq 8 \times 10^{21}\ \rm{cm^{-2}}$. }
\end{figure*}

\subsubsection{Line Luminosity Estimation}\label{subsubsec:luminosity}

To determine the line luminosity of \hcn\ and \hcop\ within the dense region ($L_{\rm X,in}$), we identified pixels that meet the column density threshold criteria for the ``dense" region.
To calculate the luminosity from the entire cloud ($L_{\rm X,tot}$), we used the outer boundary condition of column density, $N_{\rm{H_{2}}}  \geq 1.5\ee{21}\ \rm{cm^{-2}}$.
The velocity range for \hcop\ is the same as for \coo\, while for \hcn, the velocity range is broader (approximately 3-6 \kms\ on each side, according to target requirements, as reported in Table A.1) compared to \coo\ to accommodate the hyperfine structure. 
Our luminosity and uncertainty in luminosity calculations are based on the method presented in \cite{2022AJ....164..129P}. 
The luminosity for the dense region is defined as follows
\begin{equation}
        L_{\rm X,in} = D^{2}I_{\rm X}N_{\rm in}\Omega_{\rm pix} 
\end{equation}
where $N_{\rm in}$ represents the number of pixels 
accounted as ``dense" region, $\Omega_{\rm pix}$ denotes the solid angle of a pixel, $D$ is the heliocentric distance of the target in pc and $I_{\rm X}$ is the average integrated intensity for the velocity interval spanning from $v_{\rm sp,l}$ to $v_{\rm sp,u}$,  which are the lower and upper limits of the line spectral window. 
We utilize the spectra (Figure \ref{fig:spectra_figure}) that are averaged over the relevant contours  to determine the average integrated intensity.
\begin{equation}
    I_{\rm X} = \int_{v_{\rm sp, l}}^{v_{\rm sp, u}} T_X dv
\end{equation}

We calculated the uncertainty in luminosity using the following equation
\begin{equation}
    \sigma_{L} = D^{2} \delta v \sqrt{N_{ch}} \  \sigma_{T_X} \ N_{\rm in} \Omega_{\rm pix},
\end{equation}
where, $N_{ch}$ and $\delta v$ represents the number of channels within the line region and channel width, respectively, and $\sigma_{T_X}$ denotes the rms noise in the baseline of the averaged spectrum. In a similar way, we have estimated the luminosity from the entire  cloud ($L_{\rm X, tot}$) by considering the pixels satisfying $N_{\rm{H_{2}}} (\mathrm{and} \ N_{\rm{H_{2}}}^{'}) \geq 1.5 \times 10^{21}\ \rm{cm^{-2}}$.

Figure \ref{fig:spectra_figure} shows the average \hcn\ and \hcop\ spectrum for the new targets, showing both ``in" and ``total" regions.
The left column presents average spectrum from the gas-based analysis, while the right column shows the dust-based analysis. 
Sh2-128, Sh2-209, and Sh2-270 are included only in the gas-based analysis.
Red lines represent the spectra averaged over the entire cloud, with a column density threshold of $N_{\rm{H_{2}}} (\mathrm{and} \ N_{\rm{H_{2}}}^{'}) \geq 1.5 \times 10^{21}\ \rm{cm^{-2}}$ for the top (no metallicity correction) and bottom (with $Z-$correction) panels.
The blue lines in the left column represent the spectra averaged over pixels that satisfy the ``in'' condition  based on column density criterion for both the no $Z-$correction (top panel) and with $Z-$correction (bottom panel).
In the right column, the blue lines show the average spectrum from the BGPS mask region.

We tabulated the logarithmic values of the luminosity with and without metallicity correction ($L_{\rm Q}$ and $L_{\rm Q}^{'}$) in Table \ref{tab:gas_without_Z} and \ref{tab:gas_with_Z}, respectively. 
The luminosity values for the ``in" region ($L_{\rm in}$) of Sh2-209, Sh2-212, and \inncloudd\ are not included in Table \ref{tab:gas_without_Z}. 
This omission is due to the absence of pixels with $N_{\rm{H_{2}}} \geq 8 \times 10^{21}\ \rm{cm^{-2}}$ when the metallicity correction is not applied.
After applying the metallicity correction in the molecular hydrogen column density estimation, \inncloudc\ and \inncloudd\ have no pixels satisfying the ``dense" condition ($N_{\rm{H_{2}}}^{'} \geq 8 \times 10^{21}\ \rm{cm^{-2}}$), see Table \ref{tab:gas_with_Z}. 
Interestingly, despite the lack of dense pixels, all these sources exhibit significant 24 \micron\ luminosities and are associated with \hii\ regions \citep{2016ApJ...831...73V}, which clearly indicates ongoing  or recent star formation.  A lack of dense gas now can indicate that star formation is about to pause or end.

\begin{deluxetable*}{l c | c c c c | c   c c c c c}
    \tablenum{3}
    \tabletypesize{\footnotesize}
    \tablecaption{$M_{\rm dg, Gas}$ versus Luminosities (without metallicity correction) \label{tab:gas_without_Z}}
    \tablewidth{0pt}
    \tablehead{
    \colhead{Source} & \colhead{$\mathrm{Log}\ M_{\rm dg, Gas}$} & \colhead{Log $L_{\rm tot}$} & \colhead{Log $L_{\rm in}$} & \colhead{$\mathrm{Log \ \alpha_{tot}}$} & \colhead{$\mathrm{Log \ \alpha_{in}}$} &  \colhead{Log $L_{\rm tot}$} & \colhead{Log $L_{\rm in}$} & \colhead{$\mathrm{Log \ \alpha_{tot}}$} & \colhead{$\mathrm{Log \ \alpha_{in}}$} \\ 
    \cline{3-6}\cline{7-10}  \colhead{} & \colhead{(\msun)} & \multicolumn{4}{c}{\hcn} &  \multicolumn{4}{c}{\hcop}
    }
    \startdata
        & \multicolumn{6}{c}{Outer Galaxy Targets}\\
        \hline
        Sh2-128 & $3.35^{+0.11}_{-0.11}$ & $2.39^{+0.12}_{-0.11}$ & $1.77^{+0.11}_{-0.11}$  & $0.96^{+0.12}_{-0.11}$  & $1.58^{+0.11}_{-0.11}$ & $2.47^{+0.11}_{-0.11}$ & $1.65^{+0.11}_{-0.11}$ & $0.88^{+0.12}_{-0.11}$ & $1.70^{+0.11}_{-0.11}$\\
        Sh2-132 & $3.24^{+0.05}_{-0.05}$ & $2.57^{+0.05}_{-0.06}$ & $1.36^{+0.05}_{-0.06}$  & $0.67^{+0.05}_{-0.06}$  & $1.88^{+0.05}_{-0.06}$ & $2.58^{+0.05}_{-0.06}$ & $1.33^{+0.05}_{-0.06}$ & $0.66^{+0.05}_{-0.06}$ & $1.91^{+0.05}_{-0.06}$\\
        Sh2-142 & $1.48^{+0.05}_{-0.05}$ & $1.70^{+0.05}_{-0.05}$ & $-0.28^{+0.05}_{-0.05}$ & $-0.22^{+0.05}_{-0.05}$ & $1.76^{+0.05}_{-0.05}$ & $1.65^{+0.05}_{-0.05}$ &$-0.40^{+0.05}_{-0.05}$ & $-0.17^{+0.05}_{-0.05}$ & $1.88^{+0.05}_{-0.05}$ \\
        Sh2-148 & $4.46^{+0.05}_{-0.05}$ & $2.43^{+0.05}_{-0.05}$ & $2.31^{+0.05}_{-0.05}$  & $2.03^{+0.05}_{-0.05}$  & $2.15^{+0.05}_{-0.05}$ & $2.57^{+0.05}_{-0.05}$ & $2.30^{+0.05}_{-0.05}$  & $1.89^{+0.05}_{-0.05}$ & $2.09^{+0.05}_{-0.05}$\\
        Sh2-156 & $4.46^{+0.07}_{-0.07}$ & $2.82^{+0.07}_{-0.07}$ & $2.54^{+0.07}_{-0.07}$  & $1.64^{+0.07}_{-0.07}$  & $1.92^{+0.07}_{-0.07}$ & $2.81^{+0.07}_{-0.07}$ & $2.52^{+0.07}_{-0.07}$ & $1.65^{+0.07}_{-0.07}$ & $1.94^{+0.07}_{-0.07}$\\
        Sh2-209 & \nodata                & $2.09^{+0.07}_{-0.08}$ & \nodata             & \nodata & \nodata & $2.39^{+0.06}_{-0.07}$ &\nodata &\nodata &\nodata\\
        Sh2-242 & $2.97^{+0.02}_{-0.02}$ & $1.61^{+0.03}_{-0.02}$     & $1.08^{+0.02}_{-0.02}$    & $1.36^{+0.03}_{-0.02}$  & $1.89^{+0.02}_{-0.02}$ &  $1.55^{+0.03}_{-0.02}$      & $0.94^{+0.02}_{-0.02}$      &  $1.42^{+0.03}_{-0.02}$      & $2.03^{+0.02}_{-0.02}$        \\
        Sh2-266 & $3.09^{+0.09}_{-0.11}$ & $1.96^{+0.09}_{-0.11}$ & $1.25^{+0.09}_{-0.11}$  & $1.13^{+0.09}_{-0.11}$  & $1.84^{+0.09}_{-0.11}$ & $2.06^{+0.09}_{-0.11}$ & $1.30^{+0.09}_{-0.11}$      & $1.03^{+0.09}_{-0.11}$ & $1.79^{+0.09}_{-0.11}$\\
        Sh2-269 & $3.77^{+0.07}_{-0.07}$ & $2.09^{+0.07}_{-0.07}$ & $1.80^{+0.07}_{-0.07}$  & $1.68^{+0.07}_{-0.07}$  & $1.97^{+0.07}_{-0.07}$ & $2.07^{+0.07}_{-0.07}$ & $1.77^{+0.07}_{-0.07}$      & $1.70^{+0.07}_{-0.07}$ & $2.00^{+0.07}_{-0.07}$\\
        Sh2-270 & $3.62^{+0.12}_{-0.17}$ & $2.11^{+0.12}_{-0.17}$ & $1.64^{+0.12}_{-0.17}$  & $1.51^{+0.12}_{-0.17}$  & $1.98^{+0.12}_{-0.17}$ & $2.03^{+0.12}_{-0.17}$ & $1.54^{+0.12}_{-0.17}$ & $1.59^{+0.12}_{-0.17}$ & $2.08^{+0.12}_{-0.17}$\\
        \hline
        & \multicolumn{6}{c}{Outer Galaxy Targets from \cite{2022AJ....164..129P}}\\
        \hline
        Sh2-206 & $3.02^{+0.05}_{-0.05}$ & $1.88^{+0.05}_{-0.05}$ & $1.14^{+0.05}_{-0.05}$ & $1.14^{+0.05}_{-0.05}$ & $1.88^{+0.05}_{-0.05}$ & $1.84^{+0.05}_{-0.05}$ & $1.05^{+0.05}_{-0.05}$ &  $1.18^{+0.05}_{-0.05}$ & $1.97^{+0.05}_{-0.05}$ \\
        Sh2-208 & $2.58^{+0.05}_{-0.06}$ & $1.31^{+0.06}_{-0.06}$ & $0.63^{+0.06}_{-0.06}$ & $1.27^{+0.06}_{-0.06}$ & $1.95^{+0.06}_{-0.06}$ & $1.39^{+0.05}_{-0.06}$ & $0.62^{+0.05}_{-0.06}$ &  $1.19^{+0.06}_{-0.06}$ & $1.96^{+0.06}_{-0.06}$\\
        Sh2-212 & \nodata                & $1.89^{+0.15}_{-0.21}$ & \nodata         & \nodata & \nodata & $1.95^{+0.15}_{-0.21}$ & \nodata                &  \nodata & \nodata\\
        Sh2-228 & $1.84^{+0.03}_{-0.03}$ & $1.40^{+0.05}_{-0.05}$ & $-0.44^{+0.08}_{-0.09}$& $0.44^{+0.05}_{-0.05}$ & $2.28^{+0.08}_{-0.09}$ & $1.29^{+0.04}_{-0.05}$ & $-0.74^{+0.10}_{-0.13}$ & $0.55^{+0.05}_{-0.05}$ & $2.59^{+0.08}_{-0.09}$\\
        Sh2-235 & $3.89^{+0.03}_{-0.04}$ & $2.45^{+0.04}_{-0.04}$ & $2.18^{+0.03}_{-0.04}$ & $1.44^{+0.04}_{-0.04}$ & $1.71^{+0.03}_{-0.04}$ & $2.43^{+0.03}_{-0.04}$ & $2.10^{+0.03}_{-0.04}$ &  $1.46^{+0.03}_{-0.04}$ & $1.79^{+0.03}_{-0.04}$\\
        Sh2-252 & $4.33^{+0.05}_{-0.05}$ & $2.53^{+0.05}_{-0.05}$ & $2.35^{+0.05}_{-0.05}$ & $1.80^{+0.05}_{-0.05}$ & $1.98^{+0.05}_{-0.05}$ & $2.48^{+0.05}_{-0.05}$ & $2.27^{+0.05}_{-0.05}$ &  $1.85^{+0.05}_{-0.05}$ & $2.06^{+0.05}_{-0.05}$\\
        Sh2-254 & $3.91^{+0.05}_{-0.04}$ & $2.50^{+0.05}_{-0.04}$ & $2.16^{+0.05}_{-0.04}$ & $1.41^{+0.05}_{-0.04}$ & $1.75^{+0.05}_{-0.04}$ & $2.50^{+0.05}_{-0.04}$ & $2.06^{+0.05}_{-0.04}$ &  $1.41^{+0.05}_{-0.04}$ & $1.85^{+0.05}_{-0.04}$ \\
        \hline
        & \multicolumn{6}{c}{Inner Galaxy Targets from \cite{2020ApJ...894..103E}}\\
        \hline
        \inncloudb & $4.51^{+0.03}_{-0.04}$ & $3.25^{+0.03}_{-0.04}$ & $2.80^{+0.03}_{-0.04}$ & $1.26^{+0.03}_{-0.04}$ & $1.71^{+0.03}_{-0.04}$ & $3.20^{+0.03}_{-0.04}$ & $2.68^{+0.03}_{-0.04}$ & $1.31^{+0.03}_{-0.04}$ & $1.83^{+0.03}_{-0.04}$ \\
        \inncloudc & $2.72^{+0.05}_{-0.06}$ & $2.80^{+0.05}_{-0.07}$ & $0.75^{+0.06}_{-0.08}$ & $-0.08^{+0.05}_{-0.07}$ & $1.97^{+0.06}_{-0.08}$ & $2.71^{+0.05}_{-0.07}$ &$0.82^{+0.06}_{-0.07}$ & $0.01^{+0.05}_{-0.07}$ & $1.90^{+0.06}_{-0.07}$ \\
        \inncloudd & \nodata                & $2.12^{+0.03}_{-0.04}$ & \nodata                &        & \nodata & $1.94^{+0.03}_{-0.03}$ & \nodata & \nodata & \nodata\\
        \inncloude & $4.49^{+0.05}_{-0.07}$ & $3.02^{+0.05}_{-0.08}$ & $2.41^{+0.05}_{-0.07}$ & $1.47^{+0.05}_{-0.08}$ & $2.08^{+0.05}_{-0.07}$ & $2.85^{+0.06}_{-0.08}$ & $2.17^{+0.06}_{-0.08}$ & $1.64^{+0.06}_{-0.08}$ & $2.32^{+0.06}_{-0.08}$ \\
        \inncloudf & $3.30^{+0.14}_{-0.17}$ & $2.24^{+0.14}_{-0.17}$ & $1.43^{+0.14}_{-0.17}$ & $1.06^{+0.14}_{-0.17}$ & $1.87^{+0.14}_{-0.17}$ & $2.14^{+0.14}_{-0.17}$      & $1.34^{+0.14}_{-0.17}$ & $1.16^{+0.14}_{-0.17}$ & $1.96^{+0.14}_{-0.17}$ \\
        \tableline
    \enddata
    \tablecomments{1. Units of luminosities are \kkms pc$^2$.\\ 2. Units of conversion factors are $\msun \ (\rm K\ km\ s^{-1}\ pc^{2})^{-1}$. \\
    3. Sh2-142, Sh2-228, and \inncloudc\ have fewer than 9 pixels meeting the column density criterion for dense region, hence excluded for reliability.
    }
\end{deluxetable*}


\begin{deluxetable*}{l c | c c c c | c   c c c c c}
    \tablenum{4}
    \tabletypesize{\footnotesize}
    \tablecaption{$M_{\rm dg, Gas}^{'}$ versus Luminosities (with metallicity correction) \label{tab:gas_with_Z}}
    \tablewidth{0pt}
    \tablehead{
    \colhead{Source} & \colhead{$\mathrm{Log}\ M_{\rm dg, Gas}^{'}$} & \colhead{Log $L_{\rm tot}^{'}$} & \colhead{Log $L_{\rm in}^{'}$} & \colhead{$\mathrm{Log \ \alpha'_{tot}}$} & \colhead{$\mathrm{Log \ \alpha'_{in}}$} & \colhead{Log $L_{\rm tot}^{'}$} & \colhead{Log $L_{\rm in}^{'}$} & \colhead{$\mathrm{Log \ \alpha'_{tot}}$} & \colhead{$\mathrm{Log \ \alpha'_{in}}$} \\
    \cline{3-6}\cline{7-10}  \colhead{} & \colhead{(\msun)} & \multicolumn{4}{c}{\hcn} &  \multicolumn{4}{c}{\hcop}
    }
    \startdata
        & \multicolumn{6}{c}{Outer Galaxy Targets}\\
        \hline
        Sh2-128 & $4.23^{+0.11}_{-0.11}$ & $2.45^{+0.12}_{-0.12}$ & $2.21^{+0.11}_{-0.11}$  & $1.78^{+0.12}_{-0.12}$ & $2.02^{+0.11}_{-0.11}$ & $2.61^{+0.12}_{-0.11}$  & $2.14^{+0.11}_{-0.11}$ & $1.62^{+0.12}_{-0.11}$ & $2.09^{+0.11}_{-0.11}$ \\
        Sh2-132 & $3.69^{+0.05}_{-0.05}$ & $2.62^{+0.05}_{-0.05}$ & $1.77^{+0.05}_{-0.06}$  & $1.06^{+0.05}_{-0.05}$ & $1.92^{+0.05}_{-0.06}$ & $2.65^{+0.05}_{-0.06}$  & $1.68^{+0.05}_{-0.06}$ & $1.04^{+0.05}_{-0.06}$ & $2.01^{+0.05}_{-0.06}$ \\
        Sh2-142 & $2.32^{+0.05}_{-0.05}$ & $1.72^{+0.05}_{-0.05}$ & $0.47^{+0.05}_{-0.05}$  & $0.60^{+0.05}_{-0.05}$ & $1.84^{+0.05}_{-0.05}$ & $1.67^{+0.05}_{-0.05}$  &  $0.37^{+0.05}_{-0.05}$ & $0.65^{+0.05}_{-0.05}$ & $1.95^{+0.05}_{-0.05}$ \\
        Sh2-148 & $4.56^{+0.05}_{-0.05}$ & $2.42^{+0.05}_{-0.05}$ &  $2.34^{+0.05}_{-0.05}$ & $2.14^{+0.05}_{-0.05}$ & $2.22^{+0.05}_{-0.05}$ & $2.58^{+0.05}_{-0.05}$  &  $2.41^{+0.05}_{-0.05}$ & $1.98^{+0.05}_{-0.05}$ & $2.15^{+0.05}_{-0.05}$ \\
        Sh2-156 & $4.74^{+0.07}_{-0.07}$ & $2.83^{+0.07}_{-0.07}$ & $2.66^{+0.07}_{-0.07}$  & $1.91^{+0.07}_{-0.07}$ & $2.08^{+0.07}_{-0.07}$ & $2.83^{+0.07}_{-0.07}$  & $2.63^{+0.07}_{-0.07}$ & $1.91^{+0.07}_{-0.07}$ & $2.11^{+0.07}_{-0.07}$ \\
        Sh2-209 & $3.85^{+0.06}_{-0.07}$ & $2.67^{+0.07}_{-0.08}$ & $1.16^{+0.10}_{-0.12}$  & $1.18^{+0.07}_{-0.08}$ & $2.69^{+0.10}_{-0.12}$ & $2.51^{+0.07}_{-0.09}$  &  $1.90^{+0.06}_{-0.07}$ & $1.33^{+0.07}_{-0.09}$ & $1.95^{+0.06}_{-0.07}$ \\
        Sh2-242 & $2.78^{+0.02}_{-0.02}$ & $1.59^{+0.03}_{-0.02}$ & $0.97^{+0.02}_{-0.02}$  & $1.19^{+0.03}_{-0.02}$ & $1.80^{+0.02}_{-0.02}$ & $1.52^{+0.02}_{-0.02}$  &  $0.81^{+0.02}_{-0.02}$     & $1.26^{+0.02}_{-0.02}$ & $1.97^{+0.02}_{-0.02}$ \\
        Sh2-266 & $3.89^{+0.09}_{-0.11}$ & $2.06^{+0.09}_{-0.11}$ & $1.71^{+0.09}_{-0.11}$  & $1.83^{+0.09}_{-0.11}$ & $2.18^{+0.09}_{-0.11}$ & $2.20^{+0.09}_{-0.11}$  &  $1.78^{+0.09}_{-0.11}$     & $1.69^{+0.09}_{-0.11}$ & $2.11^{+0.09}_{-0.11}$ \\
        Sh2-269 & $3.90^{+0.07}_{-0.07}$ & $2.10^{+0.07}_{-0.07}$ & $1.85^{+0.07}_{-0.07}$  & $1.80^{+0.07}_{-0.07}$ & $2.05^{+0.07}_{-0.07}$ & $2.09^{+0.07}_{-0.07}$  &   $1.81^{+0.07}_{-0.07}$    & $1.81^{+0.07}_{-0.07}$ & $2.09^{+0.07}_{-0.07}$ \\
        Sh2-270 & $4.44^{+0.12}_{-0.17}$ & $2.21^{+0.12}_{-0.17}$ & $1.96^{+0.12}_{-0.17}$  & $2.23^{+0.12}_{-0.17}$ & $2.48^{+0.12}_{-0.17}$ &  $2.16^{+0.12}_{-0.17}$ & $1.86^{+0.12}_{-0.17}$ & $2.27^{+0.12}_{-0.17}$ & $2.58^{+0.12}_{-0.17}$ \\
        \hline
        & \multicolumn{6}{c}{Outer Galaxy Targets from \cite{2022AJ....164..129P}}\\
        \hline
        Sh2-206 & $3.27^{+0.05}_{-0.05}$ & $1.93^{+0.05}_{-0.05}$     & $1.27^{+0.05}_{-0.05}$ & $1.34^{+0.05}_{-0.05}$  & $2.00^{+0.05}_{-0.05}$  &  $1.89^{+0.05}_{-0.05}$     & $1.19^{+0.05}_{-0.05}$ & $1.38^{+0.05}_{-0.05}$ & $2.07^{+0.05}_{-0.05}$        \\
        Sh2-208 & $2.73^{+0.05}_{-0.06}$ & $1.33^{+0.06}_{-0.06}$     & $0.85^{+0.06}_{-0.06}$ & $1.40^{+0.06}_{-0.06}$  & $1.88^{+0.06}_{-0.06}$  &  $1.43^{+0.05}_{-0.06}$     & $0.84^{+0.05}_{-0.06}$ & $1.30^{+0.05}_{-0.06}$ & $1.89^{+0.05}_{-0.06}$        \\
        Sh2-212 & $1.88^{+0.22}_{-0.43}$ & $1.96^{+0.15}_{-0.21}$     & $0.08^{+0.15}_{-0.21}$ & $-0.08^{+0.15}_{-0.21}$ & $1.79^{+0.15}_{-0.21}$  &  $2.03^{+0.15}_{-0.21}$     & $0.04^{+0.15}_{-0.21}$ & $-0.15^{+0.15}_{-0.21}$ & $1.84^{+0.15}_{-0.21}$        \\
        Sh2-228 & $2.87^{+0.03}_{-0.03}$ & $1.42^{+0.05}_{-0.06}$     & $0.60^{+0.04}_{-0.05}$ & $1.45^{+0.05}_{-0.06}$  & $2.27^{+0.04}_{-0.05}$  &  $1.32^{+0.05}_{-0.05}$     & $0.51^{+0.04}_{-0.04}$ & $1.55^{+0.05}_{-0.05}$ & $2.36^{+0.04}_{-0.04}$        \\
        Sh2-235 & $4.03^{+0.03}_{-0.04}$ & $2.45^{+0.04}_{-0.04}$     & $2.22^{+0.03}_{-0.04}$ & $1.57^{+0.04}_{-0.04}$  & $1.80^{+0.03}_{-0.04}$  &  $2.44^{+0.04}_{-0.04}$     & $2.16^{+0.03}_{-0.04}$ & $1.58^{+0.04}_{-0.04}$ & $1.87^{+0.03}_{-0.04}$        \\
        Sh2-252 & $4.32^{+0.05}_{-0.05}$ & $2.53^{+0.05}_{-0.05}$     & $2.34^{+0.05}_{-0.05}$ & $1.79^{+0.05}_{-0.05}$  & $1.97^{+0.05}_{-0.05}$  &  $2.48^{+0.05}_{-0.05}$     & $2.26^{+0.05}_{-0.05}$ & $1.83^{+0.05}_{-0.05}$ & $2.05^{+0.05}_{-0.05}$        \\
        Sh2-254 & $4.09^{+0.05}_{-0.04}$ & $2.51^{+0.05}_{-0.04}$     & $2.23^{+0.05}_{-0.04}$ & $1.58^{+0.05}_{-0.04}$ & $1.86^{+0.05}_{-0.04}$   &  $2.51^{+0.05}_{-0.04}$     & $2.15^{+0.05}_{-0.04}$ & $1.57^{+0.05}_{-0.04}$ & $1.94^{+0.05}_{-0.04}$        \\
        \hline
        & \multicolumn{6}{c}{Inner Galaxy Targets from \cite{2020ApJ...894..103E}}\\
        \hline
        \inncloudb & $4.13^{+0.03}_{-0.04}$ & $3.23^{+0.03}_{-0.04}$  & $2.57^{+0.03}_{-0.04}$ & $0.90^{+0.03}_{-0.04}$  & $1.56^{+0.03}_{-0.04}$  &  $3.17^{+0.03}_{-0.04}$  & $2.42^{+0.03}_{-0.04}$ & $0.96^{+0.03}_{-0.04}$ & $1.71^{+0.03}_{-0.04}$ \\
        \inncloudc & \nodata                & $2.76^{+0.05}_{-0.07}$  & \nodata                & \nodata                 & \nodata                 &  $2.68^{+0.05}_{-0.07}$  & \nodata & \nodata & \nodata \\
        \inncloudd & \nodata                & $2.02^{+0.03}_{-0.03}$  & \nodata                & \nodata                 & \nodata                 &  $1.87^{+0.03}_{-0.03}$  & \nodata & \nodata & \nodata \\
        \inncloude & $4.25^{+0.05}_{-0.07}$ & $3.00^{+0.05}_{-0.08}$  & $2.23^{+0.05}_{-0.07}$ & $1.25^{+0.05}_{-0.08}$ & $2.02^{+0.05}_{-0.07}$   &  $2.83^{+0.06}_{-0.08}$  & $2.01^{+0.06}_{-0.08}$ & $1.41^{+0.06}_{-0.08}$ & $2.24^{+0.06}_{-0.08}$ \\
        \inncloudf & $3.01^{+0.14}_{-0.17}$ & $2.23^{+0.14}_{-0.17}$  & $1.19^{+0.14}_{-0.17}$ & $0.78^{+0.14}_{-0.17}$ &  $1.82^{+0.14}_{-0.17}$  &  $2.12^{+0.14}_{-0.17}$  & $1.09^{+0.14}_{-0.17}$ & $0.89^{+0.14}_{-0.17}$ & $1.92^{+0.14}_{-0.17}$ \\
        \tableline
    \enddata
    \tablecomments{1. Units of luminosities are \kkms pc$^2$.\\
    2. Units of conversion factors are $\msun \ (\rm K\ km\ s^{-1}\ pc^{2})^{-1}$. \\
    3. Sh2-212 has only 1 pixel meeting the column density criterion for dense region, hence excluded for reliability.
    } 
\end{deluxetable*}


\subsubsection{Dense Gas Mass Measurement}\label{subsubsec:mass_gas}
We have calculated the mass of dense gas by considering the pixels satisfying the ``dense" condition ($N_{\rm{H_{2}}} (\mathrm{and} \ N_{\rm{H_{2}}}^{'}) \geq 8 \times 10^{21}\ \rm{cm^{-2}}$), with and without metallicity correction. 
The calculation of dense gas mass ($M_{\rm dg, Gas}$) from the gas based analysis is performed using the following equation.

\begin{equation}
    M_{\rm dg, Gas} = \mu_{\rm H_{2}} m_{\rm H} A_{\rm pix} \Sigma N_{\rm H_{2}}
\end{equation}
where $\mu_{\rm H_{2}}$ is the mean molecular weight, which is 2.809 \citep{2022ApJ...929L..18E}, $m_{\rm H}$ is the mass of the hydrogen atom, $A_{\rm pix}$ is the area of a pixel in $\rm cm^{2}$ and $\Sigma N_{\rm H_{2}}$ is the sum of molecular hydrogen column density satisfying the condition for the ``dense" region. 
The dense gas mass values in logarithmic scale for all the targets are  listed in Table \ref{tab:gas_without_Z}.
%
However, Sh2-142, Sh2-228, and \inncloudc\ have only 2, 6 and 4 pixels satisfying the above criterion. 
We have excluded clouds with less than 9 pixels satisfying the column density criterion, providing sufficient resolution of the emission region to ensure the reliability and statistical significance of our analysis.

The metallicity-corrected mass values are obtained using the following equation 
\begin{equation}
    M_{\rm dg, Gas}^{'} = \mu_{\rm H_{2}} m_{\rm H} A_{\rm pix} \Sigma N_{\rm H_{2}}^{'}
\end{equation}
where $N_{\rm H_{2}}^{'}$ corresponds to the molecular hydrogen column density after applying the metallicity correction.
%
We have excluded Sh2-212 because only one pixel meets the ``dense" criterion.
We tabulated the logarithmic values of $M_{\rm dg, Gas}^{'}$ in Table \ref{tab:gas_with_Z}.

\begin{deluxetable*}{l c | c c c c | c c c c}
    \tablenum{5}
    \tabletypesize{\footnotesize}
    \tablecaption{Statistical summary: average mass-luminosity conversion factors from Gas Analysis\label{tab:summary_gas}}
    \tablewidth{0pt}
    \tablehead{
    \colhead{}  &   & \colhead{$\rm Log\ \alpha_{\rm in, Gas}$} &  & \colhead{$\rm Log\ \alpha_{\rm tot, Gas}$} & & &  \colhead{$\rm Log\ \alpha'_{\rm in, Gas}$}  & & \colhead{$\rm Log\ \alpha'_{\rm tot, Gas}$} 
    }
    \startdata
    \hline
    & \multicolumn{9}{c}{\hcn}\\
    \hline
    Inner Galaxy &   &    $1.89 \pm 0.18 $  &   &   $1.26 \pm 0.20 $   & & &   $1.80 \pm 0.23$       &      & $0.98 \pm 0.24$ \\
    Outer Galaxy &   &    $1.88 \pm 0.13 $  &   &   $1.39 \pm 0.35 $   & & &   $2.07 \pm 0.24$       &      & $1.55 \pm 0.41$ \\
    \hline
    & \multicolumn{9}{c}{\hcop}\\
    \hline
    Inner Galaxy &   &    $2.04 \pm 0.25 $  &   &   $1.37 \pm 0.24$    & & &   $1.96 \pm 0.27$       &   & $1.09 \pm 0.28$ \\
    Outer Galaxy &   &    $1.94 \pm 0.12 $  &   &   $1.38 \pm 0.36$    & & &   $2.08 \pm 0.17$       &   & $1.55 \pm 0.38$ \\
    \tableline
    \enddata
    \tablecomments{1. Units of conversion factors are $\msun \ (\rm K\ km\ s^{-1}\ pc^{2})^{-1}$.}
\end{deluxetable*}


\begin{figure*}
    \centering
    \includegraphics[scale=0.49]{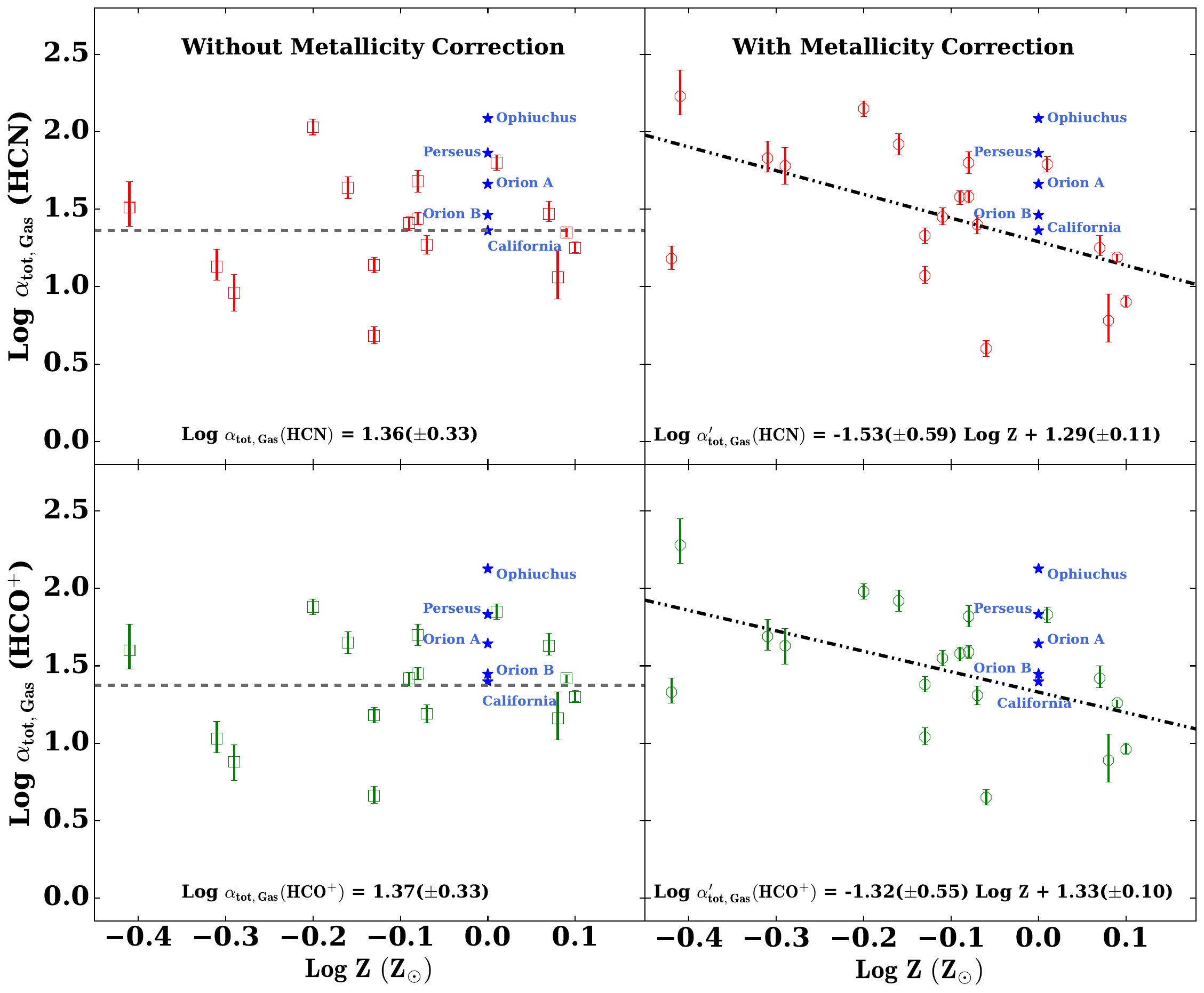}  
    \caption{Variation of $\alpha_{\rm tot, Gas}$ for \hcn\ (top panels) and \hcop\ (bottom panels) derived from gas based analysis with metallicity, with and without metallicity correction. The blue stars correspond to $\alpha_{\rm tot, Gas}$ values for local clouds with solar metallicity taken from the literature (see Section \ref{solar}).  The horizontal dashed line in the top and bottom left panels shows the mean value of $\alpha_{\rm tot, Gas}$. The dash-dotted line in the top and bottom right panels are the result of fitting the data. In both the cases local clouds are not considered.} 
    \label{fig:figure1}
\end{figure*}

\subsubsection{Mass Conversion factor: The alpha-factor ($\alpha_{Q}$) from Gas-based analysis}\label{subsubsec:alpha_gas}

In this section, we describe how we calculated the $\alpha-$factors using the gas-based analysis. 
We computed $\alpha_{\rm Q}$ by limiting the analysis to the luminosity emanating from the ``dense" molecular gas and also for the total luminosity of the bulk molecular gas.
We have estimated the mass conversion factor $\alpha_{\rm in}$ by dividing the mass of dense gas by the luminosity confined to the dense region ($\alpha_{\rm in, Gas}= M_{\rm dg, Gas}/L_{\rm in}$).
To enable a more appropriate comparison with extra-galactic observations, which encompass global cloud measurements of dense gas (\hcn, \hcop) emissions across lower extinctions and densities, we calculated $\alpha_{\rm tot, Gas}$ using the formula $\alpha_{\rm tot, Gas} = M_{\rm dg, Gas}/L_{\rm tot}$. 
In Table \ref{tab:gas_without_Z}, we have listed the values of $\alpha_{\rm in, Gas}$ and $\alpha_{\rm tot, Gas}$ in logarithmic scale (without considering the metallicity correction) for both \hcn\ and \hcop.
We have also estimated the mass conversion factors $\alpha'_{\rm in, Gas}$ and $\alpha'_{\rm tot, Gas}$ after considering the metallicity correction  in Table \ref{tab:gas_with_Z}, where $\alpha'_{\rm in, Gas}= M_{\rm dg, Gas}^{'}/L_{\rm in}^{'}$ and $\alpha'_{\rm tot, Gas}= M_{\rm dg, Gas}^{'}/L_{\rm tot}^{'}$.

We estimated the statistical parameters (mean and standard deviation) for mass conversion factors (logarithmic scale) for both the inner and outer Galaxy clouds separately. 
We have estimated the mean values, $\langle \mathrm{log}\ \alpha_{\rm in, Gas} \rangle$ and $\langle \mathrm{log}\ \alpha_{\rm tot, Gas} \rangle$,
based on 13 targets from the outer Galaxy for this analysis, namely Sh2-128, Sh2-132, Sh2-148, Sh2-156, Sh2-242, Sh2-266, Sh2-269, Sh2-270, Sh2-206, Sh2-208, Sh2-235, Sh2-252, and Sh2-254. Also, we have considered 3 targets from the inner Galaxy, namely \inncloudb, \inncloude, and \inncloudf, for statistical analysis without applying metallicity corrections.
Sh2-212, Sh2-209 and \inncloudd\ have no pixels satisfying the `dense' condition based on molecular hydrogen column density. 
Sh2-142, Sh2-228, and \inncloudc\ have fewer than 9 pixels meeting the column density criterion for a dense region, and thus we have excluded these targets to ensure reliability (see Table \ref{tab:gas_without_Z}).

For the metallicity-corrected $\alpha-$factors, we have considered 16 outer Galaxy targets  and the same 3 inner galaxy targets for statistical analysis. After metallicity correction, all but Sh2-212 satisfy the criteria for inclusion.
\inncloudc\ and \inncloudd\ have no pixels satisfying the `dense' condition based on molecular hydrogen column density (see Table \ref{tab:gas_with_Z}).
Table \ref{tab:summary_gas} presents a summary of all mean and standard deviation values obtained from the analysis of both \hcn\ and \hcop\ data. 
We calculate means in the logarithmic values to avoid over-weighting outliers and for consistency with previous work. 
For discussion purposes, we convert the mean value of the logarithm to a value.

The $\langle \mathrm{log}\ \alpha_{\rm in, Gas} (\hcn) \rangle$  value is $1.89\pm0.18$  for inner Galaxy  and  $1.88\pm0.13$ for outer Galaxy targets. The mean values in linear scale, translated from logarithmic values, are approximately 78 and 76 $\msun \ (\rm K\ km\ s^{-1}\ pc^{2})^{-1}$, respectively.
Without metallicity correction, the values for the inner and outer Galaxy are very similar.

The metallicity-corrected mean mass conversion values $\langle \mathrm{log}\ \alpha'_{\rm in, Gas} (\hcn) \rangle$ is $1.80\pm0.23$ for inner Galaxy targets, and it is $2.07\pm0.24$ for the outer Galaxy sample. 
The mean values in linear scale are approximately 63 (inner) and 117.5 (outer) $\msun \ (\rm K\ km\ s^{-1}\ pc^{2})^{-1}$, respectively. 
Correction for metallicity creates a decrease in mass conversion factor values in the inner Galaxy, and an increase in the outer Galaxy. A similar pattern applies to \hcop\ as well.

For comparison to extra-galactic observations, the line emission from the entire cloud is used. 
The averages without metallicity correction for \hcn\ in Table \ref{tab:summary_gas}, translated to linear values, are 18.2 and 24.5 $\msun \ (\rm K\ km\ s^{-1}\ pc^{2})^{-1}$ in the inner and outer Galaxy, correspondingly.
In contrast, after applying the metallicity correction, the values in linear scale are $9.5$  $\msun \ (\rm K\ km\ s^{-1}\ pc^{2})^{-1}$ for inner Galaxy clouds and $35.5$ in the outer Galaxy.
Similarly, for \hcop, the uncorrected averages are 23.4 and 23.9 $\msun \ (\rm K\ km\ s^{-1}\ pc^{2})^{-1}$, while the corrected values are 12.3 and 35.5 $\msun \ (\rm K\ km\ s^{-1}\ pc^{2})^{-1}$ for inner and outer Galaxy clouds respectively.

In this section, we estimated $\alpha_{Q}$ through a gas-based analysis. We calculated the dense gas mass using a threshold column density derived from $N_{\rm H_{2}}$ maps based on \coo\ and \cotw\ data, both with and without considering the metallicity correction. 
The mass conversion factors exhibit significant dispersion among different targets. 
However, there is an overarching increasing trend observed across the dataset with increasing galactocentric distance. 

In Figure \ref{fig:figure1}, we have plotted the $\alpha_{\rm tot, Gas}\ \rm (and \ \alpha'_{\rm tot, Gas})$ with respect to metallicity. The panels on the left make no correction for metallicity in computing the mass. Once we correct for $Z$, an  inverse relation between mass conversion factor and metallicity is apparent for both \hcn\ and \hcop.  
We performed linear least-squares polynomial fits for both species using the numpy.polyfit function in Python. The fit was obtained by minimizing the sum of the squared residuals between the observed data and the polynomial model. 
The coefficients of the polynomial, along with their associated uncertainties, were determined. 
The uncertainties were derived from the covariance matrix of the fit, which accounts for the variance and covariance of the fitted parameters. The fit corresponds to the relations: $\alpha^{'}_{\textrm{ tot, Gas}}(\textrm{HCN}) = 19.5^{+5.6}_{-4.4} Z^{(-1.53 \pm 0.59)}$ and  $\alpha^{'}_{\textrm{ tot, Gas}}(\textrm{\hcop}) = 21.4^{+5.5}_{-4.4}  Z^{(-1.32\pm0.55)}$
The implications of these relations will be discussed in \S \ref{sec:discussion}.

\subsection{Dust based Analysis}\label{sec_4.2}

The dust thermal continuum emission at millimeter wavelengths is another way to characterize the ``dense” region.
In this section, we employ the same method to estimate luminosities of \hcn\ and \hcop, but use the dust thermal continuum emission from BGPS to identify the ``in" region. 
Also, we measure the dense gas mass based on dust thermal emission where we apply the correction for the metallicity in the gas-to-dust ratio.

\subsubsection{Line Luminosity Estimation}\label{sec_4.2.1}

To define the region of dense gas, as indicated by millimeter continuum emission, we downloaded the mask maps of BGPS from Bolocat V2.1.
The masked regions, where the source is above the $2\sigma$ threshold, are used as the ``in” region (see section \ref{sec:obs}). 
Combining both inner and outer Galaxy sample, 18 targets are covered by the BGPS survey (see Table \ref{tab:cloud details}).
We convolved and regridded the BGPS masks to the same grid as the TRAO maps. 
%
%
%
%
%
Next, we calculated the luminosities of \hcn\ and \hcop\ and their uncertainties for the ``in" region ($L_{\rm in}$). 
The BGPS ``in" regions are shown by the cyan contours in Figure \ref{fig:figure0}. 
This process follows the same methodology outlined in Section \ref{subsubsec:luminosity} and involves the use of equations 6, 7, and 8.
In the BGPS-based analysis, the value of $L_{\rm in}$ remains the same with and without metallicity correction. 
This is because we cannot implement any metallicity correction in the mask file. 
The total luminosities ($L_{\rm tot}$) are the same as those obtained using the column density condition ($N_{\rm{H_{2}}} (\mathrm{and} \ N_{\rm{H_{2}}}^{'}) \geq 1.5 \times 10^{21}\ \rm{cm^{-2}}$) in Section \ref{subsubsec:luminosity}. 
The luminosity values are given in Table \ref{tab:Mdense_bgps} and \ref{tab:Mdense_bgps_Z_elia_temp} in logarithmic scale.
The average spectra of \hcn\ and \hcop\ for the entire cloud and BGPS ``in" region are shown in red and blue lines for the new targets listed in Tables \ref{tab:Mdense_bgps} and \ref{tab:Mdense_bgps_Z_elia_temp} in the right column of Figure \ref{fig:spectra_figure}.

\begin{deluxetable*}{l c | c c c c | c c c c c}
    \tablenum{6}
    \tabletypesize{\footnotesize}
    \tablecaption{$M_{\rm dg, BGPS}$ versus Luminosities (without metallicity correction) \label{tab:Mdense_bgps}}
    \tablewidth{0pt}
    \tablehead{
    \colhead{Source} & \colhead{$\mathrm{Log}\ M_{\rm dg, BGPS}$} & \colhead{Log $L_{\rm tot}$} & \colhead{Log $L_{\rm in}$} & \colhead{$\mathrm{Log \ \alpha_{tot}}$} & \colhead{$\mathrm{Log \ \alpha_{in}}$} & \colhead{Log $L_{\rm tot}$} & \colhead{Log $L_{\rm in}$} & \colhead{$\mathrm{Log \ \alpha_{tot}}$} & \colhead{$\mathrm{Log \ \alpha_{in}}$} \\
    \cline{3-6}\cline{7-10}  \colhead{} & \colhead{(\msun)} & \multicolumn{4}{c}{\hcn} &  \multicolumn{4}{c}{\hcop}
    }
    \startdata
        & \multicolumn{6}{c}{Outer Galaxy Targets}\\
        \hline
        Sh2-132 &    $3.66^{+0.08}_{-0.09}$ & $2.57^{+0.05}_{-0.06}$ & $2.13^{+0.05}_{-0.06}$ & $1.09^{+0.08}_{-0.09}$ & $1.52^{+0.08}_{-0.09}$ & $2.58^{+0.05}_{-0.06}$ & $2.13^{+0.05}_{-0.06}$ & $1.07^{+0.08}_{-0.09}$ & $1.53^{+0.08}_{-0.09}$ \\
        Sh2-142 &    $2.90^{+0.09}_{-0.11}$ & $1.70^{+0.05}_{-0.05}$ & $1.43^{+0.05}_{-0.05}$ & $1.20^{+0.09}_{-0.11}$ & $1.47^{+0.09}_{-0.11}$ & $1.65^{+0.05}_{-0.05}$ & $1.36^{+0.05}_{-0.05}$ & $1.25^{+0.09}_{-0.11}$ & $1.54^{+0.09}_{-0.11}$ \\
        Sh2-148 &    $4.07^{+0.06}_{-0.06}$ & $2.43^{+0.05}_{-0.05}$ & $2.29^{+0.05}_{-0.05}$ & $1.64^{+0.06}_{-0.07}$ & $1.78^{+0.06}_{-0.06}$ & $2.58^{+0.05}_{-0.05}$ & $2.31^{+0.05}_{-0.05}$ & $1.49^{+0.06}_{-0.06}$ & $1.76^{+0.06}_{-0.06}$ \\
        Sh2-156 &    $3.65^{+0.08}_{-0.09}$ & $2.82^{+0.07}_{-0.07}$ & $2.37^{+0.07}_{-0.07}$ & $0.83^{+0.08}_{-0.09}$ & $1.28^{+0.08}_{-0.09}$ & $2.81^{+0.07}_{-0.07}$ & $2.35^{+0.07}_{-0.07}$ & $0.84^{+0.08}_{-0.09}$ & $1.30^{+0.08}_{-0.09}$ \\
        Sh2-242 &    $3.10^{+0.05}_{-0.06}$ & $1.61^{+0.03}_{-0.03}$ & $1.23^{+0.02}_{-0.02}$ & $1.48^{+0.06}_{-0.06}$ & $1.87^{+0.06}_{-0.06}$ & $1.55^{+0.03}_{-0.02}$ & $1.11^{+0.02}_{-0.02}$ & $1.55^{+0.06}_{-0.06}$ & $1.99^{+0.06}_{-0.06}$ \\
        Sh2-266 &    $3.28^{+0.13}_{-0.19}$ & $1.96^{+0.09}_{-0.11}$ & $1.51^{+0.09}_{-0.11}$ & $1.32^{+0.13}_{-0.19}$ & $1.78^{+0.13}_{-0.19}$ & $2.07^{+0.09}_{-0.11}$ & $1.58^{+0.09}_{-0.11}$ & $1.22^{+0.13}_{-0.19}$ & $1.70^{+0.13}_{-0.19}$ \\
        Sh2-269 &    $3.91^{+0.08}_{-0.09}$ & $2.09^{+0.07}_{-0.07}$ & $1.90^{+0.07}_{-0.07}$ & $1.82^{+0.08}_{-0.09}$ & $2.01^{+0.08}_{-0.09}$ & $2.07^{+0.07}_{-0.07}$ & $1.90^{+0.07}_{-0.07}$ & $1.84^{+0.08}_{-0.09}$ & $2.01^{+0.08}_{-0.09}$ \\
        \hline
        & \multicolumn{6}{c}{Outer Galaxy Targets from \cite{2022AJ....164..129P}}\\
        \hline
        Sh2-212 &    $3.14^{+0.17}_{-0.25}$ & $1.89^{+0.15}_{-0.21}$ & $1.75^{+0.15}_{-0.21}$ & $1.25^{+0.17}_{-0.25}$ & $1.39^{+0.17}_{-0.25}$ & $1.95^{+0.15}_{-0.21}$ & $1.80^{+0.15}_{-0.21}$ & $1.19^{+0.17}_{-0.25}$ & $1.33^{+0.17}_{-0.25}$ \\
        Sh2-228 &    $2.11^{+0.07}_{-0.08}$ & $1.40^{+0.05}_{-0.05}$ & $0.84^{+0.03}_{-0.03}$ & $0.71^{+0.08}_{-0.1}$ & $1.26^{+0.07}_{-0.08}$ & $1.29^{+0.05}_{-0.05}$ & $0.89^{+0.03}_{-0.03}$ & $0.81^{+0.08}_{-0.09}$ & $1.22^{+0.07}_{-0.08}$ \\
        Sh2-235 &    $3.29^{+0.05}_{-0.06}$ & $2.45^{+0.04}_{-0.04}$ & $2.08^{+0.04}_{-0.04}$ & $0.85^{+0.05}_{-0.06}$ & $1.21^{+0.05}_{-0.06}$ & $2.43^{+0.04}_{-0.04}$ & $2.02^{+0.04}_{-0.04}$ & $0.86^{+0.05}_{-0.06}$ & $1.28^{+0.05}_{-0.06}$ \\
        Sh2-252 &    $3.62^{+0.06}_{-0.07}$ & $2.53^{+0.05}_{-0.05}$ & $2.26^{+0.05}_{-0.05}$ & $1.09^{+0.06}_{-0.07}$ & $1.37^{+0.06}_{-0.07}$ & $2.49^{+0.05}_{-0.05}$ & $2.18^{+0.05}_{-0.05}$ & $1.14^{+0.06}_{-0.07}$ & $1.44^{+0.06}_{-0.07}$ \\
        Sh2-254 &    $3.46^{+0.06}_{-0.06}$ & $2.50^{+0.05}_{-0.04}$ & $2.12^{+0.05}_{-0.04}$ & $0.96^{+0.06}_{-0.06}$ & $1.33^{+0.06}_{-0.06}$ & $2.50^{+0.05}_{-0.04}$ & $2.02^{+0.05}_{-0.04}$ & $0.96^{+0.06}_{-0.06}$ & $1.44^{+0.06}_{-0.06}$ \\
        \hline
        & \multicolumn{6}{c}{Inner Galaxy Targets from \cite{2020ApJ...894..103E}}\\
        \hline
        \inncloudb &    $4.36^{+0.05}_{-0.06}$ & $3.25^{+0.03}_{-0.04}$ & $2.97^{+0.04}_{-0.04}$ & $1.10^{+0.05}_{-0.06}$ & $1.39^{+0.05}_{-0.06}$ & $3.20^{+0.03}_{-0.04}$ & $2.88^{+0.04}_{-0.05}$ & $1.15^{+0.05}_{-0.06}$ & $1.47^{+0.06}_{-0.07}$ \\
        \inncloudc &    $3.78^{+0.07}_{-0.09}$ & $2.81^{+0.06}_{-0.07}$ & $2.41^{+0.06}_{-0.08}$ & $0.97^{+0.08}_{-0.09}$ & $1.37^{+0.08}_{-0.1}$ & $2.71^{+0.06}_{-0.07}$ & $2.37^{+0.06}_{-0.08}$ & $1.07^{+0.08}_{-0.09}$ & $1.41^{+0.08}_{-0.10}$ \\
        \inncloudd &    $3.35^{+0.07}_{-0.08}$ & $2.12^{+0.03}_{-0.03}$ & $1.51^{+0.11}_{-0.14}$ & $1.23^{+0.07}_{-0.09}$ & $1.84^{+0.12}_{-0.17}$ & $1.94^{+0.03}_{-0.03}$ & $1.45^{+0.10}_{-0.14}$ & $1.41^{+0.07}_{-0.09}$ & $1.91^{+0.12}_{-0.17}$ \\
        \inncloude &    $4.09^{+0.08}_{-0.10}$ & $3.02^{+0.05}_{-0.08}$ & $2.11^{+0.09}_{-0.13}$ & $1.07^{+0.08}_{-0.11}$ & $1.98^{+0.11}_{-0.15}$ & $2.85^{+0.06}_{-0.08}$ & $2.15^{+0.09}_{-0.12}$ & $1.23^{+0.08}_{-0.12}$ & $1.94^{+0.10}_{-0.15}$ \\
        \inncloudf &    $3.07^{+0.15}_{-0.20}$ & $2.25^{+0.14}_{-0.17}$ & $1.62^{+0.17}_{-0.25}$ & $0.82^{+0.15}_{-0.2}$ & $1.45^{+0.18}_{-0.28}$ & $2.14^{+0.14}_{-0.17}$ & $1.53^{+0.17}_{-0.27}$ & $0.93^{+0.15}_{-0.20}$ & $1.54^{+0.18}_{-0.30}$ \\
        \tableline
    \enddata
    \tablecomments{1. Units of luminosities are \kkms pc$^2$.\\
    2. Units of conversion factors are $\msun \ (\rm K\ km\ s^{-1}\ pc^{2})^{-1}$. 
    }
\end{deluxetable*}


\subsubsection{Dense Gas Mass Measurement}\label{sec_4.2.2}

We have estimated the dense gas mass from thermal dust emission ($M_{\rm dg, BGPS}$) of targets having BGPS data by following the same procedure described in \cite{2022AJ....164..129P}.
The 1.1 mm dust emission is assumed to be optically thin and  at a single temperature, and a dust opacity is assumed. 
\begin{equation}\label{eq:dustmass}
     M_{\rm dg, BGPS} = \frac{S_{1.1}D^{2}\gamma}{B_{\nu, \rm 1.1}(T_{\rm dust})\kappa_{\rm dust,1.1}} , 
\end{equation}
where $S_{1.1}$ is the integrated flux density taken from the BGPS Source catalog table\footnote{\url{https://irsa.ipac.caltech.edu/data/BOLOCAM_GPS/tables/bgps_v2.1.tbl}} \citep{2010ApJS..188..123R}, 
$B_{\nu, \rm 1.1}$ is the Planck function evaluated at $\lambda=1.1$ mm, $T_{\rm dust}$ is the dust temperature, and $\gamma$ is the gas-to-solids ratio. 
The dust opacity per gram of dust, denoted as $\kappa_{\rm dust,1.1}$, is specified as $1.14$ $\rm{cm^2 g^{-1}}$ \citep{2006ApJ...638..293E}.
Introducing the variable $\gamma$, equation \ref{eq:dustmass} can be expressed in convenient units (e.g., \citealt{2010ApJS..188..123R}), as 

\begin{equation}\label{eq:bgps_mass}
\begin{split}
        M_{\rm dg, BGPS} & = 13.1 M_{\odot}\left(\frac{D}{1 \mathrm{ kpc}}\right)^{2}\left(\frac{S_{1.1}}{1 \mathrm{Jy}}\right) \left(\frac{\gamma}{\gamma_{\odot}}\right) \\
    &   \left[ \frac{e^{(13.0 \mathrm{K}/T_{dust})}-1 }{e^{(13.0/20)}-1} \right] .
\end{split}
\end{equation}

We have estimated the dust temperature for the targets with \textit{Herschel} data using the dust temperature map  \citep{2015MNRAS.454.4282M}, which includes all the inner galaxy clouds and six outer galaxy clouds (Sh2-156, Sh2-228, Sh2-242, Sh2-252, Sh2-254, and Sh2-266). 
For the targets with available $T_{\rm dust}$ information from \textit{Herschel}, we use their individual measurements. 
For the outer galaxy targets without \textit{Herschel} data, we have applied the mean temperature of 17.6 K based on the six outer galaxy targets.
The dust temperature in the outer Galaxy sources is generally lower compared to the inner Galaxy sources. 
The mean temperature of the targets in the inner Galaxy is 20.7 K, while in the outer Galaxy it is 17.6 K. 
\citet{2021MNRAS.504.2742E} found the median temperatures of candidate \hii\ regions in the inner and outer Galaxy are 24.6 and 23.9 K, respectively. 
However, \citet{2018MNRAS.473.1059U} reported higher dust temperature in the outer Galaxy compared to inner Galaxy targets. Our limited sample is more consistent with the result of 
\citet{2021MNRAS.504.2742E}.

The conventional practice in Milky Way studies is to use a single gas-to-solids ratio ($\gamma$) as an assumption for estimating the mass and column density of molecular hydrogen from dust thermal continuum observations.   
We have assumed the local value of the gas-to-solids ratio ($\gammasun$) to be 100 \citep{1983QJRAS..24..267H}, but we discuss the uncertainties in this number and its dependence on environment in Appendix \ref{gammasun}, where we note that the mass of solids (dust plus ice) is the quantity that is about 100 for the extinction regions we consider to be dense gas. 
When not correcting for  metallicity variation, we have used $\gamma=\gammasun$ in equation \ref{eq:bgps_mass} and   computed $M_{\rm {dg, BGPS}}$ for all the targets having BGPS data. 
The values are tabulated in logarithmic scale in Table \ref{tab:Mdense_bgps}.

To convert dust column or volume densities to total (gas plus solid) values for regions far from the Solar neighborhood, we need to consider the dependence of the gas-to-solid ratio  on metallicity. Studies of other galaxies indicate that this ratio decreases with increasing metallicity
(e.g., \citealt{2020ApJ...889..150A, 2020ApJ...897..184A}). 
The dependence steepens for $Z < 0.3$, but is roughly linear, with a substantial dispersion, for the range of $Z$ values in our sample
\citep{2014A&A...563A..31R}.
In addition to galaxy-wide variations, $\gamma$ varies within galaxies.
For example, $\gamma$ increases with radius in M83, a galaxy rather similar to the Milky Way
\citep{2024ApJ...968...97L}.
Theoretical studies generally reproduce the observed super-linear decrease of $\gamma$ with increasing $Z$ below $Z \sim 0.2$ but leveling off to a roughly linear dependence for $Z$ in the range of values found in the Milky Way
\citep{2015MNRAS.447.2937H, 2017MNRAS.466..105A, 2020MNRAS.491.3844A}. 

 Although variations in $\gamma$ are rarely considered in studies of molecular clouds in the Milky Way, metal abundances are known to decrease with \rgal\ 
\citep{2022MNRAS.510.4436M}, which would lead to an increase in $\gamma$ with \rgal. Indeed,
\citet{2017A&A...606L..12G},
using a subset of the Hi-GAL data,
found that $\gamma \propto Z^{\beta}$ 
with $\beta = -1.4^{+0.3}_{-1.0}$, steeper than linear.  For consistency with our assumption that the \cotw\ abundance is linearly proportional to $Z$, and the evidence from theory and extragalactic work, we choose to define $\beta \equiv -1.0$.
For the dependence of metallicity on \rgal, we have adopted the gradient for [O/H] without correction for temperature variation ($-0.044\pm0.009$) from
\citet{2022MNRAS.510.4436M}, again for consistency with our gas-based analysis. This choice is also consistent with more recent analysis of the full Hi-GAL data set \citep{2025arXiv250114471E}.

In summary, if we have information on $Z$ for a particular region, we use 
\begin{equation}\label{eq:gamma_outer}
\gamma = \gammasun Z^{-1.0}.
\end{equation}
For outer Galaxy targets, we used this $\gamma$ vs $Z$ relation, because we have direct measurements of $Z$ for them. 

In the case that we have to estimate $Z$ from location in the Galaxy (e.g., the inner Galaxy clouds), we use the following equation:
\begin{equation}\label{eq:gamma_inner}
\gamma = \gammasun 10^{(0.044\pm0.009)(\rgal - \rgalsun)}.
\end{equation} 
The uncertainties on these relations are statistical; systematic uncertainties are certainly larger but are undetermined. For whole galaxies, 
\citet{2014A&A...563A..31R} 
found a dispersion of 0.37 dex about the mean. 
\citet{2024ApJ...968...97L}
found smaller intra-galactic dispersions ($\sim 40$\%) in 2 kpc wide radial bins in M83.
We take the latter as a crude estimate of source-to-source scatter for sources in our sample.

To obtain the metallicity-corrected dense gas mass ($M_{\rm dg, BGPS}^{'}$), we have used equations \ref{eq:gamma_outer} and \ref{eq:gamma_inner} for the $\gamma$ value in equation \ref{eq:bgps_mass} for individual targets of outer and inner Galaxy, respectively.
 The dust temperature values of the regions are used from \textit{Herschel} dust temperature maps, as discussed earlier.
The metallicity-corrected dense gas mass ($M_{\rm dg, BGPS}^{'}$) values for all the targets (including inner and outer Galaxy)  are tabulated in Table \ref{tab:Mdense_bgps_Z_elia_temp} in logarithmic scale.


\begin{deluxetable*}{l c | c c c c | c c c c c}
    \tablenum{7}
    \tabletypesize{\footnotesize}
    \tablecaption{$M_{\rm dg, BGPS}^{'}$ versus Luminosities (with metallicity correction) \label{tab:Mdense_bgps_Z_elia_temp}}
    \tablewidth{0pt}
    \tablehead{
    \colhead{Source} & \colhead{$\mathrm{Log}\ M_{\rm dg, BGPS}^{'}$} & \colhead{Log $L_{\rm tot}^{'}$} & \colhead{Log $L_{\rm in}^{'}$} & \colhead{$\mathrm{Log \ \alpha'_{tot}}$} & \colhead{$\mathrm{Log \ \alpha'_{in}}$} & \colhead{Log $L_{\rm tot}^{'}$} & \colhead{Log $L_{\rm in}^{'}$} & \colhead{$\mathrm{Log \ \alpha'_{tot}}$} & \colhead{$\mathrm{Log \ \alpha'_{in}}$} \\  
    \cline{3-6}\cline{7-10}  \colhead{} & \colhead{(\msun)} & \multicolumn{4}{c}{\hcn} &  \multicolumn{4}{c}{\hcop}
    }
    \startdata
        & \multicolumn{6}{c}{Outer Galaxy Targets}\\
        \hline
        Sh2-132 &    $3.79^{+0.08}_{-0.09}$ & $2.63^{+0.05}_{-0.06}$ & $2.13^{+0.05}_{-0.06}$ & $1.16^{+0.08}_{-0.10}$ & $1.65^{+0.08}_{-0.09}$ & $2.65^{+0.05}_{-0.06}$ & $2.13^{+0.05}_{-0.06}$ & $1.13^{+0.08}_{-0.09}$ & $1.66^{+0.08}_{-0.09}$ \\
        Sh2-142 &    $2.96^{+0.09}_{-0.11}$ & $1.72^{+0.05}_{-0.05}$ & $1.43^{+0.05}_{-0.05}$ & $1.24^{+0.09}_{-0.11}$ & $1.53^{+0.09}_{-0.11}$ & $1.67^{+0.05}_{-0.05}$ & $1.36^{+0.05}_{-0.05}$ & $1.29^{+0.09}_{-0.11}$ & $1.60^{+0.09}_{-0.11}$ \\
        Sh2-148 &    $4.27^{+0.06}_{-0.06}$ & $2.42^{+0.05}_{-0.05}$ & $2.29^{+0.05}_{-0.05}$ & $1.86^{+0.06}_{-0.07}$ & $1.98^{+0.06}_{-0.06}$ & $2.58^{+0.05}_{-0.05}$ & $2.31^{+0.05}_{-0.05}$ & $1.69^{+0.06}_{-0.06}$ & $1.96^{+0.06}_{-0.06}$ \\
        Sh2-156 &    $3.81^{+0.08}_{-0.09}$ & $2.83^{+0.07}_{-0.07}$ & $2.37^{+0.07}_{-0.07}$ & $0.98^{+0.08}_{-0.09}$ & $1.44^{+0.08}_{-0.09}$ & $2.83^{+0.07}_{-0.07}$ & $2.35^{+0.07}_{-0.07}$ & $0.98^{+0.08}_{-0.09}$ & $1.46^{+0.08}_{-0.09}$ \\
        Sh2-242 &    $3.01^{+0.05}_{-0.06}$ & $1.59^{+0.03}_{-0.02}$ & $1.23^{+0.02}_{-0.02}$ & $1.41^{+0.06}_{-0.06}$ & $1.78^{+0.06}_{-0.06}$ & $1.52^{+0.03}_{-0.02}$ & $1.11^{+0.02}_{-0.02}$ & $1.49^{+0.06}_{-0.06}$ & $1.90^{+0.06}_{-0.06}$ \\
        Sh2-266 &    $3.59^{+0.13}_{-0.19}$ & $2.06^{+0.09}_{-0.11}$ & $1.51^{+0.09}_{-0.11}$ & $1.53^{+0.14}_{-0.19}$ & $2.09^{+0.13}_{-0.19}$ & $2.20^{+0.09}_{-0.11}$ & $1.58^{+0.09}_{-0.11}$ & $1.39^{+0.13}_{-0.19}$ & $2.01^{+0.13}_{-0.19}$ \\
        Sh2-269 &    $3.99^{+0.08}_{-0.09}$ & $2.10^{+0.07}_{-0.07}$ & $1.90^{+0.07}_{-0.07}$ & $1.89^{+0.08}_{-0.09}$ & $2.09^{+0.08}_{-0.09}$ & $2.09^{+0.07}_{-0.07}$ & $1.90^{+0.07}_{-0.07}$ & $1.90^{+0.08}_{-0.09}$ & $2.09^{+0.08}_{-0.09}$ \\   
        \hline
        & \multicolumn{6}{c}{Outer Galaxy Targets from \cite{2022AJ....164..129P}}\\
        \hline
        Sh2-212 &    $3.30^{+0.17}_{-0.25}$ & $1.96^{+0.15}_{-0.21}$ & $1.75^{+0.15}_{-0.21}$ & $1.34^{+0.17}_{-0.25}$ & $1.55^{+0.17}_{-0.25}$ & $2.04^{+0.15}_{-0.21}$ & $1.80^{+0.15}_{-0.21}$ & $1.26^{+0.17}_{-0.25}$ & $1.49^{+0.17}_{-0.25}$ \\
        Sh2-228 &    $2.22^{+0.07}_{-0.08}$ & $1.42^{+0.05}_{-0.06}$ & $0.84^{+0.03}_{-0.03}$ & $0.80^{+0.08}_{-0.10}$ & $1.37^{+0.07}_{-0.08}$ & $1.32^{+0.05}_{-0.05}$ & $0.89^{+0.03}_{-0.03}$ &    $0.89^{+0.08}_{-0.10}$ & $1.33^{+0.07}_{-0.08}$ \\
        Sh2-235 &    $3.37^{+0.05}_{-0.06}$ & $2.45^{+0.04}_{-0.04}$ & $2.08^{+0.04}_{-0.04}$ & $0.92^{+0.05}_{-0.06}$ & $1.29^{+0.05}_{-0.06}$ & $2.44^{+0.04}_{-0.04}$ & $2.02^{+0.04}_{-0.04}$ & $0.93^{+0.05}_{-0.06}$ & $1.36^{+0.05}_{-0.06}$ \\
        Sh2-252 &    $3.61^{+0.06}_{-0.07}$ & $2.53^{+0.05}_{-0.05}$ & $2.26^{+0.05}_{-0.05}$ & $1.08^{+0.06}_{-0.07}$ & $1.36^{+0.06}_{-0.07}$ & $2.48^{+0.05}_{-0.05}$ & $2.18^{+0.05}_{-0.05}$ & $1.13^{+0.06}_{-0.07}$ & $1.43^{+0.06}_{-0.07}$ \\
        Sh2-254 &    $3.55^{+0.06}_{-0.06}$ & $2.51^{+0.05}_{-0.04}$ & $2.12^{+0.05}_{-0.04}$ & $1.03^{+0.06}_{-0.06}$ & $1.42^{+0.06}_{-0.06}$ & $2.51^{+0.05}_{-0.04}$ & $2.02^{+0.05}_{-0.04}$ & $1.03^{+0.06}_{-0.06}$ & $1.53^{+0.06}_{-0.06}$ \\
        \hline
        & \multicolumn{6}{c}{Inner Galaxy Targets from \cite{2020ApJ...894..103E}}\\
        \hline
        \inncloudb &    $4.26^{+0.05}_{-0.06}$ & $3.23^{+0.03}_{-0.04}$ & $2.97^{+0.04}_{-0.04}$ & $1.04^{+0.05}_{-0.06}$ & $1.30^{+0.05}_{-0.06}$ & $3.17^{+0.03}_{-0.04}$ & $2.88^{+0.04}_{-0.05}$ & $1.09^{+0.05}_{-0.06}$ & $1.38^{+0.06}_{-0.07}$ \\
        \inncloudc &    $3.65^{+0.07}_{-0.09}$ & $2.76^{+0.06}_{-0.07}$ & $2.41^{+0.06}_{-0.08}$ & $0.89^{+0.08}_{-0.09}$ & $1.24^{+0.08}_{-0.10}$ & $2.68^{+0.06}_{-0.07}$ & $2.37^{+0.06}_{-0.08}$ & $0.97^{+0.08}_{-0.09}$ & $1.28^{+0.08}_{-0.10}$ \\
        \inncloudd &    $3.21^{+0.07}_{-0.08}$ & $2.02^{+0.03}_{-0.03}$ & $1.51^{+0.11}_{-0.14}$ & $1.19^{+0.07}_{-0.09}$ & $1.70^{+0.12}_{-0.17}$ & $1.87^{+0.03}_{-0.03}$ & $1.45^{+0.10}_{-0.14}$ & $1.34^{+0.07}_{-0.09}$ & $1.77^{+0.12}_{-0.17}$ \\
        \inncloude &    $4.01^{+0.08}_{-0.10}$ & $3.00^{+0.05}_{-0.08}$ & $2.11^{+0.09}_{-0.13}$ & $1.01^{+0.08}_{-0.11}$ & $1.91^{+0.11}_{-0.15}$ & $2.84^{+0.06}_{-0.08}$ & $2.15^{+0.09}_{-0.12}$ & $1.18^{+0.08}_{-0.11}$ & $1.86^{+0.10}_{-0.15}$ \\
        \inncloudf &    $2.98^{+0.15}_{-0.20}$ & $2.23^{+0.14}_{-0.17}$ & $1.62^{+0.17}_{-0.25}$ & $0.76^{+0.15}_{-0.20}$ & $1.36^{+0.18}_{-0.28}$ & $2.12^{+0.14}_{-0.17}$ & $1.53^{+0.17}_{-0.27}$ & $0.86^{+0.15}_{-0.20}$ & $1.45^{+0.18}_{-0.30}$ \\
    \tableline
    \enddata
    \tablecomments{1. Units of luminosities are \kkms pc$^2$.\\
    2. Units of conversion factors are $\msun \ (\rm K\ km\ s^{-1}\ pc^{2})^{-1}$.\\ 
    }
\end{deluxetable*}

\begin{figure*}
    \centering
    \includegraphics[scale=0.49]{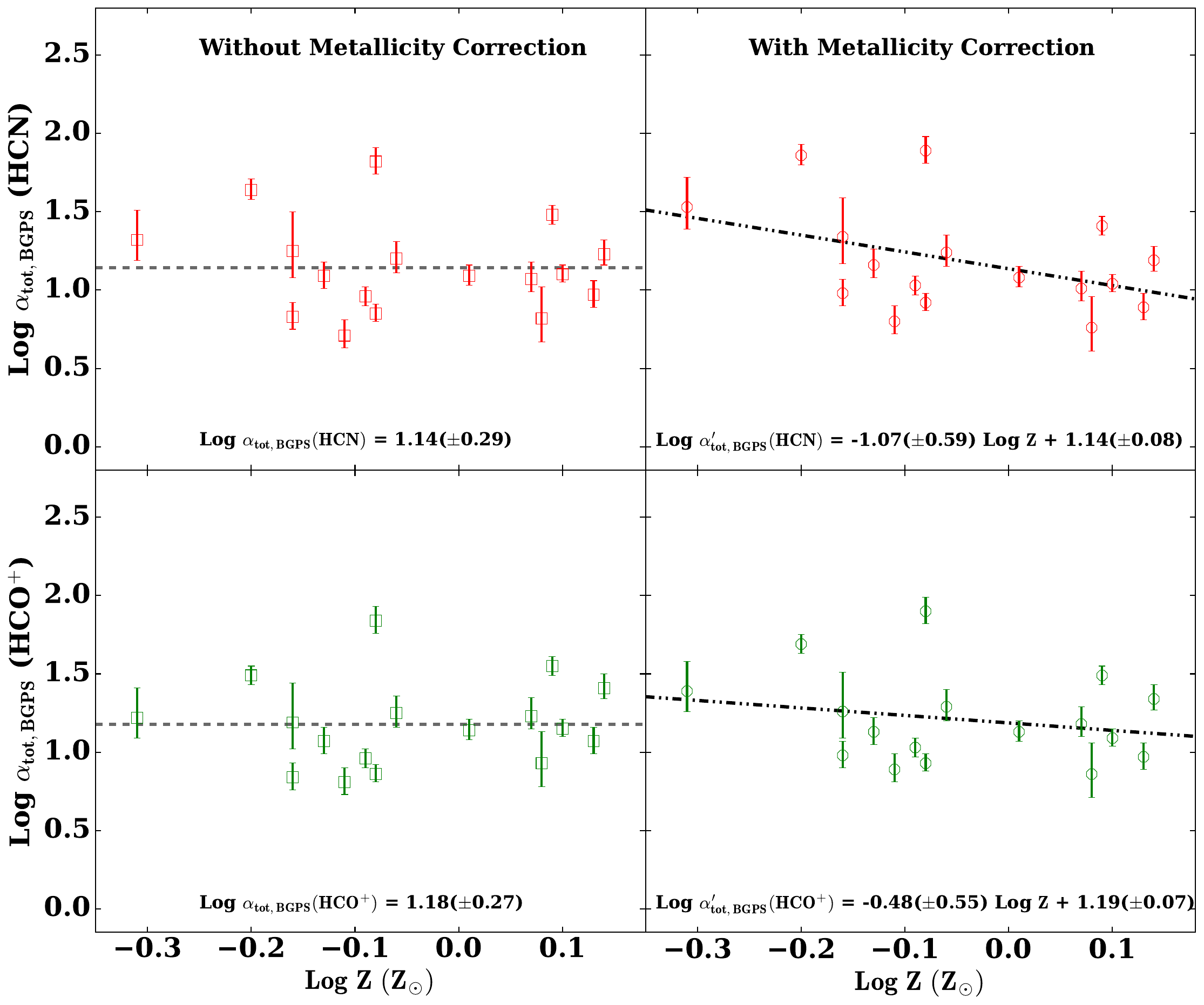} 
    \caption{Variation of $\alpha_{\rm tot, BGPS}$ for \hcn\ (top panels) and \hcop\ (bottom panels) derived from dust based analysis with metallicity, both with (right) and without (left) metallicity correction. The horizontal dashed line in the top and bottom left panels shows the mean value of $\alpha_{\rm tot, BGPS}$. The dash-dotted line in the top and bottom right panels are done by fitting.}
    \label{fig:figure2_temp}
\end{figure*}

\subsubsection{Mass Conversion factor: The alpha-factor ($\alpha_{Q}$) from Dust-based analysis} \label{sec_4.2.3}

Following Section \ref{subsubsec:alpha_gas}, we have derived mass conversion factor $\alpha_{\rm in, BGPS}$ by dividing the mass of dense gas ($M_{\rm {dg, BGPS}}$) by the \hcn\ or \hcop\ luminosity within the ``dense” region defined by BGPS mask region.  For $\alpha_{\rm tot, BGPS}$, we  divide $M_{\rm {dg, BGPS}}$ by $L_{\rm tot}$ for \hcn\ or \hcop.
The values, presented in Table \ref{tab:Mdense_bgps}, are listed in logarithmic scale without metallicity correction for both \hcn\ and \hcop.

Subsequently, we computed metallicity-corrected mass conversion factors $\alpha'_{\rm in, BGPS}$ and $\alpha'_{\rm tot, BGPS}$ for all targets in Table \ref{tab:Mdense_bgps_Z_elia_temp} using the metallicity-corrected dense gas mass ($M_{\rm dg, BGPS}^{'}$) values. 
Finally, we obtained the statistical parameters separately for inner and outer Galaxy clouds, detailed in Table \ref{tab:summary_bgps}, to discern potential variations between the inner and outer Galaxy regions.
Correction for the effects of metallicity leads to significant differences in $\alpha'_{\rm tot, BGPS}$.
For instance, the mean value of the $\alpha_{\rm tot, BGPS}(\hcn)$ in linear scale is  11 and 15.5 $\msun \ (\rm K\ km\ s^{-1}\ pc^{2})^{-1}$ for inner and outer Galaxy targets, respectively, for the case where we do not consider the metallicity correction.
The $\alpha'_{\rm tot, BGPS}$ for \hcn\  decreases by a factor of 1.15 for inner Galaxy clouds and 
increases by a factor of 1.20 for outer Galaxy targets once the correction is applied. 
In Figure \ref{fig:figure2_temp}, we have plotted the $\alpha_{\rm tot, BGPS}\ \rm (and \ \alpha'_{\rm tot, BGPS})$ with respect to metallicity for both \hcn\ and \hcop. These plots also show the inverse relation between mass conversion factor and metallicity.

\begin{deluxetable*}{l c | c c c c | c c c c}
    \tablenum{8}
    \tabletypesize{\footnotesize}
    \tablecaption{Statistical summary: average mass-luminosity conversion factors from BGPS analysis
    \label{tab:summary_bgps}}
    \tablewidth{0pt}
    \tablehead{
    \colhead{}  & & \colhead{$\rm Log\ \alpha_{\rm in, BGPS}$} &   & \colhead{$\rm Log\ \alpha_{\rm tot, BGPS}$} & & & \colhead{$\rm Log\ \alpha'_{\rm in, BGPS}$} &  & \colhead{$\rm Log\ \alpha'_{\rm tot, BGPS}$} 
    }
    \startdata
    \hline
    & \multicolumn{9}{c}{\hcn}\\
    \hline
    Inner Galaxy & & $1.61 \pm 0.28 $   &   &    $1.04 \pm 0.15 $   & & &  $1.50 \pm 0.29$   &      & $0.98 \pm 0.16$ \\
    Outer Galaxy & & $1.52 \pm 0.27 $   &   &    $1.19 \pm 0.34 $   & & &  $1.63 \pm 0.29$   &      & $1.27 \pm 0.35$ \\
    \hline
    & \multicolumn{9}{c}{\hcop}\\
    \hline
    Inner Galaxy & & $1.65 \pm 0.25 $    &  &    $1.16 \pm 0.18 $   & & & $1.55 \pm 0.25$    &      & $1.09 \pm 0.18$ \\
    Outer Galaxy & & $1.54 \pm 0.27 $    &  &    $1.18 \pm 0.31 $   & & & $1.65 \pm 0.27$    &      & $1.26 \pm 0.31$ \\
    \tableline
    \enddata
    \tablecomments{1. Units of conversion factors are $\msun \ (\rm K\ km\ s^{-1}\ pc^{2})^{-1}$.}
\end{deluxetable*}


\section{Mass-Luminosity conversion factors in the Solar neighborhood} \label{solar}

In order to comprehensively assess the mass-conversion factor across the Milky Way, we included molecular clouds within the Solar neighborhood in our analysis. 
Several recent studies have focused on elucidating the characteristics of the nearby clouds.
For instance, \cite{2023ApJ...944..197D}  surveyed the entire Perseus molecular cloud using the CfA 1.2m telescope, focusing on \hcn. 
\cite{2023A&A...679A...4S} employed the IRAM 30m telescope to investigate the Orion B Giant Molecular Cloud (GMC). 
\cite{2023A&A...679A.112T} expanded their research to cover three distinct star-forming regions --- California, Perseus, and Orion A -- employing the IRAM 30m telescope at a wavelength of 3 mm. 
\cite{2021ApJS..256...16Y,2024ApJS..271...36Y}, utilizing the TRAO 14m telescope, explored Orion A and Ophiuchus, concentrating on \hcn\ and \hcop. 
We have compiled the mass and luminosity information for these clouds from the respective studies and tabulated the data in Table \ref{tab:local}.
For Ophiuchus, we extracted mass information from \cite{2014ApJ...782..114E} and rescaled it to the updated distance value of 137 pc. 
Luminosity values for \hcn\ and \hcop\ were obtained from \citet{2024ApJS..271...36Y}, and we subsequently calculated the total mass-to-luminosity ratio ($\mathrm{\alpha_{tot}}$).
The Orion A cloud was investigated by both \cite{2023A&A...679A.112T} and \cite{2021ApJS..256...16Y,2024ApJS..271...36Y}. The luminosity values in the corrected table \citep{2024ApJS..271...36Y} are consistent with those in \citet{2023A&A...679A.112T}.
 However, \cite{2021ApJS..256...16Y} did not measure the dense-gas mass, so we have used the mass information from \cite{2023A&A...679A.112T}. 
Similarly, the Perseus molecular cloud was studied by \cite{2023A&A...679A.112T} and \cite{2023ApJ...944..197D}.
Both studies indicate higher $\mathrm{\alpha_{tot}}$ values for Perseus compared to most clouds, excluding Ophiuchus.
The $\mathrm{\alpha_{tot}(\hcn)}$ value reported by \cite{2023ApJ...944..197D} is significantly larger than that of \cite{2023A&A...679A.112T}. 
The observed discrepancy in values may be attributed to differences in the methodology employed for calculating \hcn\ luminosity.
\cite{2023ApJ...944..197D} mentioned that $\mathrm{\alpha_{tot}(\hcn) =76\ \msun \ (\rm K\ km\ s^{-1}\ pc^{2})^{-1} }$ if they consider \hcn\ outside their sampling boundary, which is consistent with \cite{2023A&A...679A.112T}.

Across the entire sample of Solar neighborhood clouds, the dense-gas mass to \hcn\ luminosity for the entire cloud varies from 23 to 122 $\msun \ (\rm K\ km\ s^{-1}\ pc^{2})^{-1}$.
The $\mathrm{\alpha_{tot}}$ values for \hcn\ and \hcop\ for the Perseus and Ophiuchus clouds are notably higher than the typical values.
In Figure \ref{fig:figure1}, we have indicated the $\alpha_{\rm tot}$ values of individual local clouds with the blue star marks.
The local clouds are not shown in Figure \ref{fig:figure2_temp} because BGPS surveys of nearby clouds were sensitive only to cores, not the larger clumps \citep{2011ApJ...741..110D}.

Most of the values of $\mathrm{\alpha_{tot}}$ are notably higher for the Solar neighborhood clouds, especially the low-mass clouds, Perseus and Ophiuchus. We have computed the average values, both with and without those two clouds, in Table \ref{tab:local}. Translated to actual values, the average without the two clouds is about 50\% bigger than our fitted relationship would predict. With those two clouds, the average is about three times larger. Neither of those two clouds is characteristic of the clouds in our sample, so the conversion factors may be affected by other properties.

\begin{deluxetable*}{l c |c c | c c |l }[h]
    \tablenum{9}
    \tabletypesize{\footnotesize}
    \tablecaption{Dense gas masses and conversion factors in molecular clouds from the Solar neighborhood \label{tab:local}} 
    \tablewidth{0pt}
    \tablehead{
    \colhead{Cloud} & \colhead{Log $\mathrm{\mdense}$}  & \colhead{Log $L_{\rm tot}$}   & \colhead{Log $\mathrm{\alpha_{tot}}$} & \colhead{Log $L_{\rm tot}$} & \colhead{Log $\mathrm{\alpha_{tot}}$} & \colhead{References}\\
    \colhead{}      & \colhead{\msun}     & \colhead{\hcn}     & \colhead{\hcn}     & \colhead{\hcop}     & \colhead{\hcop}     & \colhead{} 
    }
    \startdata
        California  &  3.68   &  2.32   &   1.36   &  2.24      &  1.45        & Tafalla et al. (2023) \\
        Orion B     &  3.49   &  2.04   &   1.46   &  2.08      &  1.40        & Santa-Maria et al. (2023) \\
        Orion A     &  4.38   &  2.72   &   1.66   &  2.73      &  1.64        & Tafalla et al. (2023) \\
                    &  4.38   &  2.62   &   1.76   &  2.61      &  1.76        &  Yun et al. (2021, 2024), Tafalla et al. (2023)\\ 
        Ophiuchus   &  3.16   &  1.08   &   2.09   &  1.03      &  2.13        & Yun et al. (2021, 2024), Evans et al. (2014)\\
        Perseus     &  3.75   &  1.89   &   1.86   &  1.91      &  1.83        & Tafalla et al. (2023) \\
                    &  3.71   &  1.74   &   1.96   & \nodata           & \nodata             & Dame \& Lada (2023) \\            
    \tableline
     Mean    &  3.69   &  1.98   &   1.71   &   1.99    &  1.70       & \\
     Median  &  3.68   &  2.04   &   1.71   &   2.08    &  1.70       & \\
     Std     &  0.45   &  0.60   &   0.30   &   0.60    &  0.30       & \\
    \tableline
     Mean$^{\dagger}$    &  3.85   &  2.34     &   1.51    &  2.33  &   1.52  & \\
     Median$^{\dagger}$  &  3.68   &  2.32     &   1.46    &  2.24  &   1.45  & \\
     Std$^{\dagger}$     &  0.47   &  0.31     &   0.18    &  0.30  &   0.16  & \\
    \tableline
    \enddata
    \tablecomments{1. Units of luminosities are \kkms pc$^2$.\\
                   2. Units of conversion factors are $\msun\ (\rm K\ km\ s^{-1}\ pc^{2})^{-1}$.\\
                   3. $^{\dagger}$We have excluded Ophiuchus and Perseus from the statistical analysis of these local clouds.  }
\end{deluxetable*}


\section{Discussion} \label{sec:discussion}

\subsection{Effect of Metallicity on Mass Conversion Factor}

The outer Milky Way is a convenient proxy for other low-metallicity systems, at the same time offering easier and clearer observations of molecular clouds than nearby galaxies.
This study explores the impact of metallicity on the mass-luminosity conversion factor of dense gas in the Milky Way, comparing the inner parts of the Galaxy, with metallicity up to 1.29 times solar metallicity, with the outer disk, where the metallicity falls as low as 0.3 times the solar metallicity. 
 
Recent studies \citep{2017A&A...604A..74S, 2017A&A...605L...5K, 2017A&A...599A..98P, 2020ApJ...894..103E, 2020MNRAS.497.1972B, 2022AJ....164..129P, 2023ApJ...944..197D, 2023A&A...679A.112T, 2023A&A...679A...4S} have highlighted that \hcn, \hcop\ emissions can originate from regions with low levels of extinction and emphasized the need to account for the entire cloud when calculating the $\alpha_{Q}$ factor for comparison with extra-galactic studies, where the line luminosity of the whole cloud is used to estimate mass. 
\cite{2004ApJ...606..271G} estimated the value for the extragalactic conversion factor $\alpha_{\rm HCN(1\to0)}=10$ $\msun\ (\rm K\ km\ s^{-1}\ pc^{2})^{-1}$ from a theoretical consideration, where luminosity from the whole cloud is considered to obtain mass.
This value has been widely adopted, although \citet{2023MNRAS.521.3348N} use a value of 14, rather than 10.
In the following section, we discuss the effect of metallicity on mass conversion factor
when the line luminosity is computed from the whole cloud ($\alpha_{\rm tot}$), which is most relevant to extragalactic comparisons. 

In Section \ref{subsubsec:alpha_gas}, we calculated $\alpha_{\rm tot, Gas}$ and $\alpha'_{\rm tot, Gas}$ for \hcn\ and \hcop\ from a gas-based analysis, without and with metallicity correction, respectively.
Figure \ref{fig:figure1} presents all the results in a four-panel diagram. 
The upper left and lower left panels of Figure \ref{fig:figure1} illustrate $\mathrm{log\ \alpha_{\rm tot, Gas}}$ plotted against $\mathrm{log\ Z}$ for \hcn\ and \hcop, respectively, without accounting for metallicity corrections in the analysis. All data points scatter around the mean value, represented by the black dashed line, revealing a `flat' distribution without significant upward or downward trends. It corresponds to a constant mass-luminosity conversion factor across the Galaxy when metallicity corrections are not considered.  
The mean value of the logarithm translates to $\mathrm{\alpha_{\rm tot,Gas}}$ for \hcn\ of 
 $22.9^{+26.1}_{-12.2}$ $\msun\ (\rm K\ km\ s^{-1}\ pc^{2})^{-1}$; for \hcop, the value is $23.4^{+26.7}_{-12.5}$ $\msun\ (\rm K\ km\ s^{-1}\ pc^{2})^{-1}$, combining both inner and outer Galaxy clouds. 
These values are around a factor of 2 higher than the value derived by \cite{2004ApJS..152...63G}, and commonly used in extragalactic studies.
We have also plotted the $\alpha_{\rm tot}$ values of the local clouds mentioned in Table \ref{tab:local} in Figure \ref{fig:figure1}. 

The scenario changes when we use the metallicity correction in the analysis outlined in section \ref{subsec:gas}. 
The targets with low metallicity (i.e., with large \rgal) show higher $\mathrm{\alpha'_{\rm tot,Gas}}$ value compared the targets with higher metallicity (or located in the inner Galaxy).
The upper right and lower right panels of Figure \ref{fig:figure1} illustrate $\mathrm{ log\ \alpha'_{\rm tot, Gas}}$ plotted against $\mathrm{ log}\ Z$ for \hcn\ and \hcop, respectively, after considering the metallicity corrections in the analysis.
Both \hcn\ and \hcop\ conversion factors demonstrate a decreasing trend with increasing metallicity, echoing the observed pattern in $\mathrm{X_{CO}}$ variation with metallicity \citep{2020ApJ...903..142G, 2024ApJ...968...97L}.
The  Pearson correlation parameter between $\mathrm{\alpha'_{tot, gas}(\hcn)}$ and $\mathrm{ log}\ Z$, the Pearson correlation coefficient  is $-0.53$, and for \hcop\ it is  $-0.51$ (see Figure \ref{fig:correlation}).
To quantitatively analyze this relationship, we fit linear functions in log-log space to derive the scaling relations. 
However, we have not included the local clouds for the fitting because of the diversity of mass measurements and the outer boundary definitions.
For \hcn, the scaling relation obtained from this dataset is 
$\mathrm{\alpha'_{tot, gas} (\hcn) \propto Z^{-1.53(\pm 0.59)}}$ and for \hcop, it is $\mathrm{\alpha'_{tot, gas}(\hcop) \propto Z^{-1.32(\pm0.55)}}$. 
While uncertain, the $Z$ dependence  for \hcn\ is stronger than linear; this would be consistent with the evidence for strong dependence of the HCN abundance on metallicity \citep{2024arXiv241104451S}.
Similarly in section \ref{sec_4.2.3}, we have calculated the mass-luminosity conversion factor $\alpha_{\rm tot, BGPS}$ by taking the ratio between the dense mass of gas obtained from BGPS dust continuum data and \hcn, \hcop\ line luminosities from the entire cloud.
The upper left and lower left panels of Figure \ref{fig:figure2_temp} illustrate $\mathrm{ log\ \alpha_{\rm tot, BGPS}}$ plotted against $\mathrm{ log}\ Z$ for \hcn\ and \hcop, respectively, without accounting for metallicity corrections in the analysis. 
The black-dashed lines correspond to the mean value of $\alpha_{\rm tot, BGPS}$ in logarithmic scale, combining inner and outer Galaxy targets.
The mean values, converted from the logarithm, are $\mathrm{\alpha_{\rm tot,BGPS}}$ for \hcn\ $13.8^{+13.1}_{-6.7}$ $\msun\ (\rm K\ km\ s^{-1}\ pc^{2})^{-1}$ and for \hcop, $15.1^{+13.0}_{-7.0}$ $\msun\ (\rm K\ km\ s^{-1}\ pc^{2})^{-1}$, combining both inner and outer Galaxy clouds.

The upper and lower right panels of Figure \ref{fig:figure2_temp} depict the variation of  $\mathrm{ log\ \alpha'_{\rm tot, BGPS}}$ with $\mathrm{ log}\ Z$ for \hcn\ and \hcop, respectively, 
after correcting for the metallicity in the dense gas mass obtained from dust continuum data.
The conversion factor for \hcn\ shows an decreasing trend with increasing metallicity, while for \hcop\ the trend is weak.
To quantify the analysis, we fitted the data as shown in the figures and the scaling relations are 
$\mathrm{\alpha'_{tot, BGPS} (\hcn)} \propto Z^{-1.07(\pm 0.59)}$ for \hcn, and
$\mathrm{\alpha'_{tot, BGPS} (\hcop)} \propto Z^{-0.48(\pm 0.55)}$ for \hcop. 
The  Pearson correlation parameter between $\mathrm{\alpha'_{tot, BGPS}(\hcn)}$ and $\mathrm{ log} \ Z$ is $-0.47$ and for \hcop\ it is $-0.28$, respectively.

The mass conversion factors derived from the gas-based analysis in Section \ref{subsec:gas} are somewhat higher than those from the BGPS-based analysis in Section \ref{sec_4.2}. This difference is caused by the fact that $M_{\rm dg, Gas}$ is on average higher than $M_{\rm dg, BGPS}$ (mean ratio is 1.76, with a median value of 1.54 and a standard deviation of 1.64, indicating substantial dispersion). This fact seems in conflict with
Figure \ref{fig:figure0}, which shows that the BGPS (cyan) contours are bigger than the gas-based dense regions  (white (or red) contours) for most (but not all) of the clouds (e.g., Sh2-132, Sh2-142, Sh2-242, Sh2-266, Sh2-269). Comparison with the $N_{\rm H_{2}}$ maps indicates that BGPS mask boundaries correspond to regions with \av\ values ranging from 3-8 magnitudes across the sample, so that the BGPS traces somewhat lower density regions on average, and thus should result in larger masses. The discrepancy is probably caused by the assumed opacity at 1.1 mm, which may be too high for these regions.

%

Despite these caveats, both analyses show that using the traditional method without correcting for metallicity in galactic clouds results in overestimating the mass conversion factor (and therefore the mass) for entire clouds in the inner part of the Galaxy (or high metallicity regions). Simultaneously, it leads to underestimating the mass conversion factor (and thus the mass) in outer Galaxy regions (or low metallicity regions).

  Without correcting for metallicity effects, no variations in the mass conversion factors with $Z$ are seen (left panels of Figures \ref{fig:figure1} and \ref{fig:figure2_temp}). Linear fits to those data showed no statistically significant slopes: $0.05 \pm0.59$, $0.32\pm0.60$, $-0.37 \pm 0.58$, $0.19 \pm0.54$ for $\alpha_{\rm tot, Gas}$(HCN), $\alpha_{\rm tot, Gas}$(\hcop), $\alpha_{\rm tot, BGPS}$(HCN), $\alpha_{\rm tot, BGPS}$(\hcop), respectively.
The most significant of these is 0.6 of the uncertainty on the slope and most are much smaller than the uncertainty. Because the mass is measured with \coo\ or dust emission, both assumed to scale with metallicity, the most likely explanation is that the HCN and \hcop\ emission also scales with metallicity, thus cancelling out. 
In the left panels, no scaling of mass determined from \coo\ or dust with $Z$ is applied, so masses at $Z > 1$ are overestimated by the factor, $Z$. If \hcn\ and \hcop\ emission increases linearly with $Z$, the luminosities will increase in the same way, obscuring any actual trend in $\alpha_{\rm tot}$. We find dependences on $Z$ that are somewhat stronger than linearly proportional, but linear with uncertainties.
Our model for how the mass depends on $Z$ is very simple, indicating the need for theoretical analysis for the $Z$ dependence of HCN and \hcop\ emission, similar to that available for CO \citep{2020ApJ...903..142G, 2022ApJ...931...28H}.

While our main focus is on $\alpha_{\rm tot}$, relevant to extra-galactic studies, we comment briefly on $\alpha_{\rm in}$ relevant to studies in the Milky Way. 
\cite{2010ApJS..188..313W} was the first to attempt to obtain   $\alpha_{\text{in}}(\text{HCN})$  for massive, dense cores in the Milky Way, which was found to be on average 20 $\msun \ (\rm K\ km\ s^{-1}\ pc^{2})^{-1}$. \cite{2017A&A...604A..74S} calculated  $\alpha_{\text{in}}(\text{HCN})$ in the range of $35–454$ for 10 heavily extincted regions with $\av>8$ mag based on \textit{Herschel} dust map, indicating a large spread in the derived values. 

Without correcting for metallicity, the average values, converted from logarithmic to linear scale, for the mass conversion factor in the dense region are 78 and 76 $\msun \ (\rm K\ km\ s^{-1}\ pc^{2})^{-1}$ for inner and outer Galaxy targets, respectively, in gas-based analysis (mentioned in Section \ref{subsubsec:alpha_gas}).
However, after applying the metallicity correction, the mean values in linear scale decreases to  63 $\msun \ (\rm K\ km\ s^{-1}\ pc^{2})^{-1}$ for inner Galaxy  targets and increases to 117.5 $\msun \ (\rm K\ km\ s^{-1}\ pc^{2})^{-1}$ for outer Galaxy targets, respectively.
The same trend is observed for \hcop.
Also, in the analysis based on BGPS, the mass conversion factor ($\alpha_{\text{in}}(\hcn)$) derived from the dense regions of the cloud decreases by a factor of 1.29 for inner Galaxy clouds, whereas for outer Galaxy targets, it increases by the same factor, after applying the metallicity correction.
 The high value of $\alpha_{\text{in}}$ observed in our study, as well as in the local clouds discussed in Section \ref{solar} can be explained by the larger mapping area conducted in these cases compared to that of \cite{2010ApJS..188..313W}.
The latter primarily focused on the clump positions within the dense regions of the cloud rather than encompassing the entire area. 

For maps of resolved regions, common in Galactic studies, the observational details, such as sensitivity and mapping area have a major impact on the conversion from observations to the mass of dense gas.

\begin{figure*}
    \centering
    \includegraphics[scale=0.55]{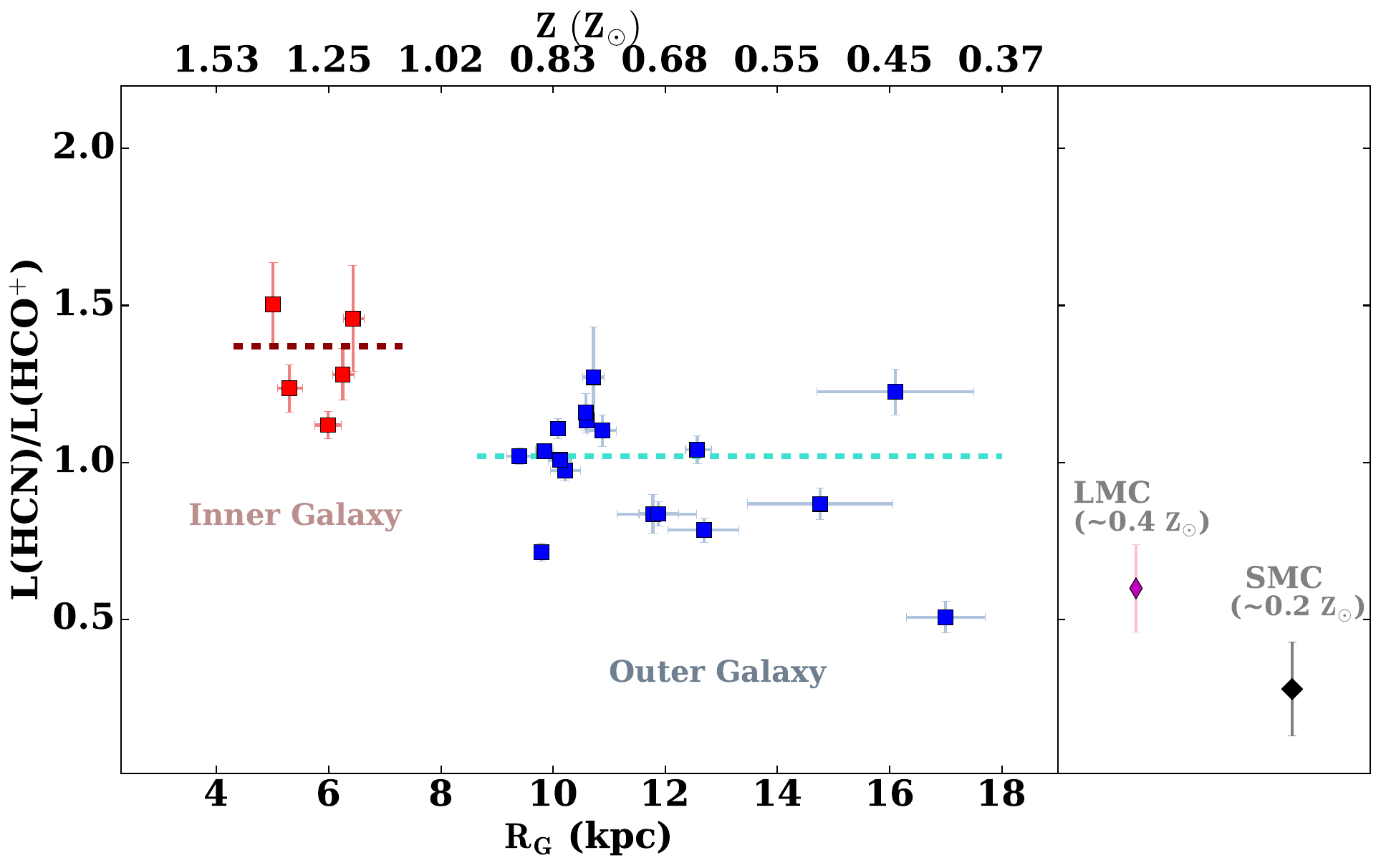} 
    \caption{Luminosity ratio of \hcn\ and \hcop\ for inner Galaxy (red squares) and outer Galaxy (blue squares) targets, as a function of \rg. The red and cyan dashed line show the median value of $L(\hcn)/L(\hcop)$ for the inner and outer Galaxy sample, respectively. The magenta and black diamonds show  the \hcn\ and \hcop\ ratio for LMC, SMC \citep{1997A&A...317..548C, 2020A&A...643A..63G}, respectively.}
    \label{fig:figure_ratio}
 \end{figure*}


 \subsubsection{Comparison between \hcn\ and \hcop\ as tracers} 
%
The bottom right panel of Figure \ref{fig:figure2_temp} shows that the luminosity to mass conversion for \hcop\ is less sensitive to metallicity than the same factor for \hcn, making it more robust against metallicity variations.
\hcn\ and \hcop\ are both linear molecules, with similar dipole moments, but the collisional cross-section of \hcop\ is significantly larger than that of \hcn. The net result is that the effective density for excitation of the \jj10\ transition of \hcop\ is about 0.12 that for \hcn\ \citep{2015PASP..127..299S}, suggesting that \hcn\ would better probe the denser gas. However, other factors like radiative trapping and other excitation mechanisms such as electron collisions and mid-infrared pumping also affect the line luminosity. Chemical abundances can also affect the observed luminosities.
The substitution of $\rm O^{+}$ for N in \hcop\ suggests that the nitrogen-to-oxygen (N/O) abundance ratio will affect the relative abundance of these molecules.
\citet{2024ApJ...973...89P} finds a steeper dependence on \rg\ for N/H ($-0.068$ dex/kpc) than the dependence for O/H ($-0.044$ dex/kpc).

These molecules have been observed also in low-metallicity galaxies  \citep{1997A&A...317..548C, 2013A&A...549A..17B, 2017A&A...597A..44B, 2020A&A...643A..63G}.
The previous studies on LMC, SMC, M33, IC10, M31 show that the $\hcn/\hcop$  intensity ratio decreases from solar to sub-solar metallicity galaxies. 
Further decreases have been observed in the Large Magellanic Cloud, where the ratio is approximately 0.6 \citep{1997A&A...317..548C}, and in the Small Magellanic Cloud, where it reaches 0.3 \citep{1998A&A...330..901C}. 

In Figure \ref{fig:figure_ratio}, we have plotted the luminosity ratio of \hcn\ and \hcop\  ($L_{\mathrm{tot}}(\hcn)/L_{\mathrm{tot}}(\hcop)$) for the entire sample including five inner Galaxy and 17 outer Galaxy targets. 
It shows that  $L_{\mathrm{tot}}(\hcn)/L_{\mathrm{tot}}(\hcop)$ decreases with increasing \rg. 
The median value of  $L_{\mathrm{tot}}(\hcn)/L_{\mathrm{tot}}(\hcop)$ is 1.37 for the inner Galaxy targets and 1.02 for the outer Galaxy sample. 
For comparison, the LMC and SMC are also shown on the right. The $L(\hcn)/L(\hcop)$ ratios for the two nearest low-metallicity dwarf galaxies are marked with magenta and black diamond points 
 \citep{1997A&A...317..548C, 2020A&A...643A..63G}, 
highlighting the trend of decreasing ratio with lower metallicity. 
In other words, the luminosity ratio of $\hcn/\hcop$  decreases by a factor of 3 as the metallicity decreases from 1.3 $\rm Z_{\odot}$ to 0.3 $\rm Z_{\odot}$, in agreement with  \cite{2017A&A...597A..44B}. 
This behavior is consistent with a strong change in the abundance ratio of \hcop\ to \hcn\ with $Z$ and with [N/H] in particular \citep{2024arXiv241104451S, 2016ApJ...818..161N, 2016ApJ...829...94N}.

\begin{figure*}
    \centering
    \includegraphics[scale=0.55]{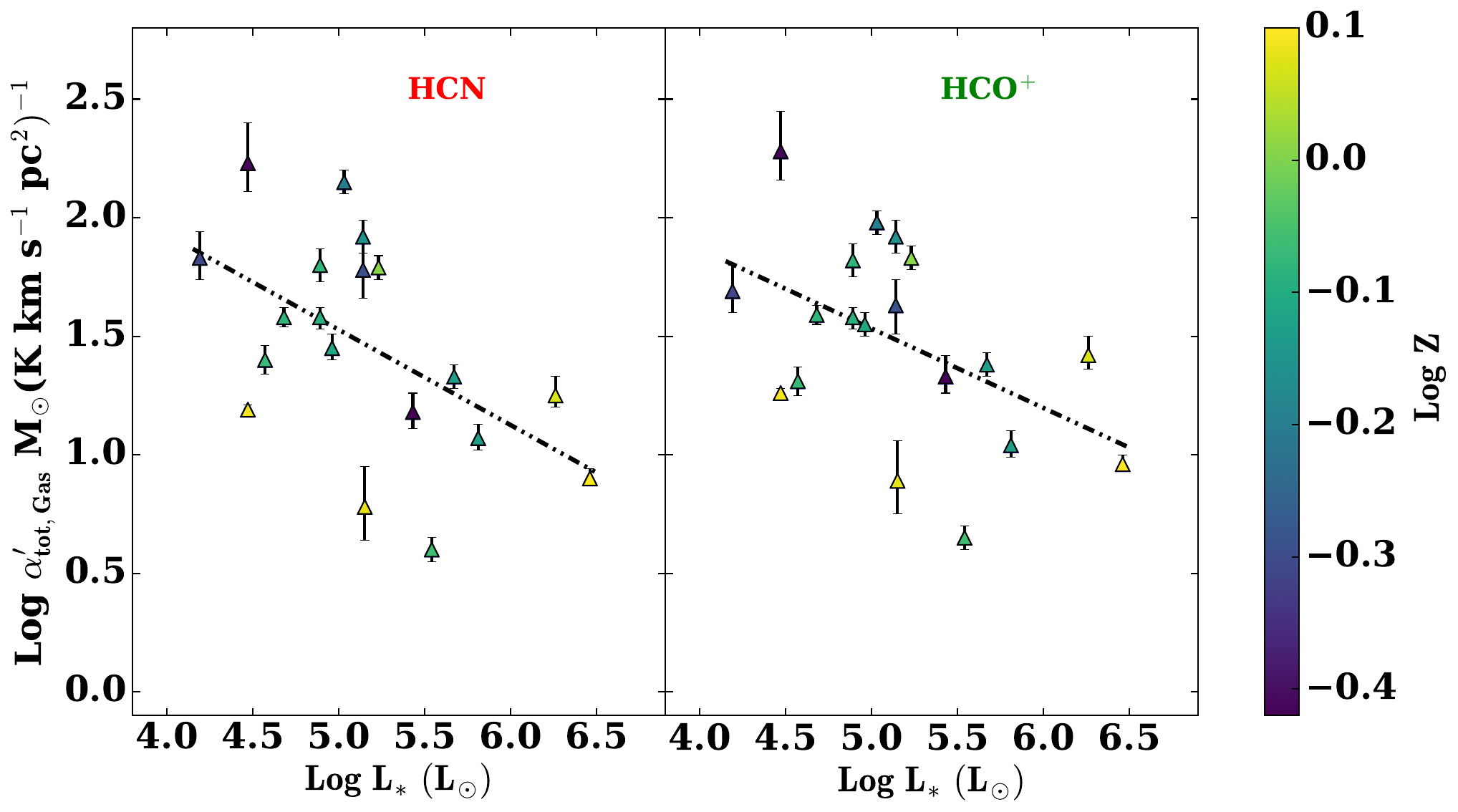}
    \caption{Variation of $\alpha_{\rm tot, Gas}$ derived from gas based analysis with bolometric luminosity from nearby massive star(s)  and the metallicity of the regions.}
    \label{fig:figure3}
 \end{figure*}

\subsection{Dispersion in the dense gas mass conversion factor}\label{dispersion}
 
The dispersion in the $\alpha_{\rm tot}$ values across different regions is quite large, even after applying the metallicity correction. 
We consider here the possible effects of radiation, as traced by the total stellar luminosity of the associated cluster. 
The ultraviolet (UV) radiation from the massive stars (OB type stars) impacts the physics and chemistry of star-forming regions significantly.
\cite{2017A&A...604A..74S} mentioned that the the conversion factors for \hcn\ and \hcop\ anti-correlate with the strength of the local UV radiation. 
 \cite{2021A&A...656A.146M} also studied the impact of UV on \hcn\ in the immediate surroundings of protostars, but they did not find much effect of UV on the overall \hcn.
\cite{2017A&A...599A..98P} showed that the molecules such as \hcn\ (also \hcop, CN, and HNC) are sensitive to the far-UV radiation produced from star formation.
In short, the effect of UV on $\alpha_{\rm tot}$ has been unclear.

We have used the $\alpha_{\rm tot}$ value obtained from gas-based analysis since all the targets are included.
For instance, Sh2-270 and Sh2-209 both are metal-poor star forming regions with 12+log[O/H] value 8.08 and 8.09, respectively. 
Despite their similarity in metallicity, the $\alpha_{\rm tot}$ values for these regions exhibit a huge difference. 
The $\alpha_{\rm tot}(\hcn)$ value for Sh2-209 is 15 $\msun\ (\rm K\ km\ s^{-1}\ pc^{2})^{-1}$, whereas $\alpha_{\rm tot}(\hcn)$ for Sh2-270 is 11 times higher than Sh2-209, 170 $\msun\ (\rm K\ km\ s^{-1}\ pc^{2})^{-1}$.
The difference in the number of massive stars among these regions is striking; Sh2-270 has only one B-type star \citep{2022MNRAS.510.4436M}, whereas Sh2-209 has at least 4 O-type and 1 B-type star \citep{2023ApJ...943..137Y}. 
This difference in the number of massive stars suggests a stronger far-ultraviolet (UV) radiation field in Sh2-209, as O-type stars are known to emit significant amounts of UV radiation. 
The UV radiation is 9 times stronger in Sh2-209 region compared to Sh2-270.
The UV can help with excitation by creating photo-ionization, and electrons excite the \hcn\ \citep{2017ApJ...841...25G}; as a result $\alpha_{\rm tot}$ will be lower for regions having a  strong UV radiation field.
We have seen a similar case for another pair of regions --- Sh2-142 and Sh2-235 ($\mathrm{12+log[O/H]\sim 8.43}$);
Sh2-142 has binary O-type stars DH Cep (O6$+$O7) \citep{2022ApJ...928...17S}; Sh2-235 has a single O-type (O9.5V) star and a few B-type stars \citep{2008MNRAS.388..729K}.
While both clouds have O stars, the early type binary will produce a lot more ionizing photons than one O9.5 star, 
resulting in Sh2-142 having seven times stronger UV radiation compared to the Sh2-235 region.
In this case, Sh2-142 has low $\alpha_{\rm tot}$ compared to Sh2-235.
Among the solar neighborhood star forming regions, Perseus and Ophiuchus have very large $\alpha_{\rm tot}$ values for \hcn\ and \hcop.
These regions lack OB stars, thereby avoiding the influence of UV radiation emitted by such massive stars. Consequently, the dominance of electrons for the excitation of \hcn\ is not significant in these areas, resulting in high $\alpha_{\rm tot}$ values for \hcn.
\cite{2023A&A...679A...4S} argue that the enhanced UV radiation favors the formation of \hcn\ and the excitation of the $J=1-0$ line at large scales for Orion B.

Because we lack information on the UV radiation for some regions, we use the total stellar luminosity  ($L_{*}$) as a proxy for external radiation,
as reported in Table \ref{tab:cloud details}.
For the outer Galaxy clouds, we have used the spectral types of the main stars (spectral types earlier than B2) from \citet{2022MNRAS.510.4436M} or from the literature on the cluster. From these, we computed the bolometric luminosity for each region from \cite{2018ApJ...864..136B}. 
We have no direct information on the stars for the inner Galaxy targets, so we assume that $L_{*}$ is the same as the far-infrared luminosity ($L_\mathrm{FIR}$). $L_\mathrm{FIR}$ was derived by combining data from the MIPSGAL (Spitzer) and Hi-GAL (\textit{Herschel}) surveys, covering wavelengths from 24 to 500 $\mu$m. A spectral energy distribution (SED) was constructed for each pixel, and its integral was calculated. 
As a first step, all maps were re-projected onto a common grid corresponding to the coarsest resolution, that of the Herschel 500~$\mu$m map. 
For each pixel, a modified blackbody function was fitted to the portion of the SED at wavelengths $\lambda \geq 160 \mu$m \citep[e.g.,][]{2013ApJ...772...45E}. 
The total SED integral was then calculated by summing two components: the integral of the observed SED between 24 and 160 $\mu$m and the analytic integral of the best-fit modified blackbody for $\lambda \geq 160 \mu$m, an extrapolation that adds the contribution from unmapped emission at $\lambda > 500 \mu$m. 
After appropriate unit conversions, the final pixel luminosity was obtained by multiplying by the factor $4 \pi D^2$, where $D$ is the heliocentric distance. 
Finally, the luminosities of all pixels lying within the defined outer boundary ($N_{\rm{H_{2}}}^{'} \geq 1.5 \times 10^{21}\ \rm{cm^{-2}} $)
were summed to obtain the total luminosity of the region. 
%

In Figure \ref{fig:figure3}, we plot the $\mathrm{ log}\ \alpha'_{\rm tot, Gas}$  against the total  bolometric luminosity ($\mathrm{ log} \ L_{*}$) of nearby massive stars. 
The color gradient in the plot represents the variation in metallicity values.
The regions with higher metallicity (i.e., the  points with yellowish color) show higher bolometric luminosity \citep{2024MNRAS.528.4746U}. 
Regions exhibiting higher radiation levels display lower $\alpha'_{\rm tot}$ values compared to clouds with comparable metallicity levels.
We have calculated the correlation parameter between $\mathrm{\alpha'_{\rm tot, Gas}(\hcn)}$ and $\mathrm{ log}\ L_{*}$; the  Pearson correlation coefficient is $-0.54$ and for \hcop\ it is $-0.49$  (Figure \ref{fig:correlation}).
The fit to the relation between $\alpha^{'}_{\rm tot, Gas}(\rm HCN)$ and $\rm  log\ L_{*}$ is (in linear scale and normalized to a typical luminosity)
$ \alpha^{'}_{\rm tot, Gas}(\rm HCN) = 33.6 (L_{*}/10^5)^{-0.40\pm0.15}$.
For \hcop, the relation is 
$\alpha^{'}_{\rm tot, Gas}(\rm \hcop) = 34.0 (L_{*}/10^5)^{-0.33\pm0.14}$.
These relations are valid only for the range of $\mathrm{log} \ L_{*}$ in this sample.

\subsection{Multiple Factors affecting the mass conversion factors} 

\begin{figure*}
    \centering
    \includegraphics[scale=0.73]{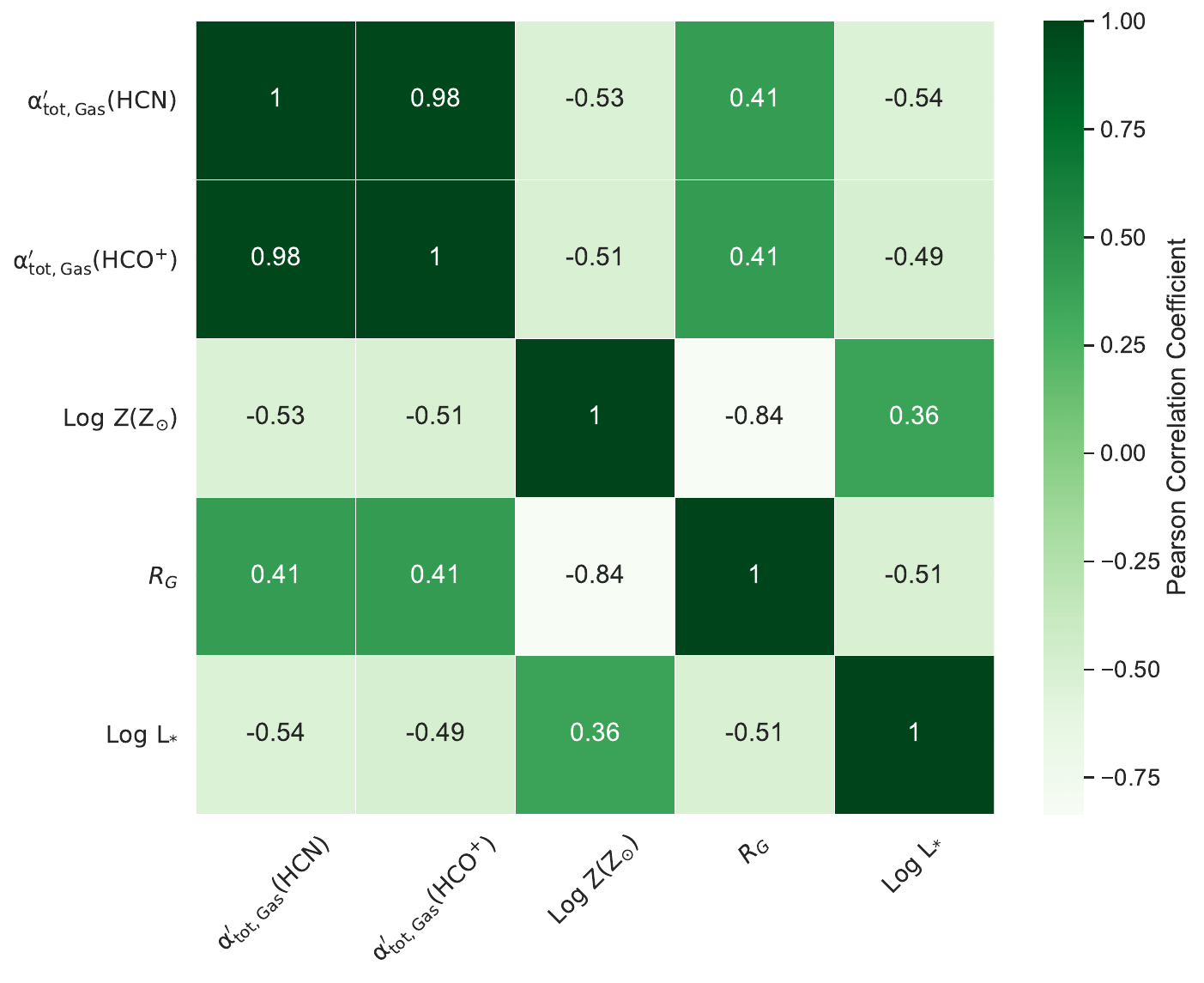}
    \caption{The Correlation Matrix: Correlation among the parameters $\alpha_{\rm tot, Gas}(\hcn)$, $\alpha_{\rm tot, Gas}(\hcop)$, $\mathrm{Log}\ Z \mathrm{(Z_{\odot}})$, \rg, and $\mathrm{Log}\ L_{*}$.   The color scale represents the Pearson correlation coefficient, as indicated by the labeled color bar.} 
    \label{fig:correlation}
\end{figure*}

In Figures \ref{fig:figure1} and \ref{fig:figure2_temp}, the mass conversion factor varies with metallicity for both \hcn\ and \hcop, though \hcop\ exhibits a weaker dependence on metallicity. 
Another contributor to the scatter in the plot is the variation in UV field strength, as illustrated in Figure \ref{fig:figure3}. 
In Figure \ref{fig:correlation}, we have shown the correlation matrix, which is a $5\times 5$ square matrix  with the same variables shown in the rows and columns.
This correlation matrix represents how different variables ($\alpha'_{\rm tot, Gas}(\hcn)$, $\alpha'_{\rm tot, Gas}(\hcop)$, $\mathrm{ log}\ Z \mathrm{(Z_{\odot}})$, \rg, and $\textrm{ log}\ L_{*}$) interact with each other.
In the correlation matrix, the Pearson correlation coefficient is calculated between pairs of variables, considering only the two variables at a time and disregarding the contribution of other variables.
Figure \ref{fig:correlation} reveals that $\alpha'_{\rm tot, Gas}(\text{HCN})$ and $\alpha'_{\rm tot, Gas}(\text{HCO}^+)$ exhibit a moderate negative correlation with both metallicity and the bolometric luminosity from external radiation.
It is noteworthy that the $\alpha'_{\rm tot}$ values for \hcn\ and \hcop\ are equally influenced by metallicity and UV radiation from massive stars.

There could be several other potential factors that can influence the mass conversion factors.
For example, the evolutionary stages of the regions may be different. 
The evolutionary stage affects the gas distribution, radiation field, and star formation feedback processes, which in turn influence both the mass of the region and its luminosity, thus altering the mass-to-luminosity conversion factor over time.
There can be other factors like the dust grain properties or the cosmic ray ionization rate in the region. 
Though these factors are potentially important, our current dataset is limited in scope and does not allow for a full exploration of these variables. This presents an exciting opportunity for future work to investigate these influences in greater detail. In particular,  observing more sources in the inner Galaxy with metallicity and well sampled maps of \hcn\ and \hcop\  is the most pressing need. 

\subsection{Impact of Metallicity on Dense Gas Mass in the Central Molecular Zone}

A longstanding puzzle in astronomy is the slow rate of star formation per unit mass of dense gas in the Central Molecular Zone (CMZ) of the Milky Way \citep{2013MNRAS.429..987L}. While there are many factors that may be at work (see \citealt{2023ASPC..534...83H} for a recent review), essentially all measures of dense gas mass (molecular line emission, dust continuum emission, etc.) are likely to be affected by metallicity. Extrapolating the relations of \citet{2022MNRAS.510.4436M} to $\rgal = 0.5$ kpc predicts $Z = 2.18$. If we extrapolate our scaling relation for \hcn\ to the CMZ,  $\alpha'_{\rm tot, Gas} (\hcn)=5.92$, and 
$\alpha'_{\rm tot, BGPS} (\hcn)= 5.99$, about 1/3 the local values. 
The nitrogen abundance in the CMZ is about 1.3 times higher than the local value \citep{2024ApJ...973...89P}. If this factor increases the \ammonia\ abundance used by \citet{2013MNRAS.429..987L} to estimate dense gas mass, it would decrease the dense gas mass estimate. Similarly, 
\citet{2018ApJ...853..171G}
used a $\gamma$ of 100 to compute surface densities from \textit{Herschel} data for Sgr B2, concluding that the threshold surface density for star formation is at least an order of magnitude larger than the local value. If $\gamma$ ratio varies according to 
equation \ref{eq:gamma_inner},
the column density estimates would decrease by a factor of about 2. The same would be true for the CMZoom study \citep{2020ApJS..249...35B} and the recent analysis of 3D-CMZ \citep{2024arXiv241017332B}, which also assumed $\gamma = 100$.  Interestingly, \citet{2025arXiv250114471E} found no deficit of star formation in the CMZ compared to that predicted from dense gas relations using the Hi-GAL catalog of dense clumps.

Consideration of the effects of metallicity on inferred cloud and clump properties in the CMZ may contribute substantially to explaining the low star formation rate per dense gas mass, along with recent evidence of higher incipient star formation rates \citep{2024ApJ...962...14H}. 
The amount of dense gas may also be overestimated if molecular line conversion factors are overestimated in regions of large velocity dispersions, as seen in the centers of other barred galaxies \citep{2024ApJ...961...42T}.

\section{Conclusions} \label{sec:conclude}
In this study, we have analyzed 17 outer Galaxy clouds, along with 5 inner Galaxy clouds from \cite{2020ApJ...894..103E}, and 5 local clouds to obtain a sample with metallicity range from $0.38\ \rm Z_{\odot} - 1.29\ \rm Z_{\odot}$.
We have investigated the impact of metallicity on the mass-luminosity conversion factor obtained from \hcn\ and \hcop, commonly used tracers of dense gas.  

\begin{enumerate}
    \item The metallicity can be  an important factor to  consider when calculating the mass of dense gas using \hcn\ or \hcop. The average value of $\alpha_{\rm tot} (\hcn)$ is $\sim 3$ times higher in the outer Galaxy than in the inner Galaxy after correcting for metallicity.

    \item We suggest to use $\alpha^{'}_{\textrm{ tot, Gas}}(\textrm{HCN}) = 19.5^{+5.6}_{-4.4} Z^{(-1.53 \pm 0.59)}$ and  $\alpha^{'}_{\textrm{ tot, Gas}}(\textrm{\hcop}) = 21.4^{+5.5}_{-4.4}  Z^{(-1.32\pm0.55)}$  for the metallicity corrected dense gas factor for extragalactic studies.

    \item Analyses of \hcn\ emission in other galaxies (e.g., \citealt{2023MNRAS.521.3348N}) may have been systematically underestimating the mass of dense gas by using smaller conversion factors in galaxies with solar metallicity. Larger factors could arise for metallicities considerably different from solar.
    
    \item Our results indicate that \hcop\ is less sensitive to metallicity than \hcn, so  \hcop\ may be the more robust tracer of dense gas. The luminosity ratio $L(\hcn)/L(\hcop)$  decreases by a factor of 3 as the metallicity drops from 1.3 $\rm Z_{\odot}$ to 0.3 $\rm Z_{\odot}$.

    \item  The dispersion in  the mass conversion factors are large, even after correction for metallicity. The effect of environment, in particular UV luminosity, appears to have an effect similar in significance to that of metallicity. This effect could be important in low-$Z$ dwarf galaxies with higher radiation fields and less dust shielding.

    \item  The mass conversion factors derived from gas based analysis and BGPS based analysis differ. 
    The gas-based $\alpha_{\rm tot}$ values are about 1.7 times the BGPS-based values, reflecting lower masses in the dust-based definition of dense gas. This is likely related to the choice of dust continuum opacity.
    
    \item The values of $\alpha_{\rm tot}$  for Solar neighborhood clouds tend to be higher than predicted by our best-fitting relations, especially for the low-mass star forming regions.

    \item  Extrapolation of the metallicity corrections to the Central Molecular Zone would result in column densities lower than the usual estimates by factors of 2 to 3, contributing to an explanation of the relatively slow star formation there.
    
\end{enumerate}

\begin{acknowledgments}
 We thank the referee for insightful comments that have improved the paper.
We thank the staff of the TRAO for support during the course of these observations. 
SP and KT-Kim would like to express their gratitude to Dongwook Kim for his assistance with the on-site observations at KASI.
SP thanks DST-INSPIRE Fellowship (No. IF180092) of the Department of Science and Technology to support the Ph.D. program. 
NJE thanks the Department of Astronomy at the University of Texas at Austin for ongoing research support. 
JJ acknowledges the financial support received through the DST-SERB grant SPG/2021/003850.
\end{acknowledgments}

%

\vspace{5mm}
\facilities{TRAO 14-m Telescope, FCRAO, BGPS, \textit{Herschel} }


\software{astropy \citep{2013A&A...558A..33A,2018AJ....156..123A,2022arXiv220614220T},
          numpy \citep{2011CSE....13b..22V},
          matplotlib \citep{2007CSE.....9...90H},
          GILDAS \citep{2018ssdd.confE..11P},
          DS9 \citep{2003ASPC..295..489J, 2000ascl.soft03002S}.}



\clearpage
\appendix
\section{TRAO Data Reduction} \label{appenA}
We follow the standard data reduction method explained in Appendix A of \cite{2022AJ....164..129P}.
The steps are as follows. We first examine the data to identify the velocity intervals with significant emission ($v_{win}$). Next, we exclude the velocity window containing line emission and subtract the second-order polynomial baseline, utilizing only the velocity range necessary to obtain a good baseline at both ends ($v_{sp}$). 
We varied the velocity range ($v_{sp}$) and baseline order to achieve the best fit (for more details see \citealt{2022AJ....164..129P}).
Following this, we generate spectral cubes in FITS format to facilitate further analysis. Details regarding the total velocity range and excluded windows can be found in Table \ref{tab:reduction}.


\begin{longtable}{c l c c c c c c }
\caption{Velocity intervals considered for baseline subtraction and integration} \label{tab:reduction}\\
\hline
\hline
Source      &       Line        & $v_{sp}$      &       $v_{win}$              \\
            &                   & (\kms)        &       (\kms)                  \\
\hline
\endfirsthead
\multicolumn{8}{c}%
{\tablename\ \thetable\ -- \textit{Continued from previous page}} \\
\hline
\hline
Source      &       Line        & $v_{sp}$      &       $v_{win}$             \\
            &                   & (\kms)        &       (\kms)                \\
\hline
\endhead
\hline \multicolumn{8}{r}{\textit{Continued on next page}} \\
\endfoot
\hline
\endlastfoot
       
    Sh2-128 & \cotw    &  -110, -35    &  -80, -65    \\
         & \coo     &  -110, -35    &  -80, -65    \\ 
         & \hcop    &  -110, -35    &  -80, -65    \\
         & \hcn     &  -110, -35    &  -85, -60    \\
     \\
     Sh2-132 & \cotw & -80, -10   & -55, -38 \\
          & \coo  & -80, -10   & -55, -38 \\
          & \hcop & -80, -10   & -55, -38 \\
          & \hcn  & -80, -10   & -58, -35 \\
     \\
     Sh2-142   & \cotw  &  -75, -10       &  -46, -34      \\
            & \coo   &  -75, -10       &  -46, -34   \\
            & \hcop  &   -75, -10       &  -46, -34  \\
            & \hcn   &   -75, -10       &  -54, -32   \\
    \\
    Sh2-148   & \cotw  &  -75, -25         &     -58, -44  \\
            & \coo   & -75, -25        &  -58, -44      \\
            & \hcop  &  -75, -25       &   -58, -44     \\
            & \hcn   &  -75, -25       &  -60, -42      \\
    \\
    Sh2-156   & \cotw  &  -90, -15      &   -60, -44   \\
            & \coo   &  -90, -15      &   -60, -44   \\
            & \hcop  &  -90, -15      &   -60, -44        \\
            & \hcn   &  -90, -15      &   -65, -40      \\
    \\
    Sh2-209    & \hcop  &  -80, -30       &   -60, -48     \\
            & \hcn   &  -80, -30       &   -64, -44     \\
    \\
    Sh2-242    & \hcop  &  -30, 30       & -5, 8        \\
            & \hcn   &   -30, 30      & -10, 10        \\
    \\
    Sh2-266   & \cotw   & -5, 70  &   25, 38     \\
            & \coo   & -5, 70  &   25, 38     \\
            & \hcop  & -5, 70  &   25, 38     \\
            & \hcn   & -5, 70  &   20, 40     \\
    \\
    Sh2-269   & \cotw   & -20, 50        & 12, 24    \\
            & \coo   & -20, 50        & 12, 24    \\
            & \hcop  & -20, 50        & 12, 24   \\
            & \hcn   & -20, 50        & 6, 27       \\
    \\
     Sh2-270   & \cotw  &  0, 50       & 22, 28     \\
            & \coo   &   0, 50      & 22, 28      \\
            & \hcop  &   0, 50      & 22, 28       \\
            & \hcn   &   0, 50      & 16, 32       \\
    \\
\label{lineprops}
\end{longtable}


\section{The Gas to Solid Ratio in the Solar Neighborhood}\label{gammasun}

Conversion from dust continuum emission to gas density requires an assumption of a gas-to-dust mass ratio. Papers often say only that they used a ``standard gas-to-dust" value.
Almost always, this phrase means that they assume that every
gram of dust corresponds to 100 grams of gas. This is a dubious assumption, even for the solar neighborhood, so we reconsider the value of the solar neighborhood gas-to-dust ratio (\gammasun) here. 

We use the latest model of dust in the diffuse ISM, but older models
do not change the argument much. The newest model 
\citep{2021ApJ...906...73H, 2023ApJ...948...55H, 2024ApJ...962...99H} accounts for many observations not adequately explained by previous models. It is based on the abundances given in Table 2 of
\citet{2021ApJ...906...73H}.
The table has values for abundances (expressed as [X/H]) for the ISM and separately for those abundances in the gas and solid phase in the diffuse ISM, including the 9 most important elements, from O to Ni. These abundances are for the local ISM so should determine the local gas-to-dust ratio, \gammasun.

If we add the information on atomic mass for the solar isotope mix
\citep{2000asqu.book.....C}, we can compute the dust to hydrogen ratio for each
element from the column labeled (X/H)$_{dust}$ in \citet{2021ApJ...906...73H} for each element and add them up.
\begin{equation}
1/\gammasun_{\rm diffuse} = \sum X/H_{dust} \times AW_X
\end{equation}
where $AW_X$ is the atomic weight of each element with a solar mix of isotopes
(e.g., 12.011 for C, 24.3 for Mg). The result is the dust to {\it hydrogen}
ratio. To get the dust-to-gas ratio, including He, we divide by the mass per H atom in amu, 
$\mu_{\rm H} = 1.404$ (Shirley et al. in prep.).
The result is that dust-to-hydrogen ratio is 1.06\ee{-2} and the dust-to-gas ratio is 7.55\ee{-3}.
Thus, the gas-to-dust ratio is 132.5. This is what we really should use for the diffuse ISM.

The plot thickens in molecular clouds, where more heavy elements condense as ices
\citep{2013ApJ...774..102W, 2014prpl.conf..363P}.
For this case, we use the column labeled (X/H)$_{gas}$ in Table 2 of
\citet{2021ApJ...906...73H}.
In the limit that   all heavy elements condense, the result for the ice-to-hydrogen ratio is 1.00\ee{-2} and the ice-to-gas ratio is 7.12\ee{-3} (or gas-to-ice ratio of 140).
So, in the extreme, ice can roughly double the mass of solids. If we add
dust and ice, we get a solid-to-gas ratio of 1.46\ee{-2} and a gas-to-solids ratio of 68.2.
 It could be even lower if the
C and O bring H with them. In the extreme case that all C is in CH$_4$ and all O is in H$_2$O, the gas to solid ratio would be 60. So, in a very icy protostellar core, the correct gas-to-solid ratio is at least 60 but probably closer to 60 than to 132.5.

What is the evidence on ice-to-dust ratios in clouds with $A_V \sim 8$ mag that characterizes most of our observations?
Figure 5 of 
\citet{2014prpl.conf..363P}
shows ratios of ices to silicates for ``intra-cloud medium", YSOs and ``extreme" (deeply embedded) YSOs. 
Even for the intra-cloud medium, they favor a ratio of about 1.3.
What range of column densities has substantial ice?
\citet{2013ApJ...774..102W} found a threshold of $A_V = 3.2\pm 0.1$ mag for water ice mantles and a super-linear growth (slope of $1.30 \pm 0.04$) with
$A_V$. 
In the absence of more detailed information for each source, we use the ice-to-dust ratio from 
\citet{2014prpl.conf..363P}
for the intra-cloud medium of 1.3.
This leads to a gas-to-solid ratio for the Solar neighborhood, $\gammasun = 101.5$.
We approximate this with $\gammasun = 100$, for consistency with common practice, but we emphasize that this is the {\it gas-to-solid} ratio, where solid includes both ice and dust, that the uncertainties are substantial, and that it depends on location in the ISM.



\bibliography{ms}{}

\begin{thebibliography}{}
\expandafter\ifx\csname natexlab\endcsname\relax\def\natexlab#1{#1}\fi
\providecommand{\url}[1]{\href{#1}{#1}}
\providecommand{\dodoi}[1]{doi:~\href{http://doi.org/#1}{\nolinkurl{#1}}}
\providecommand{\doeprint}[1]{\href{http://ascl.net/#1}{\nolinkurl{http://ascl.net/#1}}}
\providecommand{\doarXiv}[1]{\href{https://arxiv.org/abs/#1}{\nolinkurl{https://arxiv.org/abs/#1}}}

\bibitem[{{Aniano} {et~al.}(2020{\natexlab{a}}){Aniano}, {Draine}, {Hunt}, {Sandstrom}, {Calzetti}, {Kennicutt}, {Dale}, {Galametz}, {Gordon}, {Leroy}, {Smith}, {Roussel}, {Sauvage}, {Walter}, {Armus}, {Bolatto}, {Boquien}, {Crocker}, {De Looze}, {Donovan Meyer}, {Helou}, {Hinz}, {Johnson}, {Koda}, {Miller}, {Montiel}, {Murphy}, {Rela{\~n}o}, {Rix}, {Schinnerer}, {Skibba}, {Wolfire}, \& {Engelbracht}}]{2020ApJ...889..150A}
{Aniano}, G., {Draine}, B.~T., {Hunt}, L.~K., {et~al.} 2020{\natexlab{a}}, \apj, 889, 150, \dodoi{10.3847/1538-4357/ab5fdb}

\bibitem[{{Aniano} {et~al.}(2020{\natexlab{b}}){Aniano}, {Draine}, {Hunt}, {Sandstrom}, {Calzetti}, {Kennicutt}, {Dale}, {Galametz}, {Gordon}, {Leroy}, {Smith}, {Roussel}, {Sauvage}, {Walter}, {Armus}, {Bolatto}, {Boquien}, {Crocker}, {De Looze}, {DDonovan Meyer}, {Helou}, {Hinz}, {Johnson}, {Koda}, {Miller}, {Montiel}, {Murphy}, {Rela{\~n}o}, {Rix}, {Schinnerer}, {Skibba}, {Wolfire}, \& {Engelbracht}}]{2020ApJ...897..184A}
---. 2020{\natexlab{b}}, \apj, 897, 184, \dodoi{10.3847/1538-4357/aba0bb}

\bibitem[{{Aoyama} {et~al.}(2020){Aoyama}, {Hirashita}, \& {Nagamine}}]{2020MNRAS.491.3844A}
{Aoyama}, S., {Hirashita}, H., \& {Nagamine}, K. 2020, \mnras, 491, 3844, \dodoi{10.1093/mnras/stz3253}

\bibitem[{{Aoyama} {et~al.}(2017){Aoyama}, {Hou}, {Shimizu}, {Hirashita}, {Todoroki}, {Choi}, \& {Nagamine}}]{2017MNRAS.466..105A}
{Aoyama}, S., {Hou}, K.-C., {Shimizu}, I., {et~al.} 2017, \mnras, 466, 105, \dodoi{10.1093/mnras/stw3061}

\bibitem[{{Astropy Collaboration} {et~al.}(2013){Astropy Collaboration}, {Robitaille}, {Tollerud}, {Greenfield}, {Droettboom}, {Bray}, {Aldcroft}, {Davis}, {Ginsburg}, {Price-Whelan}, {Kerzendorf}, {Conley}, {Crighton}, {Barbary}, {Muna}, {Ferguson}, {Grollier}, {Parikh}, {Nair}, {Unther}, {Deil}, {Woillez}, {Conseil}, {Kramer}, {Turner}, {Singer}, {Fox}, {Weaver}, {Zabalza}, {Edwards}, {Azalee Bostroem}, {Burke}, {Casey}, {Crawford}, {Dencheva}, {Ely}, {Jenness}, {Labrie}, {Lim}, {Pierfederici}, {Pontzen}, {Ptak}, {Refsdal}, {Servillat}, \& {Streicher}}]{2013A&A...558A..33A}
{Astropy Collaboration}, {Robitaille}, T.~P., {Tollerud}, E.~J., {et~al.} 2013, \aap, 558, A33, \dodoi{10.1051/0004-6361/201322068}

\bibitem[{{Astropy Collaboration} {et~al.}(2018){Astropy Collaboration}, {Price-Whelan}, {Sip{\H{o}}cz}, {G{\"u}nther}, {Lim}, {Crawford}, {Conseil}, {Shupe}, {Craig}, {Dencheva}, {Ginsburg}, {VanderPlas}, {Bradley}, {P{\'e}rez-Su{\'a}rez}, {de Val-Borro}, {Aldcroft}, {Cruz}, {Robitaille}, {Tollerud}, {Ardelean}, {Babej}, {Bach}, {Bachetti}, {Bakanov}, {Bamford}, {Barentsen}, {Barmby}, {Baumbach}, {Berry}, {Biscani}, {Boquien}, {Bostroem}, {Bouma}, {Brammer}, {Bray}, {Breytenbach}, {Buddelmeijer}, {Burke}, {Calderone}, {Cano Rodr{\'\i}guez}, {Cara}, {Cardoso}, {Cheedella}, {Copin}, {Corrales}, {Crichton}, {D'Avella}, {Deil}, {Depagne}, {Dietrich}, {Donath}, {Droettboom}, {Earl}, {Erben}, {Fabbro}, {Ferreira}, {Finethy}, {Fox}, {Garrison}, {Gibbons}, {Goldstein}, {Gommers}, {Greco}, {Greenfield}, {Groener}, {Grollier}, {Hagen}, {Hirst}, {Homeier}, {Horton}, {Hosseinzadeh}, {Hu}, {Hunkeler}, {Ivezi{\'c}}, {Jain}, {Jenness}, {Kanarek}, {Kendrew}, {Kern}, {Kerzendorf}, {Khvalko}, {King}, {Kirkby}, {Kulkarni},
  {Kumar}, {Lee}, {Lenz}, {Littlefair}, {Ma}, {Macleod}, {Mastropietro}, {McCully}, {Montagnac}, {Morris}, {Mueller}, {Mumford}, {Muna}, {Murphy}, {Nelson}, {Nguyen}, {Ninan}, {N{\"o}the}, {Ogaz}, {Oh}, {Parejko}, {Parley}, {Pascual}, {Patil}, {Patil}, {Plunkett}, {Prochaska}, {Rastogi}, {Reddy Janga}, {Sabater}, {Sakurikar}, {Seifert}, {Sherbert}, {Sherwood-Taylor}, {Shih}, {Sick}, {Silbiger}, {Singanamalla}, {Singer}, {Sladen}, {Sooley}, {Sornarajah}, {Streicher}, {Teuben}, {Thomas}, {Tremblay}, {Turner}, {Terr{\'o}n}, {van Kerkwijk}, {de la Vega}, {Watkins}, {Weaver}, {Whitmore}, {Woillez}, {Zabalza}, \& {Astropy Contributors}}]{2018AJ....156..123A}
{Astropy Collaboration}, {Price-Whelan}, A.~M., {Sip{\H{o}}cz}, B.~M., {et~al.} 2018, \aj, 156, 123, \dodoi{10.3847/1538-3881/aabc4f}

\bibitem[{{Bailer-Jones} {et~al.}(2021){Bailer-Jones}, {Rybizki}, {Fouesneau}, {Demleitner}, \& {Andrae}}]{2021AJ....161..147B}
{Bailer-Jones}, C.~A.~L., {Rybizki}, J., {Fouesneau}, M., {Demleitner}, M., \& {Andrae}, R. 2021, \aj, 161, 147, \dodoi{10.3847/1538-3881/abd806}

\bibitem[{{Bailer-Jones} {et~al.}(2018){Bailer-Jones}, {Rybizki}, {Fouesneau}, {Mantelet}, \& {Andrae}}]{2018AJ....156...58B}
{Bailer-Jones}, C.~A.~L., {Rybizki}, J., {Fouesneau}, M., {Mantelet}, G., \& {Andrae}, R. 2018, \aj, 156, 58, \dodoi{10.3847/1538-3881/aacb21}

\bibitem[{{Balser} {et~al.}(2011){Balser}, {Rood}, {Bania}, \& {Anderson}}]{2011ApJ...738...27B}
{Balser}, D.~S., {Rood}, R.~T., {Bania}, T.~M., \& {Anderson}, L.~D. 2011, \apj, 738, 27, \dodoi{10.1088/0004-637X/738/1/27}

\bibitem[{{Barnes} {et~al.}(2020){Barnes}, {Kauffmann}, {Bigiel}, {Brinkmann}, {Colombo}, {Guzm{\'a}n}, {Kim}, {Sz{\H{u}}cs}, {Wakelam}, {Aalto}, {Albertsson}, {Evans}, {Glover}, {Goldsmith}, {Kramer}, {Menten}, {Nishimura}, {Viti}, {Watanabe}, {Weiss}, {Wienen}, {Wiesemeyer}, \& {Wyrowski}}]{2020MNRAS.497.1972B}
{Barnes}, A.~T., {Kauffmann}, J., {Bigiel}, F., {et~al.} 2020, \mnras, 497, 1972, \dodoi{10.1093/mnras/staa1814}

\bibitem[{{Battersby} {et~al.}(2020){Battersby}, {Keto}, {Walker}, {Barnes}, {Callanan}, {Ginsburg}, {Hatchfield}, {Henshaw}, {Kauffmann}, {Kruijssen}, {Longmore}, {Lu}, {Mills}, {Pillai}, {Zhang}, {Bally}, {Butterfield}, {Contreras}, {Ho}, {Ott}, {Patel}, \& {Tolls}}]{2020ApJS..249...35B}
{Battersby}, C., {Keto}, E., {Walker}, D., {et~al.} 2020, \apjs, 249, 35, \dodoi{10.3847/1538-4365/aba18e}

\bibitem[{{Battersby} {et~al.}(2024){Battersby}, {Walker}, {Barnes}, {Ginsburg}, {Lipman}, {Alboslani}, {Hatchfield}, {Bally}, {Glover}, {Henshaw}, {Immer}, {Klessen}, {Longmore}, {Mills}, {Molinari}, {Smith}, {Sormani}, {Tress}, \& {Zhang}}]{2024arXiv241017332B}
{Battersby}, C., {Walker}, D.~L., {Barnes}, A., {et~al.} 2024, arXiv e-prints, arXiv:2410.17332, \dodoi{10.48550/arXiv.2410.17332}

\bibitem[{{Bigiel} {et~al.}(2016){Bigiel}, {Leroy}, {Jim{\'e}nez-Donaire}, {Pety}, {Usero}, {Cormier}, {Bolatto}, {Garcia-Burillo}, {Colombo}, {Gonz{\'a}lez-Garc{\'\i}a}, {Hughes}, {Kepley}, {Kramer}, {Sandstrom}, {Schinnerer}, {Schruba}, {Schuster}, {Tomicic}, \& {Zschaechner}}]{2016ApJ...822L..26B}
{Bigiel}, F., {Leroy}, A.~K., {Jim{\'e}nez-Donaire}, M.~J., {et~al.} 2016, \apjl, 822, L26, \dodoi{10.3847/2041-8205/822/2/L26}

\bibitem[{{Binder} \& {Povich}(2018)}]{2018ApJ...864..136B}
{Binder}, B.~A., \& {Povich}, M.~S. 2018, \apj, 864, 136, \dodoi{10.3847/1538-4357/aad7b2}

\bibitem[{{Bolatto} {et~al.}(2013){Bolatto}, {Wolfire}, \& {Leroy}}]{2013ARA&A..51..207B}
{Bolatto}, A.~D., {Wolfire}, M., \& {Leroy}, A.~K. 2013, \araa, 51, 207, \dodoi{10.1146/annurev-astro-082812-140944}

\bibitem[{{Braine} {et~al.}(2017){Braine}, {Shimajiri}, {Andr{\'e}}, {Bontemps}, {Gao}, {Chen}, \& {Kramer}}]{2017A&A...597A..44B}
{Braine}, J., {Shimajiri}, Y., {Andr{\'e}}, P., {et~al.} 2017, \aap, 597, A44, \dodoi{10.1051/0004-6361/201629781}

\bibitem[{{Buchbender} {et~al.}(2013){Buchbender}, {Kramer}, {Gonzalez-Garcia}, {Israel}, {Garc{\'\i}a-Burillo}, {van der Werf}, {Braine}, {Rosolowsky}, {Mookerjea}, {Aalto}, {Boquien}, {Gratier}, {Henkel}, {Quintana-Lacaci}, {Verley}, \& {van der Tak}}]{2013A&A...549A..17B}
{Buchbender}, C., {Kramer}, C., {Gonzalez-Garcia}, M., {et~al.} 2013, \aap, 549, A17, \dodoi{10.1051/0004-6361/201219436}

\bibitem[{{Chen} {et~al.}(2015){Chen}, {Gao}, {Braine}, \& {Gu}}]{2015ApJ...810..140C}
{Chen}, H., {Gao}, Y., {Braine}, J., \& {Gu}, Q. 2015, \apj, 810, 140, \dodoi{10.1088/0004-637X/810/2/140}

\bibitem[{{Chiang} {et~al.}(2021){Chiang}, {Sandstrom}, {Chastenet}, {Herrera}, {Koch}, {Kreckel}, {Leroy}, {Pety}, {Schruba}, {Utomo}, \& {Williams}}]{2021ApJ...907...29C}
{Chiang}, I.-D., {Sandstrom}, K.~M., {Chastenet}, J., {et~al.} 2021, \apj, 907, 29, \dodoi{10.3847/1538-4357/abceb6}

\bibitem[{{Chin} {et~al.}(1998){Chin}, {Henkel}, {Millar}, {Whiteoak}, \& {Marx-Zimmer}}]{1998A&A...330..901C}
{Chin}, Y.~N., {Henkel}, C., {Millar}, T.~J., {Whiteoak}, J.~B., \& {Marx-Zimmer}, M. 1998, \aap, 330, 901, \dodoi{10.48550/arXiv.astro-ph/9710253}

\bibitem[{{Chin} {et~al.}(1997){Chin}, {Henkel}, {Whiteoak}, {Millar}, {Hunt}, \& {Lemme}}]{1997A&A...317..548C}
{Chin}, Y.~N., {Henkel}, C., {Whiteoak}, J.~B., {et~al.} 1997, \aap, 317, 548, \dodoi{10.48550/arXiv.astro-ph/9606081}

\bibitem[{{Cox}(2000)}]{2000asqu.book.....C}
{Cox}, A.~N. 2000, {Allen's astrophysical quantities}

\bibitem[{{Dame} \& {Lada}(2023)}]{2023ApJ...944..197D}
{Dame}, T.~M., \& {Lada}, C.~J. 2023, \apj, 944, 197, \dodoi{10.3847/1538-4357/acb438}

\bibitem[{{Dunham} {et~al.}(2011){Dunham}, {Rosolowsky}, {Evans}, {Cyganowski}, \& {Urquhart}}]{2011ApJ...741..110D}
{Dunham}, M.~K., {Rosolowsky}, E., {Evans}, Neal~J., I., {Cyganowski}, C., \& {Urquhart}, J.~S. 2011, \apj, 741, 110, \dodoi{10.1088/0004-637X/741/2/110}

\bibitem[{{Elia} {et~al.}(2025){Elia}, {Evans}, {Soler}, {Strafella}, {Schisano}, {Molinari}, {Giannetti}, \& {Patra}}]{2025arXiv250114471E}
{Elia}, D., {Evans}, II, N.~J., {Soler}, J.~D., {et~al.} 2025, arXiv e-prints, arXiv:2501.14471, \dodoi{10.48550/arXiv.2501.14471}

\bibitem[{{Elia} {et~al.}(2013){Elia}, {Molinari}, {Fukui}, {Schisano}, {Olmi}, {Veneziani}, {Hayakawa}, {Pestalozzi}, {Schneider}, {Benedettini}, {di Giorgio}, {Ikhenaode}, {Mizuno}, {Onishi}, {Pezzuto}, {Piazzo}, {Polychroni}, {Rygl}, {Yamamoto}, \& {Maruccia}}]{2013ApJ...772...45E}
{Elia}, D., {Molinari}, S., {Fukui}, Y., {et~al.} 2013, \apj, 772, 45, \dodoi{10.1088/0004-637X/772/1/45}

\bibitem[{{Elia} {et~al.}(2021){Elia}, {Merello}, {Molinari}, {Schisano}, {Zavagno}, {Russeil}, {M{\`e}ge}, {Martin}, {Olmi}, {Pestalozzi}, {Plume}, {Ragan}, {Benedettini}, {Eden}, {Moore}, {Noriega-Crespo}, {Paladini}, {Palmeirim}, {Pezzuto}, {Pilbratt}, {Rygl}, {Schilke}, {Strafella}, {Tan}, {Traficante}, {Baldeschi}, {Bally}, {di Giorgio}, {Fiorellino}, {Liu}, {Piazzo}, \& {Polychroni}}]{2021MNRAS.504.2742E}
{Elia}, D., {Merello}, M., {Molinari}, S., {et~al.} 2021, \mnras, 504, 2742, \dodoi{10.1093/mnras/stab1038}

\bibitem[{{Enoch} {et~al.}(2006){Enoch}, {Young}, {Glenn}, {Evans}, {Golwala}, {Sargent}, {Harvey}, {Aguirre}, {Goldin}, {Haig}, {Huard}, {Lange}, {Laurent}, {Maloney}, {Mauskopf}, {Rossinot}, \& {Sayers}}]{2006ApJ...638..293E}
{Enoch}, M.~L., {Young}, K.~E., {Glenn}, J., {et~al.} 2006, \apj, 638, 293, \dodoi{10.1086/498678}

\bibitem[{{Esteban} {et~al.}(2017){Esteban}, {Fang}, {Garc{\'\i}a-Rojas}, \& {Toribio San Cipriano}}]{2017MNRAS.471..987E}
{Esteban}, C., {Fang}, X., {Garc{\'\i}a-Rojas}, J., \& {Toribio San Cipriano}, L. 2017, \mnras, 471, 987, \dodoi{10.1093/mnras/stx1624}

\bibitem[{{Esteban} \& {Garc{\'\i}a-Rojas}(2018)}]{2018MNRAS.478.2315E}
{Esteban}, C., \& {Garc{\'\i}a-Rojas}, J. 2018, \mnras, 478, 2315, \dodoi{10.1093/mnras/sty1168}

\bibitem[{{Esteban} {et~al.}(2022){Esteban}, {M{\'e}ndez-Delgado}, {Garc{\'\i}a-Rojas}, \& {Arellano-C{\'o}rdova}}]{2022ApJ...931...92E}
{Esteban}, C., {M{\'e}ndez-Delgado}, J.~E., {Garc{\'\i}a-Rojas}, J., \& {Arellano-C{\'o}rdova}, K.~Z. 2022, \apj, 931, 92, \dodoi{10.3847/1538-4357/ac6b38}

\bibitem[{{Evans} {et~al.}(2014){Evans}, {Heiderman}, \& {Vutisalchavakul}}]{2014ApJ...782..114E}
{Evans}, Neal~J., I., {Heiderman}, A., \& {Vutisalchavakul}, N. 2014, \apj, 782, 114, \dodoi{10.1088/0004-637X/782/2/114}

\bibitem[{{Evans} {et~al.}(2020){Evans}, {Kim}, {Wu}, {Chao}, {Heyer}, {Liu}, {Nguyen-Lu'o'ng}, \& {Kauffmann}}]{2020ApJ...894..103E}
{Evans}, Neal~J., I., {Kim}, K.-T., {Wu}, J., {et~al.} 2020, \apj, 894, 103, \dodoi{10.3847/1538-4357/ab8938}

\bibitem[{{Evans} {et~al.}(2022){Evans}, {Kim}, \& {Ostriker}}]{2022ApJ...929L..18E}
{Evans}, N.~J., {Kim}, J.-G., \& {Ostriker}, E.~C. 2022, \apjl, 929, L18, \dodoi{10.3847/2041-8213/ac6427}

\bibitem[{{Galametz} {et~al.}(2020){Galametz}, {Schruba}, {De Breuck}, {Immer}, {Chevance}, {Galliano}, {Gusdorf}, {Lebouteiller}, {Lee}, {Madden}, {Polles}, \& {van Kempen}}]{2020A&A...643A..63G}
{Galametz}, M., {Schruba}, A., {De Breuck}, C., {et~al.} 2020, \aap, 643, A63, \dodoi{10.1051/0004-6361/202038641}

\bibitem[{{Gao} \& {Solomon}(2004{\natexlab{a}})}]{2004ApJS..152...63G}
{Gao}, Y., \& {Solomon}, P.~M. 2004{\natexlab{a}}, \apjs, 152, 63, \dodoi{10.1086/383003}

\bibitem[{{Gao} \& {Solomon}(2004{\natexlab{b}})}]{2004ApJ...606..271G}
---. 2004{\natexlab{b}}, \apj, 606, 271, \dodoi{10.1086/382999}

\bibitem[{{Giannetti} {et~al.}(2017){Giannetti}, {Leurini}, {K{\"o}nig}, {Urquhart}, {Pillai}, {Brand}, {Kauffmann}, {Wyrowski}, \& {Menten}}]{2017A&A...606L..12G}
{Giannetti}, A., {Leurini}, S., {K{\"o}nig}, C., {et~al.} 2017, \aap, 606, L12, \dodoi{10.1051/0004-6361/201731728}

\bibitem[{{Ginsburg} {et~al.}(2013){Ginsburg}, {Glenn}, {Rosolowsky}, {Ellsworth-Bowers}, {Battersby}, {Dunham}, {Merello}, {Shirley}, {Bally}, {Evans}, {Stringfellow}, \& {Aguirre}}]{2013ApJS..208...14G}
{Ginsburg}, A., {Glenn}, J., {Rosolowsky}, E., {et~al.} 2013, \apjs, 208, 14, \dodoi{10.1088/0067-0049/208/2/14}

\bibitem[{{Ginsburg} {et~al.}(2018){Ginsburg}, {Bally}, {Barnes}, {Bastian}, {Battersby}, {Beuther}, {Brogan}, {Contreras}, {Corby}, {Darling}, {De Pree}, {Galv{\'a}n-Madrid}, {Garay}, {Henshaw}, {Hunter}, {Kruijssen}, {Longmore}, {Lu}, {Meng}, {Mills}, {Ott}, {Pineda}, {S{\'a}nchez-Monge}, {Schilke}, {Schmiedeke}, {Walker}, \& {Wilner}}]{2018ApJ...853..171G}
{Ginsburg}, A., {Bally}, J., {Barnes}, A., {et~al.} 2018, \apj, 853, 171, \dodoi{10.3847/1538-4357/aaa6d4}

\bibitem[{{Glover} \& {Mac Low}(2011)}]{2011MNRAS.412..337G}
{Glover}, S.~C.~O., \& {Mac Low}, M.~M. 2011, \mnras, 412, 337, \dodoi{10.1111/j.1365-2966.2010.17907.x}

\bibitem[{{Goldsmith} \& {Kauffmann}(2017)}]{2017ApJ...841...25G}
{Goldsmith}, P.~F., \& {Kauffmann}, J. 2017, \apj, 841, 25, \dodoi{10.3847/1538-4357/aa6f12}

\bibitem[{{Gong} {et~al.}(2020){Gong}, {Ostriker}, {Kim}, \& {Kim}}]{2020ApJ...903..142G}
{Gong}, M., {Ostriker}, E.~C., {Kim}, C.-G., \& {Kim}, J.-G. 2020, \apj, 903, 142, \dodoi{10.3847/1538-4357/abbdab}

\bibitem[{{Gravity Collaboration} {et~al.}(2019){Gravity Collaboration}, {Abuter}, {Amorim}, {Baub{\"o}ck}, {Berger}, {Bonnet}, {Brandner}, {Cl{\'e}net}, {Coud{\'e} Du Foresto}, {de Zeeuw}, {Dexter}, {Duvert}, {Eckart}, {Eisenhauer}, {F{\"o}rster Schreiber}, {Garcia}, {Gao}, {Gendron}, {Genzel}, {Gerhard}, {Gillessen}, {Habibi}, {Haubois}, {Henning}, {Hippler}, {Horrobin}, {Jim{\'e}nez-Rosales}, {Jocou}, {Kervella}, {Lacour}, {Lapeyr{\`e}re}, {Le Bouquin}, {L{\'e}na}, {Ott}, {Paumard}, {Perraut}, {Perrin}, {Pfuhl}, {Rabien}, {Rodriguez Coira}, {Rousset}, {Scheithauer}, {Sternberg}, {Straub}, {Straubmeier}, {Sturm}, {Tacconi}, {Vincent}, {von Fellenberg}, {Waisberg}, {Widmann}, {Wieprecht}, {Wiezorrek}, {Woillez}, \& {Yazici}}]{2019A&A...625L..10G}
{Gravity Collaboration}, {Abuter}, R., {Amorim}, A., {et~al.} 2019, \aap, 625, L10, \dodoi{10.1051/0004-6361/201935656}

\bibitem[{{Hatchfield} {et~al.}(2024){Hatchfield}, {Battersby}, {Barnes}, {Butterfield}, {Ginsburg}, {Henshaw}, {Longmore}, {Lu}, {Svoboda}, {Walker}, {Callanan}, {Mills}, {Ho}, {Kauffmann}, {Kruijssen}, {Ott}, {Pillai}, \& {Zhang}}]{2024ApJ...962...14H}
{Hatchfield}, H.~P., {Battersby}, C., {Barnes}, A.~T., {et~al.} 2024, \apj, 962, 14, \dodoi{10.3847/1538-4357/ad10af}

\bibitem[{{Heiderman} {et~al.}(2010){Heiderman}, {Evans}, {Allen}, {Huard}, \& {Heyer}}]{2010ApJ...723.1019H}
{Heiderman}, A., {Evans}, Neal~J., I., {Allen}, L.~E., {Huard}, T., \& {Heyer}, M. 2010, \apj, 723, 1019, \dodoi{10.1088/0004-637X/723/2/1019}

\bibitem[{{Henshaw} {et~al.}(2023){Henshaw}, {Barnes}, {Battersby}, {Ginsburg}, {Sormani}, \& {Walker}}]{2023ASPC..534...83H}
{Henshaw}, J.~D., {Barnes}, A.~T., {Battersby}, C., {et~al.} 2023, in Astronomical Society of the Pacific Conference Series, Vol. 534, Protostars and Planets VII, ed. S.~{Inutsuka}, Y.~{Aikawa}, T.~{Muto}, K.~{Tomida}, \& M.~{Tamura}, 83, \dodoi{10.48550/arXiv.2203.11223}

\bibitem[{{Hensley} \& {Draine}(2021)}]{2021ApJ...906...73H}
{Hensley}, B.~S., \& {Draine}, B.~T. 2021, \apj, 906, 73, \dodoi{10.3847/1538-4357/abc8f1}

\bibitem[{{Hensley} \& {Draine}(2023)}]{2023ApJ...948...55H}
---. 2023, \apj, 948, 55, \dodoi{10.3847/1538-4357/acc4c2}

\bibitem[{{Hensley} \& {Draine}(2024)}]{2024ApJ...962...99H}
---. 2024, \apj, 962, 99, \dodoi{10.3847/1538-4357/ad205d}

\bibitem[{{Heyer} {et~al.}(2022){Heyer}, {Gregg}, {Calzetti}, {Elmegreen}, {Kennicutt}, {Adamo}, {Evans}, {Grasha}, {Lowenthal}, {Narayanan}, {Rosa-Gonzalez}, {Schloerb}, {Souccar}, {Tang}, {Teuben}, {Vega}, {Wall}, \& {Yun}}]{2022ApJ...930..170H}
{Heyer}, M., {Gregg}, B., {Calzetti}, D., {et~al.} 2022, \apj, 930, 170, \dodoi{10.3847/1538-4357/ac67ea}

\bibitem[{{Hildebrand}(1983)}]{1983QJRAS..24..267H}
{Hildebrand}, R.~H. 1983, \qjras, 24, 267

\bibitem[{{Hirashita}(2015)}]{2015MNRAS.447.2937H}
{Hirashita}, H. 2015, \mnras, 447, 2937, \dodoi{10.1093/mnras/stu2617}

\bibitem[{{Hirashita}(2023)}]{2023MNRAS.522.4612H}
---. 2023, \mnras, 522, 4612, \dodoi{10.1093/mnras/stad1286}

\bibitem[{{Hu} {et~al.}(2022){Hu}, {Schruba}, {Sternberg}, \& {van Dishoeck}}]{2022ApJ...931...28H}
{Hu}, C.-Y., {Schruba}, A., {Sternberg}, A., \& {van Dishoeck}, E.~F. 2022, \apj, 931, 28, \dodoi{10.3847/1538-4357/ac65fd}

\bibitem[{{Hunter}(2007)}]{2007CSE.....9...90H}
{Hunter}, J.~D. 2007, Computing in Science and Engineering, 9, 90, \dodoi{10.1109/MCSE.2007.55}

\bibitem[{{Jacob} {et~al.}(2020){Jacob}, {Menten}, {Wiesemeyer}, {G{\"u}sten}, {Wyrowski}, \& {Klein}}]{2020A&A...640A.125J}
{Jacob}, A.~M., {Menten}, K.~M., {Wiesemeyer}, H., {et~al.} 2020, \aap, 640, A125, \dodoi{10.1051/0004-6361/201937385}

\bibitem[{{Jeong} {et~al.}(2019){Jeong}, {Kang}, {Jung}, {Lee}, {Byun}, {Je}, {Kang}, {Lee}, \& {Lee}}]{2019JKAS...52..227J}
{Jeong}, I.-G., {Kang}, H., {Jung}, J., {et~al.} 2019, Journal of Korean Astronomical Society, 52, 227

\bibitem[{{Jiao} {et~al.}(2021){Jiao}, {Gao}, \& {Zhao}}]{2021MNRAS.504.2360J}
{Jiao}, Q., {Gao}, Y., \& {Zhao}, Y. 2021, \mnras, 504, 2360, \dodoi{10.1093/mnras/stab1035}

\bibitem[{{Jim{\'e}nez-Donaire} {et~al.}(2019){Jim{\'e}nez-Donaire}, {Bigiel}, {Leroy}, {Usero}, {Cormier}, {Puschnig}, {Gallagher}, {Kepley}, {Bolatto}, {Garc{\'\i}a-Burillo}, {Hughes}, {Kramer}, {Pety}, {Schinnerer}, {Schruba}, {Schuster}, \& {Walter}}]{2019ApJ...880..127J}
{Jim{\'e}nez-Donaire}, M.~J., {Bigiel}, F., {Leroy}, A.~K., {et~al.} 2019, \apj, 880, 127, \dodoi{10.3847/1538-4357/ab2b95}

\bibitem[{{Joye} \& {Mandel}(2003)}]{2003ASPC..295..489J}
{Joye}, W.~A., \& {Mandel}, E. 2003, in Astronomical Society of the Pacific Conference Series, Vol. 295, Astronomical Data Analysis Software and Systems XII, ed. H.~E. {Payne}, R.~I. {Jedrzejewski}, \& R.~N. {Hook}, 489

\bibitem[{{Kauffmann} {et~al.}(2017){Kauffmann}, {Goldsmith}, {Melnick}, {Tolls}, {Guzman}, \& {Menten}}]{2017A&A...605L...5K}
{Kauffmann}, J., {Goldsmith}, P.~F., {Melnick}, G., {et~al.} 2017, \aap, 605, L5, \dodoi{10.1051/0004-6361/201731123}

\bibitem[{{Kirsanova} {et~al.}(2008){Kirsanova}, {Sobolev}, {Thomasson}, {Wiebe}, {Johansson}, \& {Seleznev}}]{2008MNRAS.388..729K}
{Kirsanova}, M.~S., {Sobolev}, A.~M., {Thomasson}, M., {et~al.} 2008, \mnras, 388, 729, \dodoi{10.1111/j.1365-2966.2008.13429.x}

\bibitem[{{Lacy} {et~al.}(2017){Lacy}, {Sneden}, {Kim}, \& {Jaffe}}]{2017ApJ...838...66L}
{Lacy}, J.~H., {Sneden}, C., {Kim}, H., \& {Jaffe}, D.~T. 2017, \apj, 838, 66, \dodoi{10.3847/1538-4357/aa6247}

\bibitem[{{Lada} {et~al.}(2012){Lada}, {Forbrich}, {Lombardi}, \& {Alves}}]{2012ApJ...745..190L}
{Lada}, C.~J., {Forbrich}, J., {Lombardi}, M., \& {Alves}, J.~F. 2012, \apj, 745, 190, \dodoi{10.1088/0004-637X/745/2/190}

\bibitem[{{Lada} {et~al.}(2010){Lada}, {Lombardi}, \& {Alves}}]{2010ApJ...724..687L}
{Lada}, C.~J., {Lombardi}, M., \& {Alves}, J.~F. 2010, \apj, 724, 687, \dodoi{10.1088/0004-637X/724/1/687}

\bibitem[{{Lee} {et~al.}(2024){Lee}, {Koda}, {Hirota}, {Egusa}, \& {Heyer}}]{2024ApJ...968...97L}
{Lee}, A.~M., {Koda}, J., {Hirota}, A., {Egusa}, F., \& {Heyer}, M. 2024, \apj, 968, 97, \dodoi{10.3847/1538-4357/ad40a0}

\bibitem[{{Lemasle} {et~al.}(2018){Lemasle}, {Hajdu}, {Kovtyukh}, {Inno}, {Grebel}, {Catelan}, {Bono}, {Fran{\c{c}}ois}, {Kniazev}, {da Silva}, \& {Storm}}]{2018A&A...618A.160L}
{Lemasle}, B., {Hajdu}, G., {Kovtyukh}, V., {et~al.} 2018, \aap, 618, A160, \dodoi{10.1051/0004-6361/201834050}

\bibitem[{{Liu} {et~al.}(2015){Liu}, {Gao}, {Isaak}, {Daddi}, {Yang}, {Lu}, \& {van der Werf}}]{2015ApJ...810L..14L}
{Liu}, D., {Gao}, Y., {Isaak}, K., {et~al.} 2015, \apjl, 810, L14, \dodoi{10.1088/2041-8205/810/2/L14}

\bibitem[{{Longmore} {et~al.}(2013){Longmore}, {Bally}, {Testi}, {Purcell}, {Walsh}, {Bressert}, {Pestalozzi}, {Molinari}, {Ott}, {Cortese}, {Battersby}, {Murray}, {Lee}, {Kruijssen}, {Schisano}, \& {Elia}}]{2013MNRAS.429..987L}
{Longmore}, S.~N., {Bally}, J., {Testi}, L., {et~al.} 2013, \mnras, 429, 987, \dodoi{10.1093/mnras/sts376}

\bibitem[{{Marsh} {et~al.}(2015){Marsh}, {Whitworth}, \& {Lomax}}]{2015MNRAS.454.4282M}
{Marsh}, K.~A., {Whitworth}, A.~P., \& {Lomax}, O. 2015, \mnras, 454, 4282, \dodoi{10.1093/mnras/stv2248}

\bibitem[{{Marsh} {et~al.}(2017){Marsh}, {Whitworth}, {Lomax}, {Ragan}, {Becciani}, {Cambr{\'e}sy}, {Di Giorgio}, {Eden}, {Elia}, {Kacsuk}, {Molinari}, {Palmeirim}, {Pezzuto}, {Schneider}, {Sciacca}, \& {Vitello}}]{2017MNRAS.471.2730M}
{Marsh}, K.~A., {Whitworth}, A.~P., {Lomax}, O., {et~al.} 2017, \mnras, 471, 2730, \dodoi{10.1093/mnras/stx1723}

\bibitem[{{Mattern} {et~al.}(2024){Mattern}, {Andr{\'e}}, {Zavagno}, {Russeil}, {Roussel}, {Peretto}, {Schuller}, {Shimajiri}, {Di Francesco}, {Arzoumanian}, {Rev{\'e}ret}, \& {De Breuck}}]{2024A&A...688A.163M}
{Mattern}, M., {Andr{\'e}}, P., {Zavagno}, A., {et~al.} 2024, \aap, 688, A163, \dodoi{10.1051/0004-6361/202449908}

\bibitem[{{M{\'e}ndez-Delgado} {et~al.}(2022){M{\'e}ndez-Delgado}, {Amayo}, {Arellano-C{\'o}rdova}, {Esteban}, {Garc{\'\i}a-Rojas}, {Carigi}, \& {Delgado-Inglada}}]{2022MNRAS.510.4436M}
{M{\'e}ndez-Delgado}, J.~E., {Amayo}, A., {Arellano-C{\'o}rdova}, K.~Z., {et~al.} 2022, \mnras, 510, 4436, \dodoi{10.1093/mnras/stab3782}

\bibitem[{{Mirocha} {et~al.}(2021){Mirocha}, {Karska}, {Gronowski}, {Kristensen}, {Tychoniec}, {Harsono}, {Figueira}, {G{\l}adkowski}, \& {{\.Z}{\'o}{\l}towski}}]{2021A&A...656A.146M}
{Mirocha}, A., {Karska}, A., {Gronowski}, M., {et~al.} 2021, \aap, 656, A146, \dodoi{10.1051/0004-6361/202140833}

\bibitem[{{Molinari} {et~al.}(2010){Molinari}, {Swinyard}, {Bally}, {Barlow}, {Bernard}, {Martin}, {Moore}, {Noriega-Crespo}, {Plume}, {Testi}, {Zavagno}, {Abergel}, {Ali}, {Andr{\'e}}, {Baluteau}, {Benedettini}, {Bern{\'e}}, {Billot}, {Blommaert}, {Bontemps}, {Boulanger}, {Brand}, {Brunt}, {Burton}, {Campeggio}, {Carey}, {Caselli}, {Cesaroni}, {Cernicharo}, {Chakrabarti}, {Chrysostomou}, {Codella}, {Cohen}, {Compiegne}, {Davis}, {de Bernardis}, {de Gasperis}, {Di Francesco}, {di Giorgio}, {Elia}, {Faustini}, {Fischera}, {Fukui}, {Fuller}, {Ganga}, {Garcia-Lario}, {Giard}, {Giardino}, {Glenn}, {Goldsmith}, {Griffin}, {Hoare}, {Huang}, {Jiang}, {Joblin}, {Joncas}, {Juvela}, {Kirk}, {Lagache}, {Li}, {Lim}, {Lord}, {Lucas}, {Maiolo}, {Marengo}, {Marshall}, {Masi}, {Massi}, {Matsuura}, {Meny}, {Minier}, {Miville-Desch{\^e}nes}, {Montier}, {Motte}, {M{\"u}ller}, {Natoli}, {Neves}, {Olmi}, {Paladini}, {Paradis}, {Pestalozzi}, {Pezzuto}, {Piacentini}, {Pomar{\`e}s}, {Popescu}, {Reach}, {Richer}, {Ristorcelli},
  {Roy}, {Royer}, {Russeil}, {Saraceno}, {Sauvage}, {Schilke}, {Schneider-Bontemps}, {Schuller}, {Schultz}, {Shepherd}, {Sibthorpe}, {Smith}, {Smith}, {Spinoglio}, {Stamatellos}, {Strafella}, {Stringfellow}, {Sturm}, {Taylor}, {Thompson}, {Tuffs}, {Umana}, {Valenziano}, {Vavrek}, {Viti}, {Waelkens}, {Ward-Thompson}, {White}, {Wyrowski}, {Yorke}, \& {Zhang}}]{2010PASP..122..314M}
{Molinari}, S., {Swinyard}, B., {Bally}, J., {et~al.} 2010, \pasp, 122, 314, \dodoi{10.1086/651314}

\bibitem[{{Neumann} {et~al.}(2023){Neumann}, {Gallagher}, {Bigiel}, {Leroy}, {Barnes}, {Usero}, {den Brok}, {Belfiore}, {Be{\v{s}}li{\'c}}, {Cao}, {Chevance}, {Dale}, {Eibensteiner}, {Glover}, {Grasha}, {Henshaw}, {Jim{\'e}nez-Donaire}, {Klessen}, {Kruijssen}, {Liu}, {Meidt}, {Pety}, {Puschnig}, {Querejeta}, {Rosolowsky}, {Schinnerer}, {Schruba}, {Sormani}, {Sun}, {Teng}, \& {Williams}}]{2023MNRAS.521.3348N}
{Neumann}, L., {Gallagher}, M.~J., {Bigiel}, F., {et~al.} 2023, \mnras, 521, 3348, \dodoi{10.1093/mnras/stad424}

\bibitem[{{Nishimura} {et~al.}(2016{\natexlab{a}}){Nishimura}, {Shimonishi}, {Watanabe}, {Sakai}, {Aikawa}, {Kawamura}, \& {Yamamoto}}]{2016ApJ...818..161N}
{Nishimura}, Y., {Shimonishi}, T., {Watanabe}, Y., {et~al.} 2016{\natexlab{a}}, \apj, 818, 161, \dodoi{10.3847/0004-637X/818/2/161}

\bibitem[{{Nishimura} {et~al.}(2016{\natexlab{b}}){Nishimura}, {Shimonishi}, {Watanabe}, {Sakai}, {Aikawa}, {Kawamura}, \& {Yamamoto}}]{2016ApJ...829...94N}
---. 2016{\natexlab{b}}, \apj, 829, 94, \dodoi{10.3847/0004-637X/829/2/94}

\bibitem[{{Patra} {et~al.}(2022){Patra}, {Evans}, {Kim}, {Heyer}, {Kauffmann}, {Jose}, {Samal}, \& {Das}}]{2022AJ....164..129P}
{Patra}, S., {Evans}, Neal~J., I., {Kim}, K.-T., {et~al.} 2022, \aj, 164, 129, \dodoi{10.3847/1538-3881/ac83af}

\bibitem[{{Patra} {et~al.}(2023){Patra}, {Evans}, {Jose}, {Kim}, {Heyer}, {Kauffmann}, {Samal}, \& {Das}}]{2023pcsf.conf...63P}
{Patra}, S., {Evans}, N., {Jose}, J., {et~al.} 2023, in Physics and Chemistry of Star Formation: The Dynamical ISM Across Time and Spatial Scales, ed. V.~{Ossenkopf-Okada}, R.~{Schaaf}, I.~{Breloy}, \& J.~{Stutzki}, 63

\bibitem[{{Patra} {et~al.}(2024){Patra}, {Jose}, \& {Evans}}]{2024ApJ...970...88P}
{Patra}, S., {Jose}, J., \& {Evans}, N.~J. 2024, \apj, 970, 88, \dodoi{10.3847/1538-4357/ad4996}

\bibitem[{{Pety}(2018)}]{2018ssdd.confE..11P}
{Pety}, J. 2018, in Submillimetre Single-dish Data Reduction and Array Combination Techniques, 11, \dodoi{10.5281/zenodo.1205423}

\bibitem[{{Pety} {et~al.}(2017){Pety}, {Guzm{\'a}n}, {Orkisz}, {Liszt}, {Gerin}, {Bron}, {Bardeau}, {Goicoechea}, {Gratier}, {Le Petit}, {Levrier}, {{\"O}berg}, {Roueff}, \& {Sievers}}]{2017A&A...599A..98P}
{Pety}, J., {Guzm{\'a}n}, V.~V., {Orkisz}, J.~H., {et~al.} 2017, \aap, 599, A98, \dodoi{10.1051/0004-6361/201629862}

\bibitem[{{Pineda} {et~al.}(2024){Pineda}, {Horiuchi}, {Anderson}, {Luisi}, {Langer}, {Goldsmith}, {Kuiper}, {Fischer}, {Gong}, {Brunthaler}, {Rugel}, \& {Menten}}]{2024ApJ...973...89P}
{Pineda}, J.~L., {Horiuchi}, S., {Anderson}, L.~D., {et~al.} 2024, \apj, 973, 89, \dodoi{10.3847/1538-4357/ad615a}

\bibitem[{{Pontoppidan} {et~al.}(2014){Pontoppidan}, {Salyk}, {Bergin}, {Brittain}, {Marty}, {Mousis}, \& {{\"O}berg}}]{2014prpl.conf..363P}
{Pontoppidan}, K.~M., {Salyk}, C., {Bergin}, E.~A., {et~al.} 2014, in Protostars and Planets VI, ed. H.~{Beuther}, R.~S. {Klessen}, C.~P. {Dullemond}, \& T.~{Henning}, 363, \dodoi{10.2458/azu_uapress_9780816531240-ch016}

\bibitem[{{R{\'e}my-Ruyer} {et~al.}(2014){R{\'e}my-Ruyer}, {Madden}, {Galliano}, {Galametz}, {Takeuchi}, {Asano}, {Zhukovska}, {Lebouteiller}, {Cormier}, {Jones}, {Bocchio}, {Baes}, {Bendo}, {Boquien}, {Boselli}, {DeLooze}, {Doublier-Pritchard}, {Hughes}, {Karczewski}, \& {Spinoglio}}]{2014A&A...563A..31R}
{R{\'e}my-Ruyer}, A., {Madden}, S.~C., {Galliano}, F., {et~al.} 2014, \aap, 563, A31, \dodoi{10.1051/0004-6361/201322803}

\bibitem[{{Rosolowsky} {et~al.}(2010){Rosolowsky}, {Dunham}, {Ginsburg}, {Bradley}, {Aguirre}, {Bally}, {Battersby}, {Cyganowski}, {Dowell}, {Drosback}, {Evans}, {Glenn}, {Harvey}, {Stringfellow}, {Walawender}, \& {Williams}}]{2010ApJS..188..123R}
{Rosolowsky}, E., {Dunham}, M.~K., {Ginsburg}, A., {et~al.} 2010, \apjs, 188, 123, \dodoi{10.1088/0067-0049/188/1/123}

\bibitem[{{Russell} \& {Dopita}(1992)}]{1992ApJ...384..508R}
{Russell}, S.~C., \& {Dopita}, M.~A. 1992, \apj, 384, 508, \dodoi{10.1086/170893}

\bibitem[{{Santa-Maria} {et~al.}(2023){Santa-Maria}, {Goicoechea}, {Pety}, {Gerin}, {Orkisz}, {Le Petit}, {Einig}, {Palud}, {de Souza Magalhaes}, {Be{\v{s}}li{\'c}}, {Segal}, {Bardeau}, {Bron}, {Chainais}, {Chanussot}, {Gratier}, {Guzm{\'a}n}, {Hughes}, {Languignon}, {Levrier}, {Lis}, {Liszt}, {Le Bourlot}, {Oya}, {{\"O}berg}, {Peretto}, {Roueff}, {Roueff}, {Sievers}, {Thouvenin}, \& {Yamamoto}}]{2023A&A...679A...4S}
{Santa-Maria}, M.~G., {Goicoechea}, J.~R., {Pety}, J., {et~al.} 2023, \aap, 679, A4, \dodoi{10.1051/0004-6361/202346598}

\bibitem[{{Sharma} {et~al.}(2022){Sharma}, {Chen}, {Panwar}, {Sun}, \& {Gao}}]{2022ApJ...928...17S}
{Sharma}, T., {Chen}, W.~P., {Panwar}, N., {Sun}, Y., \& {Gao}, Y. 2022, \apj, 928, 17, \dodoi{10.3847/1538-4357/ac510b}

\bibitem[{{Shimajiri} {et~al.}(2017){Shimajiri}, {Andr{\'e}}, {Braine}, {K{\"o}nyves}, {Schneider}, {Bontemps}, {Ladjelate}, {Roy}, {Gao}, \& {Chen}}]{2017A&A...604A..74S}
{Shimajiri}, Y., {Andr{\'e}}, P., {Braine}, J., {et~al.} 2017, \aap, 604, A74, \dodoi{10.1051/0004-6361/201730633}

\bibitem[{{Shimonishi}(2024)}]{2024arXiv241104451S}
{Shimonishi}, T. 2024, arXiv e-prints, arXiv:2411.04451.
\newblock \doarXiv{2411.04451}

\bibitem[{{Shirley}(2015)}]{2015PASP..127..299S}
{Shirley}, Y.~L. 2015, \pasp, 127, 299, \dodoi{10.1086/680342}

\bibitem[{{Smithsonian Astrophysical Observatory}(2000)}]{2000ascl.soft03002S}
{Smithsonian Astrophysical Observatory}. 2000, {SAOImage DS9: A utility for displaying astronomical images in the X11 window environment}, Astrophysics Source Code Library, record ascl:0003.002.
\newblock \doeprint{0003.002}

\bibitem[{{Sun} {et~al.}(2020){Sun}, {Leroy}, {Schinnerer}, {Hughes}, {Rosolowsky}, {Querejeta}, {Schruba}, {Liu}, {Saito}, {Herrera}, {Faesi}, {Usero}, {Pety}, {Kruijssen}, {Ostriker}, {Bigiel}, {Blanc}, {Bolatto}, {Boquien}, {Chevance}, {Dale}, {Deger}, {Emsellem}, {Glover}, {Grasha}, {Groves}, {Henshaw}, {Jimenez-Donaire}, {Kim}, {Klessen}, {Kreckel}, {Lee}, {Meidt}, {Sandstrom}, {Sardone}, {Utomo}, \& {Williams}}]{2020ApJ...901L...8S}
{Sun}, J., {Leroy}, A.~K., {Schinnerer}, E., {et~al.} 2020, \apjl, 901, L8, \dodoi{10.3847/2041-8213/abb3be}

\bibitem[{{Tafalla} {et~al.}(2023){Tafalla}, {Usero}, \& {Hacar}}]{2023A&A...679A.112T}
{Tafalla}, M., {Usero}, A., \& {Hacar}, A. 2023, \aap, 679, A112, \dodoi{10.1051/0004-6361/202346136}

\bibitem[{{Tan} {et~al.}(2018){Tan}, {Gao}, {Zhang}, {Greve}, {Jiang}, {Wilson}, {Yang}, {Bemis}, {Chung}, {Matsushita}, {Shi}, {Ao}, {Brinks}, {Currie}, {Davis}, {de Grijs}, {Ho}, {Imanishi}, {Kohno}, {Lee}, {Parsons}, {Rawlings}, {Rigopoulou}, {Rosolowsky}, {Bulger}, {Chen}, {Chapman}, {Eden}, {Gear}, {Gu}, {He}, {Jiao}, {Liu}, {Liu}, {Li}, {Micha{\l}owski}, {Nguyen-Luong}, {Qiu}, {Smith}, {Violino}, {Wang}, {Wang}, {Wang}, {Yeh}, {Zhao}, \& {Zhu}}]{2018ApJ...860..165T}
{Tan}, Q.-H., {Gao}, Y., {Zhang}, Z.-Y., {et~al.} 2018, \apj, 860, 165, \dodoi{10.3847/1538-4357/aac512}

\bibitem[{{Teng} {et~al.}(2024){Teng}, {Chiang}, {Sandstrom}, {Sun}, {Leroy}, {Bolatto}, {Usero}, {Ostriker}, {Querejeta}, {Chastenet}, {Bigiel}, {Boquien}, {den Brok}, {Cao}, {Chevance}, {Chown}, {Colombo}, {Eibensteiner}, {Glover}, {Grasha}, {Henshaw}, {Jim{\'e}nez-Donaire}, {Liu}, {Murphy}, {Pan}, {Stuber}, \& {Williams}}]{2024ApJ...961...42T}
{Teng}, Y.-H., {Chiang}, I.-D., {Sandstrom}, K.~M., {et~al.} 2024, \apj, 961, 42, \dodoi{10.3847/1538-4357/ad10ae}

\bibitem[{{The Astropy Collaboration} {et~al.}(2022){The Astropy Collaboration}, {Price-Whelan}, {Lian Lim}, {Earl}, {Starkman}, {Bradley}, {Shupe}, {Patil}, {Corrales}, {Brasseur}, {N{\"o}the}, {Donath}, {Tollerud}, {Morris}, {Ginsburg}, {Vaher}, {Weaver}, {Tocknell}, {Jamieson}, {van Kerkwijk}, {Robitaille}, {Merry}, {Bachetti}, {G{\"u}nther}, {Aldcroft}, {Alvarado-Montes}, {Archibald}, {B{\'o}di}, {Bapat}, {Barentsen}, {Baz{\'a}n}, {Biswas}, {Boquien}, {Burke}, {Cara}, {Cara}, {E Conroy}, {Conseil}, {Craig}, {Cross}, {Cruz}, {D'Eugenio}, {Dencheva}, {Devillepoix}, {Dietrich}, {Davis Eigenbrot}, {Erben}, {Ferreira}, {Foreman-Mackey}, {Fox}, {Freij}, {Garg}, {Geda}, {Glattly}, {Gondhalekar}, {Gordon}, {Grant}, {Greenfield}, {Groener}, {Guest}, {Gurovich}, {Handberg}, {Hart}, {Hatfield-Dodds}, {Homeier}, {Hosseinzadeh}, {Jenness}, {Jones}, {Joseph}, {Bryce Kalmbach}, {Karamehmetoglu}, {Ka{\l}uszy{\'n}ski}, {Kelley}, {Kern}, {Kerzendorf}, {Koch}, {Kulumani}, {Lee}, {Ly}, {Ma}, {MacBride}, {Maljaars}, {Muna},
  {Murphy}, {Norman}, {O'Steen}, {Oman}, {Pacifici}, {Pascual}, {Pascual-Granado}, {Patil}, {Perren}, {Pickering}, {Rastogi}, {Roulston}, {Ryan}, {Rykoff}, {Sabater}, {Sakurikar}, {Salgado}, {Sanghi}, {Saunders}, {Savchenko}, {Schwardt}, {Seifert-Eckert}, {Shih}, {Shrey Jain}, {Shukla}, {Sick}, {Simpson}, {Singanamalla}, {Singer}, {Singhal}, {Sinha}, {Sip{\H{o}}cz}, {Spitler}, {Stansby}, {Streicher}, {{\v{S}}umak}, {Swinbank}, {Taranu}, {Tewary}, {Tremblay}, {de Val-Borro}, {Van Kooten}, {Vasovi{\'c}}, {Verma}, {Cardoso}, {Williams}, {Wilson}, {Winkel}, {Wood-Vasey}, {Xue}, {Yoachim}, {ZHANG}, \& {Zonca}}]{2022arXiv220614220T}
{The Astropy Collaboration}, {Price-Whelan}, A.~M., {Lian Lim}, P., {et~al.} 2022, arXiv e-prints, arXiv:2206.14220.
\newblock \doarXiv{2206.14220}

\bibitem[{{Urquhart} {et~al.}(2018){Urquhart}, {K{\"o}nig}, {Giannetti}, {Leurini}, {Moore}, {Eden}, {Pillai}, {Thompson}, {Braiding}, {Burton}, {Csengeri}, {Dempsey}, {Figura}, {Froebrich}, {Menten}, {Schuller}, {Smith}, \& {Wyrowski}}]{2018MNRAS.473.1059U}
{Urquhart}, J.~S., {K{\"o}nig}, C., {Giannetti}, A., {et~al.} 2018, \mnras, 473, 1059, \dodoi{10.1093/mnras/stx2258}

\bibitem[{{Urquhart} {et~al.}(2024){Urquhart}, {K{\"o}nig}, {Colombo}, {Karska}, {Wyrowski}, {Menten}, {Moore}, {Brand}, {Elia}, {Giannetti}, {Leurini}, {Figueira}, {Lee}, \& {Dumke}}]{2024MNRAS.528.4746U}
{Urquhart}, J.~S., {K{\"o}nig}, C., {Colombo}, D., {et~al.} 2024, \mnras, 528, 4746, \dodoi{10.1093/mnras/stad3983}

\bibitem[{{van der Walt} {et~al.}(2011){van der Walt}, {Colbert}, \& {Varoquaux}}]{2011CSE....13b..22V}
{van der Walt}, S., {Colbert}, S.~C., \& {Varoquaux}, G. 2011, Computing in Science and Engineering, 13, 22, \dodoi{10.1109/MCSE.2011.37}

\bibitem[{{Vutisalchavakul} {et~al.}(2016){Vutisalchavakul}, {Evans}, \& {Heyer}}]{2016ApJ...831...73V}
{Vutisalchavakul}, N., {Evans}, Neal~J., I., \& {Heyer}, M. 2016, \apj, 831, 73, \dodoi{10.3847/0004-637X/831/1/73}

\bibitem[{{Wang} {et~al.}(2018){Wang}, {Luo}, {Hou}, {Wang}, {Du}, {Qin}, \& {Han}}]{2018PASP..130k4301W}
{Wang}, L.-L., {Luo}, A.~L., {Hou}, W., {et~al.} 2018, \pasp, 130, 114301, \dodoi{10.1088/1538-3873/aadf22}

\bibitem[{{Wells} {et~al.}(2022){Wells}, {Urquhart}, {Moore}, {Browning}, {Ragan}, {Rigby}, {Eden}, \& {Thompson}}]{2022MNRAS.516.4245W}
{Wells}, M.~R.~A., {Urquhart}, J.~S., {Moore}, T.~J.~T., {et~al.} 2022, \mnras, 516, 4245, \dodoi{10.1093/mnras/stac2420}

\bibitem[{{Whittet} {et~al.}(2013){Whittet}, {Poteet}, {Chiar}, {Pagani}, {Bajaj}, {Horne}, {Shenoy}, \& {Adamson}}]{2013ApJ...774..102W}
{Whittet}, D.~C.~B., {Poteet}, C.~A., {Chiar}, J.~E., {et~al.} 2013, \apj, 774, 102, \dodoi{10.1088/0004-637X/774/2/102}

\bibitem[{{Wu} {et~al.}(2010){Wu}, {Evans}, {Shirley}, \& {Knez}}]{2010ApJS..188..313W}
{Wu}, J., {Evans}, Neal~J., I., {Shirley}, Y.~L., \& {Knez}, C. 2010, \apjs, 188, 313, \dodoi{10.1088/0067-0049/188/2/313}

\bibitem[{{Wu} {et~al.}(2005){Wu}, {Evans}, {Gao}, {Solomon}, {Shirley}, \& {Vanden Bout}}]{Wu:2005}
{Wu}, J., {Evans}, II, N.~J., {Gao}, Y., {et~al.} 2005, \apjl, 635, L173, \dodoi{10.1086/499623}

\bibitem[{{Yasui} {et~al.}(2023){Yasui}, {Kobayashi}, {Saito}, {Izumi}, \& {Ikeda}}]{2023ApJ...943..137Y}
{Yasui}, C., {Kobayashi}, N., {Saito}, M., {Izumi}, N., \& {Ikeda}, Y. 2023, \apj, 943, 137, \dodoi{10.3847/1538-4357/ac94d5}

\bibitem[{{Yun} {et~al.}(2021){Yun}, {Lee}, {Choi}, {Evans}, {Offner}, {Heyer}, {Gaches}, {Lee}, {Baek}, {Choi}, {Kang}, {Lee}, {Tatematsu}, {Yang}, {Chen}, {Lee}, {Jung}, {Lee}, \& {Cho}}]{2021ApJS..256...16Y}
{Yun}, H.-S., {Lee}, J.-E., {Choi}, Y., {et~al.} 2021, \apjs, 256, 16, \dodoi{10.3847/1538-4365/ac090e}

\bibitem[{{Yun} {et~al.}(2024){Yun}, {Lee}, {Choi}, {Evans}, {Offner}, {Heyer}, {Gaches}, {Lee}, {Baek}, {Choi}, {Kang}, {Lee}, {Tatematsu}, {Yang}, {Chen}, {Lee}, {Jung}, {Lee}, \& {Cho}}]{2024ApJS..271...36Y}
---. 2024, \apjs, 271, 36, \dodoi{10.3847/1538-4365/ad20f1}

\bibitem[{{Zhang} {et~al.}(2014){Zhang}, {Gao}, {Henkel}, {Zhao}, {Wang}, {Menten}, \& {G{\"u}sten}}]{2014ApJ...784L..31Z}
{Zhang}, Z.-Y., {Gao}, Y., {Henkel}, C., {et~al.} 2014, \apjl, 784, L31, \dodoi{10.1088/2041-8205/784/2/L31}

\end{thebibliography}
\bibliographystyle{aasjournal}



\end{document}